\input harvmac.tex
\input epsf.tex
\input amssym
\input ulem.sty
\input amssym

\input graphicx.tex


\let\includefigures=\iftrue
\let\useblackboard=\iftrue
\newfam\black

\def\figin{\epsfcheck\figin}\def\figins{\epsfcheck\figins}
\def\epsfcheck{\ifx\epsfbox\UnDeFiNeD
\message{(NO epsf.tex, FIGURES WILL BE IGNORED)}
\gdef\figin##1{\vskip2in}\gdef\figins##1{\hskip.5in}
\else\message{(FIGURES WILL BE INCLUDED)}%
\gdef\figin##1{##1}\gdef\figins##1{##1}\fi}
\def\DefWarn#1{}
\def\figinsert{\goodbreak\midinsert}
\def\ifig#1#2#3{\DefWarn#1\xdef#1{fig.~\the\figno}
\writedef{#1\leftbracket fig.\noexpand~\the\figno} %
\figinsert\figin{\centerline{#3}}\medskip\centerline{\vbox{\baselineskip12pt
\advance\hsize by -1truein\noindent\footnotefont{\bf
Fig.~\the\figno:} #2}}
\bigskip\endinsert\global\advance\figno by1}


\includefigures
\message{If you do not have epsf.tex (to include figures),}
\message{change the option at the top of the tex file.}
\input epsf
\def\figin{\epsfcheck\figin}\def\figins{\epsfcheck\figins}
\def\epsfcheck{\ifx\epsfbox\UnDeFiNeD
\message{(NO epsf.tex, FIGURES WILL BE IGNORED)}
\gdef\figin##1{\vskip2in}\gdef\figins##1{\hskip.5in}
\else\message{(FIGURES WILL BE INCLUDED)}%
\gdef\figin##1{##1}\gdef\figins##1{##1}\fi}
\def\DefWarn#1{}
\def\figinsert{\goodbreak\midinsert}
\def\ifig#1#2#3{\DefWarn#1\xdef#1{fig.~\the\figno}
\writedef{#1\leftbracket fig.\noexpand~\the\figno}%
\figinsert\figin{\centerline{#3}}\medskip\centerline{\vbox{
\baselineskip12pt\advance\hsize by -1truein
\noindent\footnotefont{\bf Fig.~\the\figno:} #2}}
\endinsert\global\advance\figno by1}
\else
\def\ifig#1#2#3{\xdef#1{fig.~\the\figno}
\writedef{#1\leftbracket fig.\noexpand~\the\figno}%
\global\advance\figno by1} \fi

\def\figin{\epsfcheck\figin}\def\figins{\epsfcheck\figins}
\def\epsfcheck{\ifx\epsfbox\UnDeFiNeD
\message{(NO epsf.tex, FIGURES WILL BE IGNORED)}
\gdef\figin##1{\vskip2in}\gdef\figins##1{\hskip.5in}
\else\message{(FIGURES WILL BE INCLUDED)}%
\gdef\figin##1{##1}\gdef\figins##1{##1}\fi}
\def\DefWarn#1{}
\def\figinsert{\goodbreak\midinsert}
\def\ifig#1#2#3{\DefWarn#1\xdef#1{fig.~\the\figno}
\writedef{#1\leftbracket fig.\noexpand~\the\figno} %
\figinsert\figin{\centerline{#3}}\medskip\centerline{\vbox{\baselineskip12pt
\advance\hsize by -1truein\noindent\footnotefont{\bf
Fig.~\the\figno:} #2}}
\bigskip\endinsert\global\advance\figno by1}

\def \eps {\epsilon}
\def \la {\langle}
\def \ra {\rangle}
\def \p {\partial}

\def \e {\epsilon}
\def \b {\bar}
\def \t {\widetilde}

\def\G{{\cal G}}

\lref\IgiZZ{
  K.~Igi,
  ``pi-N Scattering Length and Singularities in the Complex J Plane,''
Phys.\ Rev.\ Lett.\  {\bf 9}, 76 (1962)..
}

\lref\CaronHuotVEP{
  S.~Caron-Huot,
  ``Analyticity in Spin in Conformal Theories,''
JHEP {\bf 1709}, 078 (2017).
[arXiv:1703.00278 [hep-th]].
}

\lref\LogunovDY{
  A.~A.~Logunov, L.~D.~Soloviev and A.~N.~Tavkhelidze,
  ``Dispersion sum rules and high-energy scattering,''
Phys.\ Lett.\  {\bf 24B}, 181 (1967).
}

\lref\IgiZZA{
  K.~Igi and S.~Matsuda,
  ``New Sum Rules and Singularities in the Complex J Plane,''
Phys.\ Rev.\ Lett.\  {\bf 18}, 625 (1967).
}

\lref\DolenJR{
  R.~Dolen, D.~Horn and C.~Schmid,
  ``Finite energy sum rules and their application to pi N charge exchange,''
Phys.\ Rev.\  {\bf 166}, 1768 (1968).
}

\lref\PappadopuloJK{
  D.~Pappadopulo, S.~Rychkov, J.~Espin and R.~Rattazzi,
  ``OPE Convergence in Conformal Field Theory,''
Phys.\ Rev.\ D {\bf 86}, 105043 (2012).
[arXiv:1208.6449 [hep-th]].
}

\lref\VenezianoYB{
  G.~Veneziano,
  ``Construction of a crossing - symmetric, Regge behaved amplitude for linearly rising trajectories,''
Nuovo Cim.\ A {\bf 57}, 190 (1968)..
}

\lref\QiaoXIF{
  J.~Qiao and S.~Rychkov,
  ``A tauberian theorem for the conformal bootstrap,''
JHEP {\bf 1712}, 119 (2017).
[arXiv:1709.00008 [hep-th]].
}

\lref\CornalbaXM{
  L.~Cornalba, M.~S.~Costa, J.~Penedones and R.~Schiappa,
  ``Eikonal Approximation in AdS/CFT: Conformal Partial Waves and Finite N Four-Point Functions,''
Nucl.\ Phys.\ B {\bf 767}, 327 (2007).
[hep-th/0611123].
}

\lref\KulaxiziIXA{
  M.~Kulaxizi, A.~Parnachev and A.~Zhiboedov,
  ``Bulk Phase Shift, CFT Regge Limit and Einstein Gravity,''
[arXiv:1705.02934 [hep-th]].
}

\lref\HarlowNY{
  D.~Harlow, J.~Maltz and E.~Witten,
  ``Analytic Continuation of Liouville Theory,''
JHEP {\bf 1112}, 071 (2011).
[arXiv:1108.4417 [hep-th]].
}

\lref\QiaoXIF{
  J.~Qiao and S.~Rychkov,
  ``A tauberian theorem for the conformal bootstrap,''
JHEP {\bf 1712}, 119 (2017).
[arXiv:1709.00008 [hep-th]].
}

\lref\IliesiuFAO{
  L.~Iliesiu, M.~Kolo?lu, R.~Mahajan, E.~Perlmutter and D.~Simmons-Duffin,
  ``The Conformal Bootstrap at Finite Temperature,''
[arXiv:1802.10266 [hep-th]].
}

\lref\SimmonsDuffinNUB{
  D.~Simmons-Duffin, D.~Stanford and E.~Witten,
  ``A spacetime derivation of the Lorentzian OPE inversion formula,''
[arXiv:1711.03816 [hep-th]].
}

\lref\FitzpatrickDM{
  A.~L.~Fitzpatrick and J.~Kaplan,
  ``Unitarity and the Holographic S-Matrix,''
JHEP {\bf 1210}, 032 (2012).
[arXiv:1112.4845 [hep-th]].
}

\lref\RychkovLCA{
  S.~Rychkov and P.~Yvernay,
  ``Remarks on the Convergence Properties of the Conformal Block Expansion,''
Phys.\ Lett.\ B {\bf 753}, 682 (2016).
[arXiv:1510.08486 [hep-th]].
}

\lref\HeemskerkPN{
  I.~Heemskerk, J.~Penedones, J.~Polchinski and J.~Sully,
  ``Holography from Conformal Field Theory,''
JHEP {\bf 0910}, 079 (2009).
[arXiv:0907.0151 [hep-th]].
}

\lref\JafferisZNA{
  D.~Jafferis, B.~Mukhametzhanov and A.~Zhiboedov,
JHEP {\bf 1805}, 043 (2018).
[arXiv:1710.11161 [hep-th]].
}

\lref\KomargodskiEK{
  Z.~Komargodski and A.~Zhiboedov,
  ``Convexity and Liberation at Large Spin,''
JHEP {\bf 1311}, 140 (2013).
[arXiv:1212.4103 [hep-th]].
}

\lref\FitzpatrickYX{
  A.~L.~Fitzpatrick, J.~Kaplan, D.~Poland and D.~Simmons-Duffin,
  ``The Analytic Bootstrap and AdS Superhorizon Locality,''
JHEP {\bf 1312}, 004 (2013).
[arXiv:1212.3616 [hep-th]].
}

\lref\Postnikov{
A.~G.~Postnikov,
``Tauberian theory and its applications,''
  American Mathematical Soc. ,
  144, 1980
  }

\lref\Subhankulov{
  M.~A.~Subhankulov, F.~I.~An,
  ``Complex tauberian theorems for the one-sided and two-sided Stieltjes transform,''
1974 Math. USSR Izv. 8 145.
}

\lref\SubhankulovR{
  M.~A.~Subhankulov,
  ``Tauberian Theorems with Remainder'', ({\it in Russian})
1976, Izdat. Nauka, Moscow.
}

\lref\Korevaar{
  J.~Korevaar,
  ``Tauberian theory: a century of development.,''
Springer, 2004. 497 p.
}

\lref\Erdelyi{
  A.~Erd\'{e}lyi et.al.,
  ``Higher Transcendental Functions", vol. 1, McGraw Hill, New York, 1953.
}

\lref\RychkovIQZ{
  S.~Rychkov,
  ``EPFL Lectures on Conformal Field Theory in D $\geq$ 3 Dimensions,''
[arXiv:1601.05000 [hep-th]].
}

\lref\SimmonsDuffinGJK{
  D.~Simmons-Duffin,
  ``The Conformal Bootstrap,''
[arXiv:1602.07982 [hep-th]].
}

\lref\PolandEPD{
  D.~Poland, S.~Rychkov and A.~Vichi,
  ``The Conformal Bootstrap: Theory, Numerical Techniques, and Applications,''
[arXiv:1805.04405 [hep-th]].
}

\lref\AldayNJK{
  L.~F.~Alday,
  ``Large Spin Perturbation Theory for Conformal Field Theories,''
Phys.\ Rev.\ Lett.\  {\bf 119}, no. 11, 111601 (2017).
[arXiv:1611.01500 [hep-th]].
}

\lref\MazacQEV{
  D.~Mazac,
  ``Analytic bounds and emergence of AdS$_{2}$ physics from the conformal bootstrap,''
JHEP {\bf 1704}, 146 (2017).
[arXiv:1611.10060 [hep-th]].
}

\lref\MazacMDX{
  D.~Mazac and M.~F.~Paulos,
  ``The Analytic Functional Bootstrap I: 1D CFTs and 2D S-Matrices,''
[arXiv:1803.10233 [hep-th]].
}

\lref\GribovNW{
  V.~N.~Gribov,
  ``The theory of complex angular momenta: Gribov lectures on theoretical physics,''
  Cambridge University Press, 2007
}

\lref\CaronHuotNS{
  S.~Caron-Huot,
  ``Asymptotics of thermal spectral functions,''
Phys.\ Rev.\ D {\bf 79}, 125009 (2009).
[arXiv:0903.3958 [hep-ph]].
}

\lref\HellermanNRA{
  S.~Hellerman, D.~Orlando, S.~Reffert and M.~Watanabe,
  ``On the CFT Operator Spectrum at Large Global Charge,''
JHEP {\bf 1512}, 071 (2015).
[arXiv:1505.01537 [hep-th]].
}

\lref\MoninJMO{
  A.~Monin, D.~Pirtskhalava, R.~Rattazzi and F.~K.~Seibold,
  ``Semiclassics, Goldstone Bosons and CFT data,''
JHEP {\bf 1706}, 011 (2017).
[arXiv:1611.02912 [hep-th]].
}

\lref\AlvarezGaumeVFF{
  L.~Alvarez-Gaume, O.~Loukas, D.~Orlando and S.~Reffert,
  ``Compensating strong coupling with large charge,''
JHEP {\bf 1704}, 059 (2017).
[arXiv:1610.04495 [hep-th]].
}

\lref\GopakumarCPB{
  R.~Gopakumar, A.~Kaviraj, K.~Sen and A.~Sinha,
  ``A Mellin space approach to the conformal bootstrap,''
JHEP {\bf 1705}, 027 (2017).
[arXiv:1611.08407 [hep-th]].
}

\lref\FerraraYT{
  S.~Ferrara, A.~F.~Grillo and R.~Gatto,
  ``Tensor representations of conformal algebra and conformally covariant operator product expansion,''
Annals Phys.\  {\bf 76}, 161 (1973)..
}

\lref\PolyakovGS{
  A.~M.~Polyakov,
  ``Nonhamiltonian approach to conformal quantum field theory,''
Zh.\ Eksp.\ Teor.\ Fiz.\  {\bf 66}, 23 (1974), [Sov.\ Phys.\ JETP {\bf 39}, 9 (1974)]..
}

\lref\MaldacenaRE{
  J.~M.~Maldacena,
  ``The Large N limit of superconformal field theories and supergravity,''
Int.\ J.\ Theor.\ Phys.\  {\bf 38}, 1113 (1999), [Adv.\ Theor.\ Math.\ Phys.\  {\bf 2}, 231 (1998)].
[hep-th/9711200].
}

\lref\WittenQJ{
  E.~Witten,
  ``Anti-de Sitter space and holography,''
Adv.\ Theor.\ Math.\ Phys.\  {\bf 2}, 253 (1998).
[hep-th/9802150].
}

\lref\GubserBC{
  S.~S.~Gubser, I.~R.~Klebanov and A.~M.~Polyakov,
  ``Gauge theory correlators from noncritical string theory,''
Phys.\ Lett.\ B {\bf 428}, 105 (1998).
[hep-th/9802109].
}

\lref\HogervorstSFD{
  M.~Hogervorst and B.~C.~van Rees,
  ``Crossing symmetry in alpha space,''
JHEP {\bf 1711}, 193 (2017).
[arXiv:1702.08471 [hep-th]].
}

\lref\BelavinVU{
  A.~A.~Belavin, A.~M.~Polyakov and A.~B.~Zamolodchikov,
  ``Infinite Conformal Symmetry in Two-Dimensional Quantum Field Theory,''
Nucl.\ Phys.\ B {\bf 241}, 333 (1984)..
}

\lref\KatzRLA{
  E.~Katz, S.~Sachdev, E.~S.~Sørensen and W.~Witczak-Krempa,
  ``Conformal field theories at nonzero temperature: Operator product expansions, Monte Carlo, and holography,''
Phys.\ Rev.\ B {\bf 90}, no. 24, 245109 (2014).
[arXiv:1409.3841 [cond-mat.str-el]].
}

\lref\HartnollAPF{
  S.~A.~Hartnoll, A.~Lucas and S.~Sachdev,
    ``Holographic quantum matter,'' 
  [1612.07324   [hep-th]].
}

\lref\ShifmanJV{
  M.~A.~Shifman,
  ``Quark hadron duality,''
[hep-ph/0009131].
}

\lref\WittenKH{
  E.~Witten,
  ``Baryons in the 1/n Expansion,''
Nucl.\ Phys.\ B {\bf 160}, 57 (1979).
}

\lref\OoguriPDQ{
  H.~Ooguri and C.~Vafa,
  ``Non-supersymmetric AdS and the Swampland,''
Adv.\ Theor.\ Math.\ Phys.\  {\bf 21}, 1787 (2017).
[arXiv:1610.01533 [hep-th]].
}

\lref\HeemskerkPN{
  I.~Heemskerk, J.~Penedones, J.~Polchinski and J.~Sully,
  ``Holography from Conformal Field Theory,''
JHEP {\bf 0910}, 079 (2009).
[arXiv:0907.0151 [hep-th]].
}

\lref\AldayEWA{
  L.~F.~Alday and A.~Zhiboedov,
  ``An Algebraic Approach to the Analytic Bootstrap,''
JHEP {\bf 1704}, 157 (2017).
[arXiv:1510.08091 [hep-th]].
}

\lref\SimmonsDuffinWLQ{
  D.~Simmons-Duffin,
  ``The Lightcone Bootstrap and the Spectrum of the 3d Ising CFT,''
JHEP {\bf 1703}, 086 (2017).
[arXiv:1612.08471 [hep-th]].
}

\lref\GiddingsGJ{
  S.~B.~Giddings and R.~A.~Porto,
  ``The Gravitational S-matrix,''
Phys.\ Rev.\ D {\bf 81}, 025002 (2010).
[arXiv:0908.0004 [hep-th]].
}

\lref\FitzpatrickHU{
  A.~L.~Fitzpatrick and J.~Kaplan,
  ``Analyticity and the Holographic S-Matrix,''
JHEP {\bf 1210}, 127 (2012).
[arXiv:1111.6972 [hep-th]].
}

\lref\DasCNV{
  D.~Das, S.~Datta and S.~Pal,
  ``Modular crossings, OPE coefficients and black holes,''
[arXiv:1712.01842 [hep-th]].
}

\lref\LashkariVGJ{
  N.~Lashkari, A.~Dymarsky and H.~Liu,
  ``Eigenstate Thermalization Hypothesis in Conformal Field Theory,''
J.\ Stat.\ Mech.\  {\bf 1803}, no. 3, 033101 (2018).
[arXiv:1610.00302 [hep-th]].
}

\lref\CardyIE{
  J.~L.~Cardy,
  ``Operator Content of Two-Dimensional Conformally Invariant Theories,''
Nucl.\ Phys.\ B {\bf 270}, 186 (1986)..
}

\lref\HellermanBU{
  S.~Hellerman,
  ``A Universal Inequality for CFT and Quantum Gravity,''
JHEP {\bf 1108}, 130 (2011).
[arXiv:0902.2790 [hep-th]].
}

\lref\CollierCLS{
  S.~Collier, Y.~H.~Lin and X.~Yin,
  ``Modular Bootstrap Revisited,''
[arXiv:1608.06241 [hep-th]].
}

\lref\RattazziPE{
  R.~Rattazzi, V.~S.~Rychkov, E.~Tonni and A.~Vichi,
  ``Bounding scalar operator dimensions in 4D CFT,''
JHEP {\bf 0812}, 031 (2008).
[arXiv:0807.0004 [hep-th]].
}

\lref\AldayMF{
  L.~F.~Alday and J.~M.~Maldacena,
  ``Comments on operators with large spin,''
JHEP {\bf 0711}, 019 (2007).
[arXiv:0708.0672 [hep-th]].
}

\lref\KosBKA{
  F.~Kos, D.~Poland and D.~Simmons-Duffin,
  ``Bootstrapping Mixed Correlators in the 3D Ising Model,''
JHEP {\bf 1411}, 109 (2014).
[arXiv:1406.4858 [hep-th]].
}

\lref\KosYSD{
  F.~Kos, D.~Poland, D.~Simmons-Duffin and A.~Vichi,
  ``Precision Islands in the Ising and $O(N)$ Models,''
JHEP {\bf 1608}, 036 (2016).
[arXiv:1603.04436 [hep-th]].
}

\lref\VladimirovXX{
  V.~S.~Vladimirov and B.~I.~Zavyalov,
  ``Tauberian Theorem In Quantum Field Theory,''
Theor.\ Math.\ Phys.\  {\bf 40}, 660 (1980)..
}

\lref\GromovHUT{
  N.~Gromov, V.~Kazakov and G.~Korchemsky,
  ``Exact Correlation Functions in Conformal Fishnet Theory,''
[arXiv:1808.02688 [hep-th]].
}

\Title{
\vbox{\baselineskip6pt
}}
{\vbox{
\centerline{Analytic Euclidean Bootstrap}
}}

\bigskip
\centerline{Baur Mukhametzhanov$^{\Delta}$\footnote{$^\dagger$}{mukhamet@physics.harvard.edu} and Alexander Zhiboedov$^{\Delta, J}$}
\bigskip
\centerline{\it $^{\Delta}$  Department of Physics, Harvard University, Cambridge, MA 02138, USA}
\centerline{\it $^{J}$  CERN, Theoretical Physics Department, 1211 Geneva 23, Switzerland
}

\vskip .2in 

\noindent
We solve crossing equations analytically in the deep Euclidean regime. Large scaling dimension $\Delta$ tails of the weighted spectral density of primary operators of given spin in one channel are matched to the Euclidean OPE data in the other channel. Subleading $1\over \Delta$ tails are systematically captured by including more operators in the Euclidean OPE in the dual channel. We use dispersion relations for conformal partial waves in the complex $\Delta$ plane, the Lorentzian inversion formula and complex tauberian theorems to derive this result.  We check our formulas in a few examples (for CFTs and scattering amplitudes) and find perfect agreement. Moreover, in these examples we observe that the large $\Delta$ expansion works very well already for small $\Delta \sim 1$. We make predictions for the 3d Ising model.  Our analysis of dispersion relations via complex tauberian theorems is very general and could be useful in many other contexts.

\Date{August 2018}

\listtoc\writetoc
\vskip .1in \noindent

\break

\newsec{Introduction}

Crossing equations express associativity of the operator product expansion (OPE) \refs{\FerraraYT, \PolyakovGS}. They are nonperturbative consistency conditions on the CFT data (spectrum of local operators and their three-point functions). Extracting physical information from crossing equations is not an easy task. But in the last decade, starting with a seminal paper \RattazziPE, significant progress in this direction was achieved both numerically and analytically, for a review and references see e.g. \refs{\RychkovIQZ, \SimmonsDuffinGJK, \PolandEPD}. In this paper we solve crossing equations analytically in the deep Euclidean regime.

Most analytic computations become possible when there is an expansion parameter in the problem (see however \refs{\MazacQEV, \MazacMDX}). One such parameter is spin $J$ \refs{\AldayMF,\FitzpatrickYX, \KomargodskiEK}. Large spin expansion arises from solving crossing equations in the vicinity of a light cone. In this case it is possible to isolate families of operators that are dual to each other on both sides of the crossing equation and match their spectral data.  Other examples include expansions in small coupling constant \refs{\AldayNJK,\GopakumarCPB}, large central charge (large $N$) \HeemskerkPN\ and large charge \refs{\JafferisZNA}. All of these, however, require some additional ``non-universal'' properties of CFTs.

There is yet another universal quantum number in the problem, namely scaling dimension $\Delta$. It is natural to ask if it is possible to solve crossing equations by expanding in ${1 \over \Delta}$.\foot{ Of course, given the fact that the CFT data is a set of numbers, it is not a priori clear in what sense one can expand in ${1 \over \Delta}$. In this paper we explain the precise meaning of such an expansion.} Such an expansion naturally arises when analyzing crossing equations in the deep Euclidean regime. This question was first addressed in \refs{\PappadopuloJK,\RychkovLCA} . The basic idea is very simple: light operators in one channel map to some cumulative property of the high energy (scaling dimension) OPE data tails in the other channel. In this way a universal high-energy asymptotic of the integrated spectral density of operators was derived in \PappadopuloJK. In this paper we develop this idea further.

Let us briefly review the results of \PappadopuloJK. Consider a four-point function of identical scalar primary operators $\phi (x)$. Let us introduce an integrated spectral density of operators that appear in the OPE of two $\phi$'s

\eqn\spectraldensity{\eqalign{
F(E) &\equiv \int_0^{E} d E' f(E') , \cr
f(E) &\equiv \sum_{k} \rho_k \delta(E-E_k) ,
}}
where the sum is over all states present in the theory, {\it both primaries and descendants of arbitrary spin}. $E_k$ stands for the scaling dimension of the exchanged state. The coefficients $\rho_k$ could be read off the OPE expansion of the correlator (see \PappadopuloJK\ for additional details) and are essentially given by the squares of the three-point functions. Unitarity implies that $\rho_k \geq 0$. It was shown in \PappadopuloJK\ that $F(E)$ has a universal asymptotic\foot{Notation $a \sim b$ stands for ${a \over b} \to 1$ in the corresponding limit.}  
\eqn\universalasym{
F(E) \sim {E^{2 \Delta_\phi} \over \Gamma(2 \Delta_\phi+1)} \ , ~~~ (E \to \infty) .
}
This rigorous result follows from unitarity and the leading contribution of the unit operator in the crossed channel Euclidean OPE via the so-called Hardy-Littlewood tauberian theorem.

There are two natural questions regarding \universalasym. First, is it possible to write a formula similar to \universalasym\ for primary operators of given spin only? In other words, can we disentangle the contribution of operators of different spin as well as of primaries and descendants. Second, can we systematically compute corrections to \universalasym\ by including contributions of extra operators in the crossed channel? The answer to both questions is affirmative and is the subject of the present paper.

Let us introduce a weighted spectral density $\rho_J^{OPE}( \Delta)$ of {\it primary operators of given spin} 
\eqn\spectraldensity{\eqalign{
\rho_J^{OPE}( \Delta) &\equiv \sum_k {p_{\Delta_k,J}  \over K_{J , \Delta_k} } \delta( \Delta - \Delta_k) , \cr
K_{J , \Delta} &= {\Gamma(\Delta -1) \over 2 \pi^2 \Gamma(\Delta - {d \over 2})} { \Gamma({\Delta + J \over 2})^4 \over \Gamma(\Delta + J) \Gamma(\Delta+J-1)} ,
}}
where $p_{\Delta,J}$ stands for the squares of OPE coefficients with the standard normalization for conformal blocks.

Note that we have an additional factor ${1 \over K_{J, \Delta}}$ compared to the standard normalization of the three-point functions. It is, of course, a matter of choice how to normalize three-point functions. However, we will find that there is a canonical choice dictated by the behavior of conformal partial waves at large complex $\Delta$, which leads to \spectraldensity.

The generalization of \universalasym\ to primaries of given spin takes the form
\eqn\result{\eqalign{
&\int_0^\Delta d  \Delta' \ \rho_J^{OPE}(\Delta')  \sim  f_J   {\Delta^{4\left(\delta_\phi-{1\over 2} \right)} \over 4\left(\delta_\phi-{1\over 2} \right)} \ , ~~~ \delta_\phi  > {1 \over 2}, \qquad (\Delta \to \infty) \ , ~~~\cr
&f_J = [1+(-1)^J] \pi^2 {2^{2J+d-4\delta_\phi +2} \Gamma\left( J + {d\over 2} \right) \over \Gamma(J+1) \Gamma(\Delta_\phi)^2\Gamma( \delta_\phi )^2 } \ , \qquad \Delta_\phi  = {d-2 \over 2} + \delta_\phi.
 }}
For $\delta_{\phi} < {1 \over 2}$ we have to consider higher moments of $\rho_J^{OPE}(\Delta)$ to which similar formulas apply.\foot{For $\delta_\phi = {1\over 2}$ we have $\int_0^\Delta d  \Delta' \ \rho_J^{OPE}(\Delta') \sim f_J \log \Delta$.} This will be discussed in details in section 5.\foot{Strictly speaking, we derived the formula \result\ only for $J>1$. However, we observed in a few simple examples that it holds down to $J=0$. It would be interesting to systematically understand the status of \result\ for $J=0,1$ in a generic CFT. To do that one should include arc contributions in the Lorentzian inversion formula.}

At large $\Delta$ (and fixed $J$): ${1 \over K_{J, \Delta}} \sim \pi 4^{\Delta + J -1} \Delta^{1-d/2}$. Therefore, the contribution of heavy operators in \spectraldensity\ is exponentially enhanced. What \result\ roughly states is that after we multiply three-point couplings by this universal exponentially growing pre-factor, they behave polynomially in $\Delta$. The actual power is controlled by the Euclidean OPE in the dual channel. The asymptotic behavior \result\ is completely rigorous and holds in any unitary CFT. 

A second natural question to ask about \universalasym\ and \result\ is regarding the corrections to this leading behavior. This question was briefly addressed in \PappadopuloJK\ where it was noticed that application of Hardy-Littlewood theorem in a real domain leads only to very weak logarithmic bounds on the correction to the integrated spectral density. In this paper we point out that the situation changes if we note that the OPE expansion is valid in a complex domain of the corresponding cross ratios. In this case one can apply more powerful {\it complex tauberian theorems} to the problem at hand \refs{\Korevaar, \Postnikov ,\SubhankulovR, \Subhankulov}. As a result one can develop a systematic ${1 \over \Delta}$ expansion for certain moments of the integrated spectral density. We discuss corrections to \universalasym\ and the corresponding tauberian theorem in section 2.

In sections 3-4 we develop CFT dispersion relations for conformal partial waves $c_J(\Delta)$. They are meromorphic polynomially bounded functions of $\Delta$ which encode the OPE data in the structure of their singularities: they have poles at the dimensions of operators appearing in the OPE with the residues given by squares of the OPE coefficients as well as an infinite set of kinematic poles. By the standard argument we write Cauchy integral in the complex $\Delta$ plane and deform the contour. This provides us with the desired dispersion relation: $c_J(\Delta)$ at some complex $\Delta$ is related to an integral of the weighted spectral density \spectraldensity\ with an appropriate kernel plus a contribution of kinematic poles of $c_J(\Delta)$. We argue that at large $\Delta$ away from the real axis both $c_J(\Delta)$ and the contribution of kinematic poles can be computed by the OPE in the crossed channel via the Lorentzian inversion formula \CaronHuotVEP.  Kinematic poles produce terms of two types at large $\Delta$. Universal terms computable by the Euclidean OPE. And non-universal terms (not computable by the OPE) that are mapped to contributions of individual operators in the weighted spectral density \spectraldensity\ and, therefore, required for consistency.

In section 5 we use CFT dispersion relations to derive asymptotics of the integrated weighted spectral density of the type \result. The crucial mathematical result that we use is a so-called complex tauberian theorem for Stieltjes transform. Dispersion relations hold for complex $\Delta$ and the corresponding complex tauberian theorem leads to a systematic $1\over \Delta $ expansion for moments of the weighted spectral density. The main result of this analysis is the formula (5.11) for moments of the weighted spectral density defined in (5.10). It systematically maps $1\over \Delta$ expansion of the weighted spectral density moments in one channel to the Euclidean OPE in the crossed channel.

In section 6 we proceed by studying large $ \Delta$ expansion in a series of examples. To our knowledge (and surprise) this question was never addressed in the existing literature. The two basic CFT examples we consider are generalized free fields and the 2d Ising model. In all cases we find that the corrections to \result\ are power-like and that relations \result\ work extremely well already for small $\Delta$. We also make predictions for the 3d Ising model. 

Finally, we point out that our analysis of dispersion relations is very general and might be useful beyond CFTs. In section 7 we discuss an application to meromorphic scattering amplitudes. As an example, we study Veneziano amplitude and again find perfect agreement with our predictions.

We prove complex tauberian theorems for Laplace transform in Appendix A and for Stieltjes transform in Appendix B. 

As a historical remark, let us mention that a similar analysis appeared in the context of hadronic scattering amplitudes in the late 60's. It was found that experimental data for the pion-nucleon scattering exhibits a set of resonances at low energies and Regge behavior at high energies. Dolen, Horn and Schmid used analyticity of the scattering amplitude to derive the so-called finite energy sum rules (FESR) which showed that resonances at low energy and Regge behavior are dual to each other \DolenJR\ (see \refs{\IgiZZ, \IgiZZA, \LogunovDY} for earlier works), namely one should not add them up to avoid double counting. This was an example of crossing confirmed by the experimental data. Inspired by this observation very soon after Veneziano wrote down the celebrated amplitude \VenezianoYB. We review this reasoning in section 7. We show that for meromorphic amplitudes a rigorous way to use FESR is via complex tauberian theorems. Our analysis grew out of an attempt to understand FESR for meromorphic amplitudes.

\newsec{Euclidean Crossing and Tauberian Theorems For Laplace Transform}

In this section we analyze crossing equations in the Euclidean kinematics. We start by reviewing the argument of \PappadopuloJK\ and then slightly generalize it. Consider a four-point function of identical scalar primary operators
 \eqn\fourid{\eqalign{
 \la \phi(x_4) \phi(x_3) \phi(x_2) \phi(x_1)\ra  &= {{\cal G}(z, \bar z) \over x_{12}^{2 \Delta_\phi} x_{34}^{2 \Delta_{\phi}}} \ , \cr
 u ={x_{12}^2 x_{34}^2 \over x_{13}^2 x_{24}^2} = z \bar z, ~~~  v&={x_{14}^2 x_{23}^2 \over x_{13}^2 x_{24}^2} = (1-z) (1- \bar z) \ .
 }}
If we set $x_1  =0$, $x_4 = \infty$ the correlation function above becomes the radial quantization matrix element $\la \phi | \phi(x_3) \phi(x_2) | \phi \ra$. 
We also set $\vec x_2 = r_2 \vec n_2$ and $\vec x_2 = r_3 \vec n_3$ in terms of the coordinates on the plane
 \eqn\plane{
 d s_{ R^d}^2 = d r^2 + r^2 d \Omega_{d-1}^2 \ .
 }
and ${\vec n}_i$ are unit vectors.
 
Consider the OPE expansion of ${\cal G}(z,\bar z)$ in the $\phi(x_1) \times \phi(x_2)$ channel ($s$-channel). It takes the following form 
\eqn\rewrite{
{\cal G}(z, \bar z) = \sum_{{\cal O}} f_{\phi \phi {\cal O}}^2 \sum_{n=0}^{\infty} e^{- E_n \beta} \la {\cal O}, n, \vec n_3 |  {\cal O}, n, \vec n_2 \ra, ~~~ E_n = \Delta_{{\cal O}} + n \ .
 }
The first sum in \rewrite\ is over primaries, while the second one is over descendants. Squares of the three-point functions between primaries are $f_{\phi \phi {\cal O}}^2$, and  the states $| {\cal O}, n, \vec n \ra$ are $n$-th level descendants of $|{\cal O}\ra$ properly contracted with $\vec n$. We also introduced $r = e^{\tau}$, which is the standard time coordinate on the cylinder and $\beta = \tau_3 - \tau_2$ for the time difference.

 In the conformal frame above the cross ratios take the form
 \eqn\crossratios{\eqalign{
 u &= z \bar z  = \left({r_2 \over r_3}\right)^2 = e^{-2 \beta} ,\cr
 v &=(1-z)(1-\bar z) = 1 + \left({r_2 \over r_3}\right)^2 - 2 \left({r_2 \over r_3}\right) \cos \alpha \ ,
 }}
 where $\alpha$ is the angle between $\vec n_2$ and $\vec n_3$. From \crossratios\ it follows that $z = e^{- \tau} e^{i \alpha}$, $\bar z = e^{- \tau} e^{-i \alpha} $.
 
 Upon setting $\vec n_2 = \vec n_3$, or $\alpha = 0$, we get the following expansion for the correlator
 \eqn\expansion{\eqalign{
 {\cal L}(\beta) &= \int_0^{\infty} d E \ f(E) e^{- E \beta} , \cr
 f(E)  &=\sum_{k} \rho_k \delta(E - E_k) , ~~~ \rho_k \geq 0 \ ,
 }}
 where $\rho_k$ are positive coefficients that can be computed using \rewrite. We can easily compute the $\beta \to 0$  limit of the correlator using the Euclidean OPE in the dual channel $\phi(x_2) \times \phi(x_3)$ ($t$-channel). The leading contribution comes from the unit operator. We therefore get\foot{ Throughout the paper we often write $O(x)$ to estimate the magnitude of different quantities. Recall that 
 \eqn\bigO{
f(x) = O( g(x) ), ~~~ x \to \infty \quad (x \to a)
}
iff there exist numbers $M, x_0$ ($M,\delta$) s.t. 
\eqn\defO{
|f(x)| < M |g(x)|, \qquad \forall ~ x > x_0 \quad (\forall ~ |x-a| < \delta)
}
}
 \eqn\coupleofleadingterms{
  {\cal L}(\beta) = \beta^{- 2 \Delta_{\phi}} \left[ 1 + O(\beta^{\Delta_0}) \right] , \qquad \beta \to 0
 }
 where the corrections come both from expanding to higher orders the contribution of the unit operator $\left( {u \over v} \right)^{\Delta_{\phi} }= \left( {e^{- \beta} \over 1 - e^{- \beta}} \right)^{2 \Delta_{\phi}}$, as well as from heavier operators.
 
 As explained in \PappadopuloJK, \expansion\ fixes the high energy behavior of the integrated spectral density $F(E)$ to be
 \eqn\resulttauberian{\eqalign{
 F(E) &\equiv \int_0^E d E' \ f(E') \ , \cr
 F(E) &= {E^{2 \Delta_{\phi}} \over \Gamma( 2 \Delta_\phi + 1) } \left( 1 + O \left({1 \over \log E} \right) \right)\ .
 }}
 This result, which crucially relies on the positivity of $f(E)$, is known as Hardy-Littlewood tauberian theorem. In general, a class of theorems which relate asymptotics of two different methods of integration (or summation) (${\cal L}(\beta)$ and $F(E)$ in the present case) are called {\it tauberian theorems}. The result \resulttauberian\ follows from \coupleofleadingterms\ for {\it real} $\beta$. A common feature of real tauberian theorems is that corrections to the leading asymptotic are only logarithmically suppressed, as is the case in \resulttauberian.\foot{For a detailed discussion of real tauberian theorems see, for example, \Korevaar. In particular, Chapter VII for the discussion of the remainders. See also appendix E in \QiaoXIF\ for an elementary proof of the leading asymptotic in \resulttauberian. }
 
 This situation changes if the condition \coupleofleadingterms\ is valid in a {\it complex} domain. In this case the corrections are more constrained, as we will shortly explain. An intuitive reason for weaker bounds in real tauberian theorems is that in a complex plane one can have two integral transforms with different analytic properties, which have the same asymptotic on a real line. Then the remainder term in a real tauberian theorem would be dictated by the integral with the worst analytic properties. 
 
 \subsec{Complex Tauberian Theorem}
 
In the formulas above we kept $\beta$ real. In the context of CFTs, however, we know that the $t$-channel OPE expansion is valid not only along the real line $z = \bar z = e^{- \beta}$, but in the  complex domain $|\beta| \ll 1$. A natural question is if this stronger condition leads to stronger tauberian theorems that are relevant for CFTs. It is indeed the case as we describe below.
 
In what follows it will be useful to introduce a notion of Cauchy moments
\eqn\Cauchymoment{\eqalign{
F_{m}(E) &\equiv {1 \over (m-1)!} \int_0^E d E' (E-E')^{m-1} f(E') \ \cr
&= \int_0^E d E_{m-1} \int_0^{E_{m-1}} d E_{m-2}  \ ... \int_0^{E_1} d E_0 \ f(E_0) \ .
}}
These are obtained by a repeated integration, which is equivalent to the first line via integration by parts. 

Imagine also that we know the OPE expansion of the correlator up to an arbitrary order in $\beta$, namely
\eqn\OPEexpansion{
 {\cal L}(\beta) = \int_0^{\infty} d E \ f(E) e^{- E \beta}  = {1 \over  \beta^{2 \Delta_{\phi} } } \sum_{\Delta_i} c_{\Delta_i} \beta^{ \Delta_i}  + ... \qquad \beta \to 0 \ ,
}
where we can imagine re-expanding the usual $t$-channel OPE to an arbitrary high order in $\beta$. We can integrate $m$ times by parts under the $E$ integral to get 
\eqn\byparts{
 {\cal L}(\beta) =\beta^{m} \int_0^{\infty} d E \ F_m(E) e^{- E \beta} ,
}
where we used $F_m(0) = 0$. It is then possible to prove the following statement:

\vskip .1in \noindent

\noindent {\bf Claim:} Given the expansion \OPEexpansion\ is valid in the complex domain $|\beta | \ll 1$ and $f(E)$ is a positive density, the Cauchy moments \Cauchymoment\ $m \geq 1$  satisfy
\eqn\complextauberian{
F_m (E) =E^{2 \Delta_{\phi}} \left(  \sum_{\Delta_i <  m} {c_{\Delta_i} E^{m - \Delta_i - 1} \over \Gamma(2 \Delta_{\phi} - \Delta_i + m)} + O\left({1 \over E} \right) \right) .
}

\noindent The formula \complextauberian\ constitutes the statement of {\it the complex tauberian theorem for Laplace transform}, which we prove in appendix A.

\vskip .1in \noindent

This result therefore holds in a generic unitary CFT. We review the proof of \complextauberian\ in appendix A. It is a particular example of more general complex tauberian theorems proved in \SubhankulovR. It is also easy to see that \complextauberian\ cannot be improved. The basic ingredient that makes \complextauberian\ possible is the validity of the asymptotic behavior \OPEexpansion\ in the complex $\beta$-plane as we approach $\beta = 0$ (and, of course, positivity of $f(E)$).

The improvement compared to the real tauberian theorem \resulttauberian\ is two-fold. First, by considering higher order $m$ Cauchy moments \Cauchymoment\ we can probe subleading operators in the $t$-channel OPE. Note that taking higher moments does not increase the error which always stays the same $O(E^{2 \Delta_{\phi} -1})$! Second, the remainder is suppressed by a power ${1 \over E}$ instead of the logarithm ${1 \over \log E}$. 

Intuitively, repeated integration in \Cauchymoment\ enhances smooth power-like tails and leaves intact oscillating pieces of the type $\sin c E$. What \complextauberian\ tells us that this naive picture is actually universal and the $t$-channel Euclidean OPE is encoded in the Cauchy moments of the $s$-channel OPE data. All the non-universal pieces, in particular the ones that encode the discreteness of the spectrum, enter only in the remainder term $O(E^{2 \Delta_{\phi} -1})$. Analytic properties of correlation functions guarantee that these non-universal pieces are not enhanced upon a repeated integration \Cauchymoment.

Let us illustrate the discussion above with a couple of simple examples.

 \subsec{Example 1}
 
 Let us demonstrate that the estimate of the remainder in the real tauberian theorem \resulttauberian\ cannot be improved. Consider a positive spectral density $f(E)$
 \eqn\subkhint{
 f(E) =   1 + \sin[ (\log E)^2 ] \geq 0 \ , \qquad E \geq 1 \ .
 }
 One can check that for real $\beta$ we have
 \eqn\subkhintB{\eqalign{
 {\cal L}(\beta) &=  {1 \over \beta} \left( 1 + \beta (c_0-1) + ... \right)  =   {1 \over \beta} \left( 1 + O(\beta) \right) , \cr
 F(E) &=E\left( 1 - {1 \over 2} {\cos[ (\log E)^2 ] \over \log E} + ... \right) = E\left( 1 + O \left({1 \over \log E} \right) \right) \ ,
 }}
 where the explicit form of $c_0$ can be found in appendix C, where we discuss the evaluation of this integral.  The result \subkhintB\ shows that the estimate of the remainder term in \resulttauberian\ is optimal.

Next, one can check that the asymptotic \subkhintB\ for ${\cal L}(\beta)$ does not hold in the vicinity of $\beta=0$ in the complex plane. We have not found analytically the leading behavior of the integral above in the complex domain, but observed numerically that it is qualitatively consistent with the following simple model. Consider a function $\beta^{\pi -1} \cos (\log \beta)^2 L(\beta)$, where $L(\beta)$ is a function of slow variation (namely $\lim_{\beta \to 0} {L(\lambda \beta) \over L(\beta} = 1$ for $\lambda>0$). The asymptotic behavior of this function depends on ${\rm arg}[\beta]$ and is given by $\beta^{\pi - 2 {\rm arg}[\beta] - 1} L(|\beta|)$. In particular, for imaginary $\beta$ the leading power becomes ${1 \over \beta}$. This is the reason why we get a weaker bound on the remainder in $F(E)$. 

 \subsec{Example 2}
 
 Consider now a simple example where the complex tauberian theorem \complextauberian\ is applicable. We consider the spectral density to be 
 \eqn\spectralextwo{
 f(E) = \sum_{n=0}^{\infty} \delta(E - n) \ .
 }
 We can explicitly compute the Laplace transform
 \eqn\laplacetwo{
 {\cal L}(\beta) = {1 \over 1 - e^{- \beta}} = {1 \over \beta} \left( 1 + {1 \over 2} \beta + {\beta \over 12} \beta^2 + ... \right) \ .
 }
 and Cauchy moments
 \eqn\CauchyExTwo{
  F_m(E) =E \left( {E^{m-1} \over \Gamma(m)} +{E^{m-2} \over 2 \Gamma(m-1)} + {E^{m-3} \over 12 \Gamma(m-2)} +  ... + O\left({1 \over E} \right) \right) \ .
 }
 The error term in this example is a function of the fractional part of $E$, namely $E-[E]$. Since $0 \leq E-[E]<1$, it is indeed $O(1)$ for any $E$. This is an example of \complextauberian\ with $\Delta_{\phi} = {1 \over 2}$.

\newsec{Dispersion Relations for Conformal Partial Waves}

In this section we derive dispersion relations for conformal partial waves in the complex $\Delta$ plane. These dispersion relations allow us to study separately the contribution of primary operators of given spin in the $s$-channel. We then analyze these dispersion relations in the limit $|\Delta| \gg 1$. We find that the large $\Delta$ behavior away from the real axis of the conformal partial waves is controlled by the $t$-channel OPE data.\foot{In a sense,  $s$-channel operators of dimension $\Delta$ probe $t$-channel distances ${1 \over \Delta}$.} In section 5 we will use these dispersion relations together with complex tauberian theorems to arrive at our final result (5.11).

\subsec{Conformal Partial Waves and Lorentzian Inversion Formula}

Consider a four-point correlator of identical scalar primary operators $\phi$ in $d \geq 2$ dimensions
\eqn\fourpt{
\la \phi(0) \phi(z, \b z) \phi(1) \phi(\infty) \ra = (z \b z)^{-\Delta_\phi} {\cal G} (z, \b z) ,
}
where, as usual, we used conformal invariance to put four points in a plane. The relation to the conformal cross ratios is
\eqn\conformalcrossratios{
u = {x_{12}^2 x_{34}^2 \over x_{13}^2 x_{24}^2} = z \b z , \qquad
v = {x_{14}^2 x_{23}^2 \over x_{13}^2 x_{24}^2} = (1-z) (1 - \b z)\ .
}
The OPE expansions in different channels are given by \foot{We mostly follow the conventions of \CaronHuotVEP, except for conformal blocks and three-point functions, which we write with a more conventional ordering of dimension and spin $G_{\Delta, J}, p_{\Delta, J}$, but keep the same normalization for them as in \CaronHuotVEP.}
\eqn\OPE{\eqalign{
&{\rm s-channel:} \qquad {\cal G}(z, \b z) = \sum_{\Delta, J} p_{\Delta, J} G_{\Delta, J}(z, \b z) \cr 
&{\rm t-channel:} \qquad {\cal G}(z, \b z) =  \left[ z \b z \over (1 - z)(1-\b z) \right]^{\Delta_\phi} \sum_{\Delta, J} p_{\Delta, J} G_{\Delta, J}(1-z, 1-\b z)
}}
where $p_{\Delta, J} \geq 0$ are squares of the OPE coefficients. We choose the normalization of conformal blocks as
\eqn\confblocknorm{
G_{\Delta, J}(z, \b z ) \sim z^{\Delta - J \over 2} \b z^{\Delta + J \over 2} , \qquad z \ll \b z \ll1 
}
In two and four dimensions the exact expressions are known and given by hypergeometric functions
\eqn\conftwofour{\eqalign{
&G_{\Delta, J}(z, \b z ) = {1\over 1+\delta_{J,0} } \left[ k_{\Delta + J}(z)k_{\Delta - J}(\b z) + k_{\Delta - J}(z)k_{\Delta + J}(\b z) \right], \qquad ~~~d=2, \cr
&G_{\Delta,J}(z, \b z) = {z \b z \over z - \b z} \left[ k_{\Delta+J}(z) k_{\Delta - J -2} (\b z) - k_{\Delta-J-2}(z) k_{\Delta + J } (\b z) \right] , \qquad d=4, \cr 
&k_{\alpha}(x) = x^{\alpha/ 2} F\left( {\alpha / 2}, {\alpha / 2}, \alpha, x \right). 
}}

Alternatively, we can expand the four-point function \fourpt\ into an orthogonal basis of eigenfunctions $F_{\Delta, J}$ of the Casimir operator. The single-valued functions $F_{\Delta, J}$, also called partial waves, are given by a linear combination of conformal block plus ``shadow''
\eqn\partialwave{
F_{\Delta, J} =  K_{J,\Delta} G_{\Delta, J} + K_{J, d-\Delta} G_{d-\Delta, J} 
}
where we defined following \CaronHuotVEP\
\eqn\Kdef{
K_{J, \Delta} = {\Gamma(\Delta -1) \over \Gamma\left( \Delta - {d\over 2} \right) } \kappa_{J+\Delta}, \qquad \kappa_{\beta} = {\Gamma\left(\beta \over 2 \right)^4 \over 2\pi^2 \Gamma(\beta -1 ) \Gamma(\beta) }   \ .
}
A complete set of square-integrable functions\foot{See e.g. appendix A in \SimmonsDuffinNUB. } is spanned by all $F_{\Delta, J}$ with integer spins $J$ and complex dimensions $\Delta = {d\over 2} + i\nu$, where $\nu$ is real and $\nu \geq 0$. Therefore, we can expand the four-point function \fourpt\ as\foot{More precisely, one should also add non-normalizable modes, coming from $s$-channel scalar operators with $\Delta \leq d/2$, as described in appendix B in \SimmonsDuffinNUB. Here, we will be interested in $s$-channel operators with $\Delta > d/2$, so we ignore these contributions. }
\eqn\partialwavedec{\eqalign{
\G(z, \b z) & = \sum_{J=0}^\infty \int_{d/2 }^{d/2 + i \infty} {d\Delta \over 2\pi i } ~  c_J(\Delta)  F_{\Delta, J}(z,\b z)  = \cr 
& = \sum_{J=0}^\infty \int_{d/2-i\infty }^{d/2 + i \infty} {d\Delta \over 2\pi i } ~  c_J(\Delta) K_{J,\Delta}  G_{\Delta, J}(z,\b z)
}}
Due to the shadow symmetry of the partial wave $F_{d- \Delta,J} = F_{\Delta, J}$ the partial wave coefficients $c_J(\Delta)$ are also shadow symmetric\foot{Our $c_J(\Delta)$ is related to $c(J,\Delta)$ in \CaronHuotVEP\ by $ c_J(\Delta) \equiv  {c\left( J, \Delta \right) \over K_{J, \Delta  } }$.}
\eqn\csymm{
c_J(\Delta) = c_J(d - \Delta) .
}

To relate the partial wave decomposition \partialwavedec\ to the $s$-channel OPE expansion \OPE, we can deform the contour in \partialwavedec\ to the real axis. The OPE expansion is reproduced if $c_J(\Delta)$ has poles at dimensions of operators appearing in the $s$-channel OPE with residues related to OPE coefficients. The precise relation was given in \CaronHuotVEP\
\eqn\threetopartial{
{p_{\Delta, J} \over K_{J,\Delta}} = - {\rm Res}_{\Delta' \to \Delta} ~ 
\cases{ 
		c_J(\Delta'), \qquad \qquad \qquad ~~~~~~~~~~~~~~~~~~~~~~~~~~~~~~~~~~~~~~
\Delta  \neq \Delta_n \cr  
		c_J(\Delta') - r_{J,\Delta'} {K_{\Delta' +1-d,J+d-1} \over K_{J,\Delta'} } c_{\Delta' + 1 -d}( J+d-1)	,\quad \Delta = \Delta_n 
	} 
}
where we introduced\foot{For non-identical external operators there will be poles for all integer $n$.}
\eqn\Deltan{
\Delta_n = J+d+n, \qquad n = 1,3,5,\dots
}
and $r_{J,\Delta}$ is defined by ($x = \Delta - J-d + 2$)
\eqn\rdJdef{
r_{J,\Delta} = {\Gamma (\Delta -1) \Gamma (\Delta +2-d) 
\over 
\Gamma \left(\Delta -{d\over 2}\right) \Gamma \left(\Delta -{d-2\over 2}\right)} 
{\Gamma \left(J+{d-2\over 2}\right) \Gamma \left(J+{d\over 2}\right)
\over 
\Gamma (J+1) \Gamma (J+d-2)}
{\Gamma (2-x) \Gamma \left({x\over 2}\right)^2 
\over 
 \Gamma (x) \Gamma \left({2-x\over 2}\right)^2}}

 The reason there is an extra term in \threetopartial\ when $\Delta = \Delta_n$ is that the conformal blocks in \partialwavedec\ have poles on the real $\Delta$ axis. Their contributions are cancelled by extra poles of $c_J(\Delta)$, which do not correspond to operators appearing in the OPE. These unphysical extra poles are explicitly subtracted in the second line of \threetopartial, as explained in \CaronHuotVEP.

Since the functions $F_{\Delta, J}$ form an orthogonal basis, one can invert \partialwavedec\ and obtain the Euclidean inversion formula. Further, it is possible to deform the contour of integration in $z, \b z$ to Lorentzian kinematics, which entails Caron-Huot's inversion formula \CaronHuotVEP.\foot{Recall that we write $G_{\Delta, J}$ for conformal blocks instead of the convention in \CaronHuotVEP\ $G_{J,\Delta}$.} In the case of identical external operators it is given by
\eqn\inversionf{
c_J(\Delta) ={1\over 2}  \delta_{J,even} {\Gamma\left( \Delta - {d\over 2} \right) \over  \Gamma(\Delta - 1) } 
\int_0^1 dz d\b z ~ \mu(z, \b z) G_{J+d-1, \Delta + 1-d}(z, \b z ) ~{\rm dDisc}~ \G (z, \b z)
}
where $\mu(z, \b z)$ is the orthogonality measure of partial waves $F_{\Delta, J}$
\eqn\Casimirmeasure{
\mu(z, \b z) = {1\over (z \b z)^2} \left| z - \b z \over z \b z \right|^{d-2} 
}
 The double-discontinuity is defined \CaronHuotVEP\ by 
\eqn\dDiscdef{\eqalign{
&{\rm dDisc}~ \G(z, \b z) = \G_{eucl}( \rho, \b \rho ) - {1\over 2} \G(\rho , \b \rho- i0) - {1\over 2} \G(\rho, \b \rho + i0) \ , \cr 
& \G_{eucl}(\rho, \b \rho) = \G(\rho, 1/\b \rho), \qquad z = {4\rho \over (1 + \rho)^2} \ .
}}
In a generic CFT the derivation of \inversionf\ applies only to $J>1$.

To recapitulate, $c_J(\Delta)$ are meromorphic shadow symmetric functions. They have poles at the positions of physical operators appearing in the $s$-channel OPE as well as a series of extra (kinematical) poles as dictated by \threetopartial . For $J>1$ $c_J(\Delta)$ could be computed from the double discontinuity of the correlator using the inversion formula \inversionf. In the next subsection we will show that a direct consequence of \inversionf\ is polynomial boundedness of $c_J(\Delta)$ at large $\Delta$. This allows us to write and study dispersion relations for $c_J(\Delta)$.

\subsec{Polynomial Boundedness}

\noindent We will be interested in the behavior of $c_J(\Delta)$ at large $\Delta$ in the complex plane and will observe that this limit in \inversionf\ is controlled by the Euclidean OPE expansion of the correlator in the $t$-channel \OPE. 

The simplest example of the use of \inversionf\ is to consider the unit operator in the $t$-channel
\eqn\tid{
\G(z, \b z) = \left( z \b z \over (1-z) (1-\b z) \right)^{\Delta_{\phi}} + \dots
}
It gives a contribution to the double-discontinuity
\eqn\dDiscGFF{
{\rm dDisc}~ {\cal G}(z, \b z) = 2 \sin^2 (\pi \Delta_\phi) \left( z \b z \over (1-z) (1-\b z) \right)^{\Delta_{\phi}}  \ . 
}
We plug \dDiscGFF\ into the inversion formula \inversionf. The corresponding partial wave coefficients are given by 
\eqn\cGFFd{\eqalign{
&c_J^{GFF}(\Delta) =
[1 + (-1)^J]  \pi^2 {  \Gamma\left( J + {d\over 2} \right) \Gamma({d\over 2}-\Delta_\phi)^2 \over  \Gamma(J+1) \Gamma(\Delta_\phi)^2}   \times \cr 
&\times   { \Gamma({\Delta +J})  \Gamma\left( -\Delta +2\Delta_\phi +J \over 2 \right)  \over 
\Gamma\left( \Delta + J \over 2 \right)^2 \Gamma\left( {\Delta +J  \over 2} - \Delta_\phi +{d\over 2} \right) }  \times 
{ \Gamma({\tilde \Delta +J})  \Gamma\left( -\tilde \Delta +2\Delta_\phi +J \over 2 \right)  \over 
\Gamma\left( \tilde \Delta + J \over 2 \right)^2 \Gamma\left( {\tilde \Delta +J  \over 2} - \Delta_\phi +{d\over 2} \right) }  
}}
where $\t \Delta \equiv d-\Delta$ is the shadow transform of $\Delta$. Incidentally, this is also the exact answer for the generalized free field theory (GFF). Indeed, the first two terms in the GFF 4-point function ${\cal G}_{GFF} = 1 + (z \b z)^{\Delta_\phi} + \left[z \b z\over (1-z)(1-\b z)\right]^{\Delta_\phi}$ do not contribute to the double discontinuity.

The partial wave \cGFFd\ has simple poles at $\Delta = 2\Delta_\phi + J +2n$ corresponding to double-trace operators in GFF,\foot{These are physical only in the theory of generalized free fields. In a generic CFT they are spurious and their proper treatment is the subject of the analytic Lorentzian bootstrap.} at $\Delta = \Delta_n$ \Deltan\  and at their shadows. The poles at $\Delta = \Delta_n$ cancel in the physical combination \threetopartial, as expected.

In $d=2,4$ the conformal blocks are known explicitly \conftwofour\ and the integrals in the inversion formula \inversionf\ can be taken explicitly.\foot{This can be done using Euler type integral representation of the hypergeometric function. } For general $d$ we simply guessed the formula \cGFFd\ by requiring that unphysical poles at $\Delta = \Delta_n$ cancel in the combination \threetopartial\ and that the residues reproduce correct 3-point functions of GFF \FitzpatrickDM.\foot{See formula (43) in that paper. To translate to our normalization of 3-point functions, one has to divide their formula (43) by a factor ${\cal C}_{\Delta_1}{\cal C}_{\Delta_1}$. Further, for identical operators one has to add a permutation term in their formula (37), which leads to an extra factor of $2$ in (43).  }

Let us understand how $c_J(\Delta)$ behaves at large $\Delta$. In the upper half-plane we have from \cGFFd\
 \eqn\clargeDelta{
c_J(i\Delta) = d_J  \Delta^{4\delta_\phi -3} + \dots  , \qquad 
\Delta \to + \infty, \quad |\arg \Delta| < {\pi \over 2} \ ,
}
where we introduced 
\eqn\deltadef{
\delta_\phi = \Delta_\phi - {d-2 \over 2} \geq 0 , \qquad d_J =   [1+(-1)^J] \pi{  2^{2J+d-4\delta_\phi +1 } \Gamma\left( J + {d\over 2} \right) \Gamma({d\over 2}-\Delta_\phi)^2 \over   \Gamma(J+1) \Gamma(\Delta_\phi)^2 } \ .
}
The bound on $\delta_\phi$ is the usual unitarity bound. In the lower-half plane the asymptotic can be obtained from \clargeDelta\ by shadow symmetry. 
The crucial observation is that including heavier operators in the $t$-channel would give rise to suppressed contributions in \clargeDelta. This is because they come with higher powers of $(1-z)$. Equivalently, we can expand in $1-z \over z$ in the $t$-channel. Then extra powers of $1-z \over z$ result in decreasing $\Delta_\phi$ in \dDiscGFF\ and, therefore, a suppressed contribution in \clargeDelta. This argument is rigorous for scalar operators in the $t$-channel, for which $1-z$ and $1-\b z$ come in the same powers. For spinning operators it is slightly less obvious and we will discuss it further in section 4. 

 The power-like expansion \clargeDelta\ breaks down close to the real axis. One way to see it is to observe that there are nonperturbative corrections to \clargeDelta\ coming from the expansion of $\Gamma$-functions in \cGFFd. These non-perturbative corrections become negligible for large ${\rm Im} (\Delta)$. We will assume that this is completely general and the power-like expansion of conformal partial waves, that we get by inserting the $t$-channel expansion into the inversion formula, is valid in the complex plane as soon as ${\rm Im}(\Delta) \gtrsim |\Delta|^\eps$ for any positive fixed $\e$.

Below, when analyzing dispersion relations, we will also need the asymptotic behavior of \cGFFd\ at large $J$ and fixed $\Delta$. We have from \cGFFd\
\eqn\clargeJ{
c_J(\Delta) = [1+(-1)^J]\pi{ 2^{2J + d-4\delta_\phi+1} \Gamma\left( {d\over 2} - \Delta_\phi \right)^2 \over \Gamma(\Delta_\phi)^2}  J^{4\delta_\phi + {d\over 2} - 4} + \dots , \qquad J \to \infty \ .
}
As above, including heavier operators leads to terms in \clargeJ\ which are suppressed by further powers of ${1 \over J}$. 

Let us emphasize that the large spin expansion of \clargeJ\ is different from the usual analytic Lorentzian bootstrap discussions. The latter corresponds to keeping the twist $\Delta - J$ fixed while taking the large spin limit. Here we are exploring the unphysical regime of large $J$ and fixed $\Delta$. The claim is that this limit is controlled by the Euclidean rather than Lorentzian OPE.

\subsec{Dispersion Relations}

\noindent Having a meromorphic and polynomially bounded \clargeDelta\ function $c_J(\Delta)$, it is natural to write down a dispersion relation. Consider a Cauchy integral
\eqn\disp{
c_J(\Delta)=  \oint { d \Delta' \over 2\pi i} ~ {c_J(\Delta') \over \Delta' - \Delta} \ , 
}
where the contour goes around $\Delta' = \Delta$ counterclockwise. This dispersion relation was briefly considered in \HogervorstSFD , see section 2.5.1 in that paper.

\ifig\usualcontour{Dispersion relations in $\Delta$ plane. We consider the contour integral \disp\ and deform the contour in the usual way.} {\epsfxsize1.5in\epsfbox{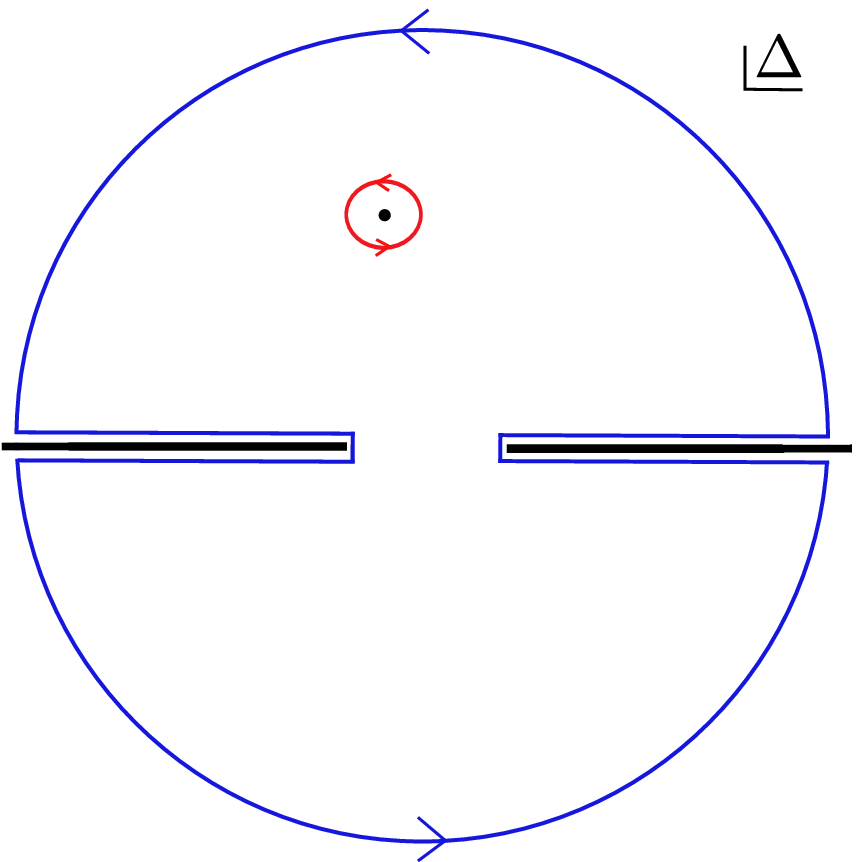}}
 
We can deform the contour as indicated on the \usualcontour. The arcs at infinity can be dropped if $c_J(\Delta) \to 0 $ as $|\Delta| \to \infty$. Otherwise, we can write a dispersion relation with subtractions by taking an appropriate number of derivatives of \disp\ 
\eqn\dispder{
{1\over N!} \p_\Delta^N c_J(\Delta) = \oint { d \Delta' \over 2\pi i} ~ {c_J(\Delta') \over (\Delta' - \Delta)^{N+1}} 
} 
Since $c_J(\Delta) \sim \Delta^{4\delta_\phi -3}$ as $\Delta \to \infty$, the contribution of arcs vanishes if we choose $N$ as follows
\eqn\nchoice{
N = \cases{  [4\delta_\phi]-2, \qquad \delta_\phi \geq {3\over 4}   \cr  0, \qquad \qquad  0 \leq \delta_\phi <  {3\over 4} }
}
For simplicity, let us first describe the dispersion relation \disp\ for $0 \leq \delta_\phi <{3\over 4}$, from which it will be trivial to generalize to the dispersion relation with subtractions \dispder\ for $\delta_\phi \geq {3\over 4}$.

For $0 \leq \delta_\phi <{3\over 4}$ we can use the equation \disp. Deforming the contour in \disp\ to the real axis, dropping the arcs at infinity and using the shadow symmetry \csymm, we obtain
\eqn\dispp{
c_J(d/2 + i\nu) = \int_0^\infty d\nu' {2\nu' \over \nu'^2 + \nu^2} \rho_J(d/2 + \nu')
}
where we shifted variables as $\Delta = {d\over 2} + i\nu, \Delta' = {d\over 2} + \nu'$ and introduced a spectral density 
\eqn\spec{
\rho_J(\Delta) = - \sum_{\Delta_{pole}} \delta(\Delta - \Delta_{pole}) {\rm Res}_{\Delta \to \Delta_{pole}}~ c_J(\Delta )
}
where the sum is over {\it all poles} of the partial wave coefficient $c_J(\Delta)$. The partial wave coefficient, however, includes not only the OPE poles, but also extra poles \Deltan\ at $\Delta_n = J+d+n$, while the OPE data is encoded in the combinations \threetopartial. Thus, we can relate \spec\ to the spectral density of OPE coefficients $\rho_J^{OPE}(\Delta)$ by
\eqn\specOPE{
\rho_J(\Delta) = \rho_J^{OPE}(\Delta) - \rho_J^{extra}(\Delta)  \ ,
}
where we defined 
\eqn\specOPEdef{\eqalign{
&\rho_J^{OPE}(\Delta) \equiv \sum_{\Delta'} {p_{\Delta', J} \over K_{J,\Delta'} } \delta(\Delta - \Delta') \ , \cr 
& \rho_J^{extra}(\Delta) \equiv \sum_{n=0}^\infty \delta_{n,odd} \delta(\Delta - \Delta_n) {K_{J+n+1, J+d-1} \over K_{J, J+d+n}} {\rm Res}_{\Delta \to \Delta_n}\left(r_{J,\Delta} \right)  c_{J+n+1}(J+d -1 )  ,
}}
and the sum in $\rho_J^{OPE}$ is only over primary operators of spin $J$ appearing in the OPE. We also used that $K_{J+n+1, J+d-1} \over K_{J, J+d+n}$ and $c_{J+n+1}(J+d-1)$ are non-singular, so that Res acts only on $r_{J,\Delta}$. Inserting \specOPE\ into \dispp\ we arrive at the desired dispersion relation
\eqn\disppp{
\int_0^\infty d\nu' \rho_J^{OPE}(d/2 +\nu') {2\nu' \over \nu'^2 + \nu^2} 
 = c_J(d/2 + i\nu)  + {\rm extra }
 }
 where we defined
 \eqn\extradef{
{\rm extra} =  \sum_{n=0}^\infty \delta_{n,odd} {2(J+n)+d \over \left( J+{d\over 2} +n \right)^2 + \nu^2}  
{K_{J+n+1, J+d-1} \over K_{J, J+d+n}} {\rm Res}_{\Delta \to \Delta_n}\left(r_{J,\Delta} \right) 
c_{J+n+ 1}(J+d -1 ) 
}
and using definitions \Kdef, \rdJdef\ we can also compute
\eqn\KKr{
{K_{J+n+1, J+d-1} \over K_{J, J+d+n}} {\rm Res}_{\Delta \to \Delta_n}\left(r_{J,\Delta} \right)  =  {(-1)^{n+1}\over n+1} \left[ { \Gamma\left( {n\over 2} + 1 \right)  \over  \Gamma\left( - {n\over 2} \right) \Gamma(n+1) } \right]^2 { (J+1)_{n+1} \over  \left(J+{d\over 2} \right)_{n+1} } 
}
where the Pochhammer symbol is $(a)_b = {\Gamma(a+b) \over \Gamma(a)}$. For the remainder of this section we will be preoccupied with studying the dispersion relation \disppp. It can be considered as a reformulation of $s=t$ crossing. Indeed, the LHS of \disppp\ contains $s$-channel OPE data of primary operators with fixed spin $J$. The kernel $2\nu' \over \nu'^2 + \nu^2$ is centered around operators with dimension $ \Delta' \sim  {d\over 2} + \nu $. The RHS of \disppp\ can be thought of as $t$-channel data if we use $t$-channel OPE and the inversion formula \inversionf\ to compute $c_J(\Delta)$ entering the RHS of \disppp. 

As it is usual with the crossing equations, we cannot solve \disppp\ for general values of parameters $(\nu, J)$. So let us consider a limit when only a few light operators dominate in one of the channels to make predictions for the other channel. We will take $\nu \gg 1$. In this limit the RHS of \disppp\ will be dominated by light operators in the $t$-channel. This is reminiscent of the fact that large momentum corresponds to short distances in Fourier transforms. We now explain how it happens and find the asymptotic of the RHS of \disppp\ at large $\nu$.

The asymptotic $\nu \gg 1$ of the first term in the RHS of \disppp\ was already found in \clargeDelta\ and was indeed controlled by the identity operator in the $t$-channel. The asymptotic of the second term in the RHS of \disppp\ is more subtle and we will study it in the next subsection.

For completeness let us also write down the most general dispersion relation that involves $N$ subtractions. Starting from \dispder\ and going through the same steps we have
\eqn\dispthree{\eqalign{
&\int_0^\infty d\nu' \rho_J^{OPE}(d/2 +\nu') 
\left[ {e^{{\pi i \over 2} N}\over (\nu' - i\nu)^{N+1}} +c.c.\right] = {1\over N!}  \p_\nu^N c_J(d/2 + i\nu)  + \cr 
&  +\sum_{n=0}^\infty \left[ {e^{{\pi i \over 2} N}\over \left(\Delta_n - {d\over 2}  - i\nu \right)^{N+1}} + c.c. \right] 
{\rm Res}_{\Delta \to \Delta_n}(r_{J,\Delta}) {K_{J+n+1, J+d-1} \over K_{J, J+d+n}}  c_{J+n+1}(J+d -1 ) \ .
}}
This equation can also be obtained by taking $N$ $\nu$-derivatives of \disppp, but unlike \disppp\ it is valid for any $\delta_\phi$ if we chose $N$ as in \nchoice.

\subsec{Contribution of Extra Poles}

To make the dispersion relations \disppp, \dispthree\ useful we need to say something about the contribution of extra poles \extradef. Here we compute the large $n$ tails of the sum \extradef. These tails are fixed by the Euclidean $t$-channel OPE, since the large $n$ asymptotics of $c_{J+n+1}(J+d-1)$  is controlled by the large spin expansion \clargeJ. The tails generate generic non-integer powers in the large $\nu$ expansion. In contrast, any fixed $n$ term in \extradef\ is non-universal. It generates terms of the type ${1 \over \nu^{2k}}$ at large $\nu$. These are mapped to the contribution of individual operators in the $s$-channel.

Let us see how this works in details. We write the contribution of extra poles \extradef\ as
\eqn\extra{
{\rm extra} = \sum_{n=0}^\infty \delta_{n,odd} E_n \left( {1\over \nu_n + i \nu} + {1\over \nu_n - i \nu} \right) \ ,
}
where 
\eqn\nun{\eqalign{
&\Delta_n = {d\over 2} + \nu_n = J+d+n, \cr 
&E_n =  {1\over n+1} \left[ { \Gamma\left( {n\over 2} + 1 \right)  \over  \Gamma\left( - {n\over 2} \right) \Gamma(n+1) } \right]^2 { (J+1)_{n+1} \over  \left(J+{d\over 2} \right)_{n+1} } c_{J+n+1}(J+d-1) \ .
}}
We are interested in the large $n$ tails of the sum, so we expand each term as
\eqn\An{
E_n = \sum_{\chi} n^{\gamma_\chi - 1} \sum_{j=0}^\infty {e_j^{(\chi)} \over n^j} \equiv \sum_{\chi} E_n^{(\chi)} \ ,
}
where we also sum over contributions of primary operators $\chi$ in the $t$-channel to the inversion formula \inversionf\ for $c_{J+n+1}(J+d-1)$. For example, using \clargeJ, for the unit operator we have
\eqn\idalpha{
\chi = \hat 1: \qquad \gamma_{\hat 1} = 4\delta_\phi - 2, \qquad e_0^{(\hat 1)} = 2^{-\gamma_{\hat 1} +2J+d+1} {\Gamma\left(J+{d\over 2} \right) \over \Gamma(J+1)} {\Gamma\left( {d\over 2} -\Delta_{\phi} \right) \over \Gamma(\Delta_{\phi} ) }
}
Let us compute the contribution of a single $t$-channel primary operator $\chi$
\eqn\extraa{\eqalign{
&{\rm extra}(\chi)  \equiv \sum_{n=0}^\infty \delta_{n,odd} E_n^{(\chi)} \left( {1\over \nu_n + i \nu} + {1\over \nu_n - i \nu} \right) =\sum_{j=0}^\infty e_j^{(\chi)} \sum_{k=0}^\infty {(2k+1)^{\gamma_\chi - 1 -j} \over \nu_{2k+1} + i \nu} + c.c. = \cr 
&=\sum_{j=0}^\infty e_j^{(\chi)} \sum_{k=0}^\infty (2k+1)^{\gamma_\chi - 1 -j} \int_0^\infty dt ~ e^{-t( \nu_{2k+1} + i \nu)}  + c.c. = \cr 
& = 2\sum_{j=0}^\infty e_j^{(\chi)} \int_0^\infty dt ~ \cos(t \nu) e^{-t\left( J + {d\over 2} +1\right)} \sum_{k=0}^\infty (2k+1)^{\gamma_\chi - 1 -j} e^{-2t k}
}}
The sum over $k$ is given by the so called Lerch transcendent
\eqn\sumLerch{
\sum_{k=0}^\infty (2k+1)^{\gamma_\chi - 1 -j} e^{-2t k} = 2^{\gamma_\chi - 1 -j} \Phi(e^{-2t}, -\gamma_\chi + 1 +j, 1/2)
}
where by definition
\eqn\Lerchdef{
\Phi(z,s,a) \equiv \sum_{k=0}^\infty {z^k \over (k+a)^s}
}
For large $\nu$ the integral in \extraa\ localizes to $t=0$. Lerch transcendant \sumLerch\ has a useful expansion around this point \Erdelyi
\eqn\Lerchexp{
\Phi(z,s,a) = z^{-a} \left( \Gamma(1-s) (-\log z)^{s-1} + \sum_{k=0}^\infty \zeta(s-k,a) {(\log z)^k \over k!} \right)
} 
where $\zeta(s,a) = \sum_{k=0}^\infty (a+k)^{-s}$ is Hurwitz zeta function. Using this expansion we get
\eqn\extrab{\eqalign{
&{\rm extra}(\chi) = \sum_{j=0}^\infty e_j^{(\chi)} \Gamma(\gamma_\chi-j) \left( J + {d\over 2} \right)^{\gamma_\chi - 1 -j} {\cal F}_{j-\gamma_\chi} \left(\nu\over J+{d\over 2} \right) + \cr 
&+\sum_{k=0}^\infty \t e_k^{(\chi)} {(-2)^k \over k!}  \left( J + {d\over 2} \right)^{-(k+1)} {\cal F}_{k}\left(\nu\over J+{d\over 2} \right)
}}
where we defined
\eqn\betadef{\eqalign{
&\t e_k^{(\chi)} = \sum_{j=0}^\infty e_j^{(\chi)} 2^{\gamma_\chi-j} \zeta(-\gamma_\chi+1+j-k, 1/2) \cr 
&{\cal F}_{s-1}(\nu) = \int_0^\infty dt ~ \cos(t \nu) e^{-t}  t^{s-1} = { \Gamma(s) \cos\left( s \arctan \nu \right) \over (1+ \nu^2 )^{s/2} }
}}
The function ${\cal F}_s(\nu)$ can be expanded at large $\nu$ by changing the integration variable in \betadef\ to $x = t\nu$ and expanding the exponent. The result is
\eqn\Fexp{
{\cal F}_{s-1}(\nu) = \sum_{n=0}^{\infty} {(-1)^n \over n!} \Gamma(n+s) \cos\left(  {\pi \over 2} (n+s) \right) {1\over \nu^{s+n} }  
}
The expansion \Fexp\ makes it clear that the first line in \extrab\ gives the universal tails controlled by the $t$-channel which are generically non-integer powers of $1\over \nu$. This happens because heavier operators in the $t$-channel have smaller $\gamma_\chi$\foot{This can be explicitly checked in $d=2,4$ by computations outlined in section 4.} and therefore give suppressed contributions in \extrab. The second line in \extrab\ makes it clear that each operator contributes to all integer powers of $1\over \nu$.  Using \Fexp\ we can write an expansion at large $\nu$
\eqn\extrafrac{\eqalign{
& {\rm extra}(\chi) = \cr 
&\nu^{\gamma_{\chi} - 1} \sum_{n=0}^\infty {(-1)^n\over \nu^n} \left( J+{d\over 2} \right)^{n} \Gamma(n-\gamma_\chi +1) \cos {\pi \over 2} (n-\gamma_\chi+1)  
\sum_{j=0}^n e_j^{(\chi)} {(-1)^j \Gamma(\gamma_\chi - j )\over \left( J+{d\over 2} \right)^{j} (n-j)!}   \cr 
&+(\rm{even ~ integer ~ powers}) \ ,
 }}
where $e_j^{(\chi)}$ and $\gamma_\chi$ are defined by the expansion \An. Note that due to shadow symmetry we expect all the odd $n$ powers in the sum to cancel. In all concrete computations it is indeed what happens. 
Let us quote the leading universal term in \extrafrac\ from the contribution of the unit operator \idalpha, which we will use later 
\eqn\extraunit{
{\rm extra}(\hat 1) = - {1 \over \cos 2 \pi \delta_{\phi}} d_J \nu^{4 \delta_\phi - 3} + ...  \ , 
}
where $d_J$ is given by \deltadef .

For $\delta_\phi \geq {3\over 4} $, when we have to use the dispersion relation with subtractions, we can obtain the asymptotics of the sum on the RHS of \dispthree\ by taking $N$ $\nu$-derivatives of \extrafrac.

The $t$-channel information \extrafrac, though non-trivial, is impossible to interpret in terms of individual primary operators and three-point functions in the $s$-channel. The kernel $2\nu' \over \nu'^2 + \nu^2$ in \disppp\ is centered around $\nu$, but has tails going to arbitrarily large $\nu'$. However, it turns out that for the large $\nu$ asymptotic it is possible to get rid of this tail and replace the kernel by the indicator function 
\eqn\indicator{
\theta(0 \leq \nu' \leq \nu)= \cases{ 1, \quad 0 \leq \nu' \leq \nu \cr 0, \quad {\rm otherwise}}
}
using a certain tauberian theorem. This will give us a more direct probe of the spectral data in the s-channel and will be the topic of section 5.

\newsec{Adding Extra Operators to The Large $\Delta$ Expansion}

So far we have explicitly discussed only the contribution of the unit operator in the $t$-channel. Let us briefly generalize the discussion to an arbitrary operator in the $t$-channel.

The basic feature of the large $\nu$ expansion is that non-analytic tails ${1 \over \nu^\alpha}$ with generically non-integer $\alpha$ are controlled by the $t$-channel OPE and are, thus, computable. The analytic terms ${1 \over \nu^{2 k} }$ with integer $k$, on the other hand, come from individual operators in the $s$-channel and are non-universal. 

Let us consider the problem of adding an extra operator in the $t$-channel. It is convenient not to distinguish between the primaries and descendants. The contribution of a state with quantum numbers $(h, \bar h)$ to the correlator is given by 
\eqn\tidB{
\G(z, \b z) = \left( z \b z \over (1-z) (1-\b z) \right)^{\Delta_\phi} (1-z)^h (1-\bar z)^{\bar h}.
}
Its contribution to the double discontinuity is given by
\eqn\dDiscGFFB{
{\rm dDisc}~ {\cal G}(z, \b z) = 2 \sin^2 (\pi (\Delta_\phi - \bar h) ) \left( z \b z \over (1-z) (1-\b z) \right)^{\Delta_\phi} (1-z)^h (1-\bar z)^{\bar h} .
}
We would like to plug this into the inversion formula and study the result in the large $\Delta$ limit. To be able to compute integrals in the inversion formula explicitly it will be convenient to expand in $1-z \over z$ instead of $1-z$
\eqn\rewritinpo{\eqalign{
(1-z)^h &= \left( {1-z \over z} \right)^h \left(1 + {1-z \over z} \right)^{-h} = \cr
&=\left( {1-z \over z} \right)^h \sum_{n=0}^\infty {\Gamma(1-h) \over \Gamma(n+1) \Gamma(1-h-n)} \left( {1-z \over z} \right)^n 
\ .
}}
We will see that this expansion translates into ${1 \over \Delta}$ expansion. 

Let us consider the case $d=2$. The case of $d=4$ is similar. We believe the same picture persists in all $d$, even though we have not proved that. In the inversion formula we are interested in the following integral
\eqn\baseint{\eqalign{ 
I^{p_0}_{p_1}(h) &= \int_0^1 {d z \over z^2} k_h (z) z^{p_0}  (1-z)^{p_1} = { \Gamma(p_0 + {h \over 2} -1 )\Gamma(p_1+1) \over \Gamma(p_0 + p_1 +{h \over 2})} \times \cr
& \times  \ _3 F_2 \left({h \over 2}, {h \over 2}, p_0+{h \over 2}-1; h, p_0 + p_1 + {h \over 2}  ; 1 \right) \ . 
}}
In terms of $I^{p_0}_{p_1}(h)$ we get the following contribution to the partial wave from \dDiscGFFB
\eqn\contrib{\eqalign{
c_J^{h,\b h}(\Delta) =  \sin^2 (\pi (\Delta_\phi - \bar h) )  \left[  I^{\Delta_\phi}_{h-\Delta_\phi}(\Delta + J) I^{\Delta_\phi}_{\bar h - \Delta_\phi}(\t \Delta + J) + (\Delta \to \t \Delta) \right]
}}
where $\t \Delta = 2-\Delta$ is the shadow of $\Delta$ in 2d. We are interested in the large $h$ limit of $I^{p_0}_{p_1}(h)$. For this purpose it is convenient to introduce a simpler integral
\eqn\simplerint{
I_{p}(h) = \int_0^1 {d z \over z^2} k_h (z) \left( {z \over 1-z} \right)^p = { \Gamma(h)  \Gamma({h \over 2} +p -1) \Gamma(1-p)^2\over \Gamma({h \over 2})^2 \Gamma({h \over 2}-p +1)} .
}
In terms of this simpler integral and using \rewritinpo\ we can write 
\eqn\relationformula{
I^{p_0}_{p_1}(h) = \sum_{n=0}^{\infty} I_{-p_1-n}(h) {(-1)^n \Gamma(p_0+p_1+n)\over \Gamma(p_0+p_1) \Gamma(n+1)}.
}
Expanding \simplerint\ at large $h$ the formula \relationformula\ provides us with an expansion at large $h$ for $I_{p_1}^{p_0}(h)$. Now we can readily compute the large $\Delta$ expansion of $c_J^{h,\b h}(\Delta)$ \contrib. The leading term takes the form
\eqn\largedexp{
c_J^{h, \b h}(1+i \nu) =\pi 4^{2+h + \bar h + J - 2 \Delta_\phi } \nu^{4 \delta_\phi - 3 - 2(h + \bar h)} {\cos(\pi(h - \bar h)) \Gamma(1+h-\Delta_\phi)^2 \over \Gamma(\Delta_\phi - \bar h)^2} + ... \ .
}
Given a primary operator in the $t$-channel with dimension $\Delta_{\chi}$ and spin $J_{\chi}$ its $t$-channel conformal block involves terms $h = { \Delta_{\chi} \mp J_{\chi} \over 2}$ and $\bar h = {\Delta_{\chi} \pm J_{\chi} \over 2}$ together with corrections which are integer powers $(1-z)^n (1-\bar z)^m$. 

One can wonder about the convergence of the $t$-channel OPE after we applied the inversion formula. If one naively expands \largedexp\ at large $\Delta_{\chi}$ one gets that it diverges very quickly $e^{\Delta_\chi \log \Delta_\chi}$. This however only signifies that the large $\nu$ and large $\Delta_\chi$ limits do not commute. Taking the full answer \contrib\ (instead of the first term in \relationformula)  one can check that it behaves at large $\Delta_\chi$ like a power. The situation is similar to the one in the context of the large spin expansion \refs{\AldayEWA,\SimmonsDuffinWLQ}. 

From the discussion above the contribution of an operator $\chi$ to the large $\nu$ expansion takes the form
\eqn\primarycontr{
c_J^{h, \bar h}({d / 2}+i \nu) =\nu^{4 \delta_\phi - 3 - 2\Delta_{\chi}} \sum_{n=0}^{\infty} {\alpha_n^{h, \b h} \over \nu^{2 n}} ,
}
where all coefficients are computable using the Lorentzian inversion formula. Moreover, the contribution of heavier operators $\chi$ in the $t$-channel is suppressed at large $\nu$, as can be seen from \primarycontr. 

\subsec{Large $\nu$ Summary}

Based on the discussion in the previous subsection we end up with the following dispersion relation at large $\nu$
\eqn\spectraldens{
\int_0^{\infty} d \nu' \ \rho_J^{OPE} \left({d \over 2} + \nu' \right) {2 \nu' \nu \over  \nu'^2 + \nu^2} = \sum_{\chi} \sum_{n=0}^\infty \alpha_n^{(\chi)} \nu^{- \delta_\chi - 2n} + \sum_{k=1}^{\infty} a_k \nu^{- 2 k+1}
}
where $\chi$ are $t$-channel operators contributing to the RHS of the dispersion relation \disppp\ via the inversion formula. The numbers $\alpha_n^{(\chi)}$ and $\delta_\chi$ are computable using $t$-channel OPE, while $a_k$ receive contributions from all operators in the $t$-channel and, therefore, are non-universal. The same conclusion holds for dispersion relations with subtractions. We would like to use  \spectraldens\ to make some predictions about the asymptotic behavior of the spectral density itself. This is the subject of tauberian theorems. We discuss this in the next section. 

Collecting formulas \clargeDelta, \extraunit\ we find the leading contribution of the unit operator to the RHS of the dispersion relation
\eqn\unitfull{
\chi = \hat 1: \qquad \alpha_0^{(\hat 1)} = d_J \left( 1 - {1\over \cos 2\pi \delta_\phi} \right), \qquad \delta_{\hat 1} = -4\delta_\phi +2
}
where $d_J$ is defined in \deltadef. From \primarycontr\ we conclude that in 2d the powers $\delta_\chi$ are related to the operators $\chi$ in the $t$-channel OPE as follows
\eqn\relation{
\delta_{\chi} = -4 \delta_\phi + 2 + 2\Delta_{\chi} }
One can check that the same formula holds in 4d. Note that the LHS of \spectraldens\ has an extra factor of $\nu$ compared to \disppp. This is the source of an extra $+1$ in \relation.

\newsec{Asymptotics of Spectral Densities}

We would like to use dispersion relations to make predictions about the integrated weighted spectral density. This is done via a complex tauberian theorem. It will be again useful to introduce Cauchy moments 
\eqn\CauchymomentB{\eqalign{
F_{m}^{J}(\nu) &\equiv {1 \over (m-1)!} \int_0^\nu d \nu' (\nu-\nu')^{m-1} \rho_{J}^{OPE}({d / 2} + \nu') \ .
}}
These are also computed by the repeated integration of the weighted spectral density $\rho_J^{OPE}(\nu)$ as in \Cauchymoment. Now let us consider the dispersion relation \spectraldens
\eqn\dispexp{
\int_0^{\infty} d \nu' \ \rho_J^{OPE} \left({d / 2} + \nu' \right) {2 \nu' \nu \over  \nu'^2 + \nu^2} = R(\nu) =  \sum_{i} \alpha_i \nu^{- \delta_i}  + \sum_{k=1}^{\infty} a_k \nu^{- 2 k+1}
}
where we simply use the notation $\delta_i$ with the index $i$ containing information about both $\chi$ and $n$ in \spectraldens\ (which receives contribution from the $n$-th level descendants). We also arrange the powers such that $\delta_{i+1} > \delta_i$. As we discussed above, there are two types of terms: computable using the $t$-channel OPE (first sum on the RHS of \dispexp) and terms that are sensitive to the details of the spectrum (second sum on the RHS of \dispexp). 

 To go from dispersion relations to statements about the Cauchy moments, we need to understand what is the region of validity of \dispexp\ in the complex $\nu$-plane, or, equivalently, in the $\Delta$-plane. Recall the relation between the two: $\Delta = {d\over 2} + i \nu$. As we discussed in section 3, the large $\Delta$ (or $\nu$) expansion breaks down close to the real axis, where partial waves have poles at the locations of primary operators. In general, we do not know what exactly this region is. However, on general grounds we expect that the smooth polynomial behavior emerges as soon as ${\rm Im}[\Delta]\gg 1$. This is also what we observed in concrete examples and from plugging separate $t$-channel operators in the inversion formula. In all these cases the corrections are suppressed by $e^{- {\rm Im}[\Delta]}$. In particular, we assume that \dispexp\ is valid for $|\Delta| \gg 1$ and ${\rm Im}[\Delta] \gtrsim |\Delta|^\eps$ where $\eps$ is an arbitrary but fixed number.

Given that \dispexp\ is valid in the complex $\nu$ region described above, the Cauchy moments have the following asymptotic at large $\nu$ (see appendix B for a proof):
\eqn\ebstilt{\eqalign{
F_m^{J}(\nu) &= \sum_{i} \alpha_i \beta_{m}(\delta_i) \nu^{m-\delta_i-1} + \sum_{k=1}^m b_{k} {\nu^{m-k} \over (m-k)!} +O(\nu^{- \delta_{\hat 1} - 1}) ,  \cr
\beta_{m}(\delta) &={\cos {\pi \delta \over 2} \over \pi} {\Gamma(-\delta) \over \Gamma(m- \delta)}  \ ,
}}
where $b_{k}$ are computable if $R(\nu)$ is known exactly, but cannot be computed based solely on the large $\nu$ data. The error term is defined by the smallest $\delta_i$ in \dispexp, which is $\delta_{\hat 1}$.\foot{Strictly speaking, in appendix B we prove \ebstilt\ with the error estimate being $O(\nu^{- \delta_{\hat 1} - 1+\eps})$, where $\eps>0$ is arbitrarily small and fixed. Therefore, we leave a possibility of multiplying the error estimate by a function $\Lambda(\nu)$ growing slower than a power. It will be implicit in what follows.}
The formula \ebstilt\ constitutes the statement of {\it the complex tauberian theorem for Stieltjes transform}, which we prove in appendix B.

Note that $\beta_{m}(\delta)$ has poles at even integer $\delta$.  The prescription in this case is to keep the regular piece in $\beta_{m}(\delta) \nu^{m-\delta-1}$ which produces $\nu^{m - \delta - 1} \log \nu$ with computable coefficient and throw away the pole (see appendix B).

Heuristically, the coefficients of the first sum in \ebstilt\ can be found by taking a naive power-law ansatz for $\rho_J^{OPE}(\nu)$, as in appendix B.5, and computing Cauchy moments. The proof of \ebstilt, however, is much more subtle. In particular, it is crucial that we consider $m$-th Cauchy moment \CauchymomentB\ in order to capture, roughly speaking, $m$ subleading terms in \ebstilt.  Again, the key ingredients are positivity of $\rho_J^{OPE} $ and validity of the large $\nu$ expansion in the complex domain. 

Finally, notice that in \dispexp\ we used dispersion relations without subtractions. One can check that subtractions do not affect the result (see appendix B). 

We would like to stress that \ebstilt\ is a rather non-trivial consequence of \dispexp. For example, consider the first Cauchy moment $F_1^J$. For a discrete operator spectrum it is a discontinuous function with a ``staircase" shape (e.g. see fig.6). It is remarkable that features of this staircase are captured by a smooth function on the RHS of \ebstilt.

\subsec{Leading Asymptotic}

Given the result \ebstilt\ let us discuss the leading asymptotic for the integrated weighted spectral density. Recall that introducing $\delta_{\phi} = \Delta_\phi- {d-2 \over 2}$ the leading asymptotic in \ebstilt\ comes from the unit operator \unitfull. Setting $m=1$ in \ebstilt\ we get
\eqn\leadingasymExA{\eqalign{
\int_0^\nu d \nu' \rho_{J}^{OPE}({d /2} + \nu') &= f_J   {\nu^{4\delta_\phi-2} \over 4\delta_\phi-2} + b_1 +O(\nu^{4\delta_\phi-3}) \ , \cr
f_J &= [1+(-1)^J]\pi^2 {2^{2J+d-4\delta_\phi +2} \Gamma\left( J + {d\over 2} \right) \over \Gamma(J+1) \Gamma(\Delta_\phi)^2\Gamma( \delta_\phi )^2 } \ ,
}}
where we used \unitfull. Note that $\nu^{4\delta_\phi-2}$ dominates over $b_1$ only for $\delta_{\phi}  > {1 \over 2}$, i.e. for operators with $\Delta_{\phi} > {d - 1 \over 2}$. 

For $\delta_{\phi} < {1 \over 2}$ the constant term in \leadingasymExA\ will dominate. Therefore, we consider the combination $F_2^{J}(\nu) - \nu F_1^{J}(\nu)$ in which $b_1$ cancels and get 
\eqn\leadingasymExB{\eqalign{
\int_0^\nu d \nu' \nu' \rho_{J}^{OPE}({d / 2} + \nu') &=f_J   {\nu^{4\delta_\phi-1} \over 4\delta_\phi-1} + b_2 + O(\nu^{4\delta_\phi-2}) \ .
}}
Again, the first term dominates for $\delta_{\phi} > {1 \over 4}$, i.e. $\Delta_{\phi} > {d-{3\over 2} \over 2}$.

For $0<\delta_{\phi} < {1 \over 4}$ we can consider $F_3^{J}(\nu) - \nu F_2^{J}(\nu) + {\nu^2 \over 2} F_1^J(\nu)$ to remove $b_1,b_2$ and obtain
\eqn\leadingasymExC{\eqalign{
\int_0^\nu d \nu' \nu'^2 \rho_{J}^{OPE}({d / 2} + \nu') &=f_J   {\nu^{4\delta_\phi} \over 4\delta_\phi} + b_3 + O(\nu^{4\delta_\phi-1}) \ .
}}

The choice of particular combinations of $F_m^J$ that we used to obtain \leadingasymExB, \leadingasymExC\ will become clearer in the next subsection, when we will discuss how to systematically remove non-universal terms in \ebstilt\ containing coefficients $b_k$. For $\delta_{\phi} = {1\over 2}$ the prescription in \leadingasymExA\ is to keep the regular piece ${\nu^{4\delta_\phi-2} \over 4\delta_\phi-2} \to \log \nu$. Similarly for $\delta_\phi = {1\over 4}$ in \leadingasymExB. 

To summarize, the leading asymptotic of the integrated weighted spectral density in any unitary CFT is given by formulas \leadingasymExA, \leadingasymExB, \leadingasymExC\ depending on the scaling dimension of the external operator.

\subsec{Systematic Corrections}

There is still one last problem to be addressed in the formula \ebstilt.  Indeed, in the large $\nu$ expansion \ebstilt\ we encountered integer powers of $\nu$ that are not controlled by OPE. Therefore, it is more practical to take certain linear combinations of Cauchy moments \CauchymomentB\ such that these integer powers cancel. Depending on how many integer powers we would like to cancel we can consider combinations
\eqn\combination{
G_{m,k}^{J}(\nu) = F_m^{J}(\nu) + p_1 \nu F_{m-1}^{J}(\nu) + ... + p_{m-k} \nu^k F_{m-k}^{J}(\nu) \ ,
}
where $k$ is the number of integer power terms that we want to remove. Using \ebstilt\ we get the following set of equations for $p_i$ 
\eqn\pequat{\eqalign{
&{1 \over (m-1)!} + {p_1 \over (m-2)!}  + ... + {p_k \over (m-k-1)!} = 0 , \cr
&...  \cr
&{1 \over (m-k)!} + {p_1 \over (m-k-2)!}  + ... + {p_k \over (m-2k)!} = 0 \ ,
}}
which ensure that terms $\nu^{m-1}$, $\nu^{m-2}$, ... , $\nu^{m-k}$ cancel. The solution of the equations \pequat\ takes the form
\eqn\solutionforp{
p_j = (-1)^j {k! \over j! (k-j)!} {\Gamma(m-j) \over \Gamma(m)} \ .
}
In terms of the weighted spectral density the moments $G_{m,k}^{J}(\nu)$ take the form
\eqn\combin{\eqalign{
G_{m,k}^{J}(\nu) &={(-1)^k \over (m-1)!} \int_0^\nu d \nu' \ \nu'^k(\nu - \nu')^{m-k-1}  \rho_{J}^{OPE}({d / 2} + \nu')   \ ,
}}
where $k$ is the number of integer power terms that we want to subtract. Note that the maximal error comes from the last term in \combination\ since all $F_m^J$ have the same error term \ebstilt. Namely $G_{m,k}^{J}$ has a remainder term $O(\nu^{k + 4 \delta_\phi - 3 })$. Nevertheless, by taking an appropriate $m$ we can access as many terms in the $t$-channel OPE as we like. 

In the discussion above for the leading asymptotic the combination $F_2^{J}(\nu) - \nu F_1^{J}(\nu)$ giving \leadingasymExB\ is nothing but $G_{2,1}^{J}(\nu)$. Similarly, the second moment \leadingasymExC\ is nothing but  $G_{3,2}^{J}(\nu)$.

Further, we can access subleading terms in the $t$-channel OPE by considering higher $G_{m,k}^J$ moments. These have the error term $\delta G_{m,k}^{J} = O(\nu^{k+4\delta_\phi-3})$ and non-universal integer-power terms with the maximal power $\nu^{m-k-1}$. On the other hand, they enhance the OPE terms by a factor $\nu^{m-1}$. By taking $m$ to be large enough we can always extract arbitrary number of OPE controlled tails.  In particular, given a term $\alpha_i \nu^{- \delta_i}$ in \dispexp\ its contribution to $G_{m,k}^{J}$ is
\eqn\contribution{\eqalign{
G_{m,k}^{J} (\nu)&= \sum_i  \left( {\cos {\pi \delta_i \over 2} \over \pi } {\Gamma(1-m)\Gamma(k-\delta_i) \over  \Gamma(1+k-m)  \Gamma(m-\delta_i)} \right) \alpha_i \nu^{m-1-\delta_i} \cr
&+ b_{k+1} \nu^{m-k-1} + \dots + b_m + O(\nu^{k-\delta_{\hat 1} -1}) \ , ~~~ m>k \ .
}}
where the leading contribution comes from the unit operator \unitfull\ and all $\alpha_i, \delta_i$ can be computed from the $t$-channel expansion.

Let us discuss some properties of the pre-factor $\left( {\cos {\pi \delta_i \over 2} \over \pi } {\Gamma(1-m)\Gamma(k-\delta) \over \Gamma(1+k-m) \Gamma(m-\delta)  } \right)$ which enters \contribution. For $m>k$ it has poles for even integer $\delta$'s 
\eqn\polesker{
k \leq \delta_p = 2 p < m \ , 
}
which correspond to the contribution of operators with dimensions $\Delta_p = 2 \delta_\phi + p - 1$. As above, the right prescription is to throw away the pole and keep the finite term with $\log \nu$. Moreover, due to these poles the contribution of operators with dimensions close to $\Delta_p$ gets enhanced.  Finally, at large $\delta$ we have $\left( {\cos {\pi \delta \over 2} \over \pi } {\Gamma(1-m)\Gamma(k-\delta) \over \Gamma(m-\delta) \Gamma(1+k-m) } \right) \sim {1 \over \delta^{m-k}}$.

Let us emphasize that even though we derived \ebstilt\ for dispersion relations without subtractions \CauchymomentB, it holds for the most general case \dispthree\ as we show in appendix B. 

Therefore, we can systematically access the $t$-channel OPE data by studying the moments of the $s$-channel weighted spectral density. This fact embodied in the formula \contribution\ is the main result of our paper.

\newsec{Examples}

In this section we test the formula \contribution\ in Generalized Free Field theory (GFF) and 2d Ising. In particular, we will see that in these examples the large $\Delta$ expansion will turn out to work well already for small values $\Delta$. We will also make predictions for 3d Ising.

\subsec{Generalized Free Field}

The simplest example where we can test our claims is Generalized Free Field theory. Indeed, in this case $c_{J}(\Delta)$ is explicitly known \cGFFd. It is a meromorphic function with all the expected properties. One can write dispersion relations for it and check the corresponding complex tauberian theorems. 

One slightly non-trivial fact in this case is that the large $\nu$ expansion of ${\rm extra}$ contribution is simply related to $c_{J}^{GFF}({d / 2} + i \nu)$\eqn\relationextraGFF{
{\rm extra} =- {1 \over \cos 2 \pi \delta_\phi} c_{J}^{GFF}({d / 2} + i \nu) + ({\rm even \ integer \ powers} ) \ .
} 
We have not derived this result to all orders, but checked analytically first few terms in the large $\nu$ expansion. Therefore, using \relationextraGFF\ we can easily make predictions to an arbitrarily high order for the moments \combin.\foot{In $d=2$ we observed relations similar to \relationextraGFF\ for generic operators. We have not tried to generalize \relationextraGFF\ to arbitrary $d$. }

The first few terms in the expansion of $c_J^{GFF}({d \over 2} + i \nu)$ take the form
\eqn\fewtermsintheexp{\eqalign{
c_J^{GFF}({d / 2} + i \nu) &= d_J \nu^{4 \delta_\phi - 3} \left( 1 + {\hat \alpha_1 \over \nu^2} + {\hat \alpha_2 \over \nu^4} + ... \right) , \cr
\hat \alpha_1 &= {8 \over 3} \delta_{\phi}^3 - 8 \delta_{\phi}^2 + \delta_{\phi} \left(2 J^2 + 2 (d-2) J +{d \over 2} (d-4) +{28 \over 3} \right) \cr
&-{3 \over 2} J^2 + {J \over 2} (7 - 3d) +{d \over 8} (14-3 d) - 4 \ ,
}}
and similarly for higher $\hat \alpha_i$ which can be trivially computed by expanding \cGFFd .

Let us now plot a few moments to see how the formula \contribution\ plays out. For concreteness we set $d =3$, $J=0$, $\delta_{\phi} = {5 \over 8}$.

\ifig\gffzeromoment{$G_{1,0}^{J=0}$ for GFF as a function of $\nu$. Parameters are chosen to be $d =3$, $J=0$, $\delta_{\phi} = {5 \over 8}$. 
Based on \leadingasymExA\ we expect the leading term to be ${8 \Gamma(-{1 \over 4})^2 \over \sqrt \pi}\sqrt{\nu} + b_1$. We fit the constant to be $b_1 \approx 79.4$. The asymptotic formula works very well down to $\nu=0$.
} {\epsfxsize3in\epsfbox{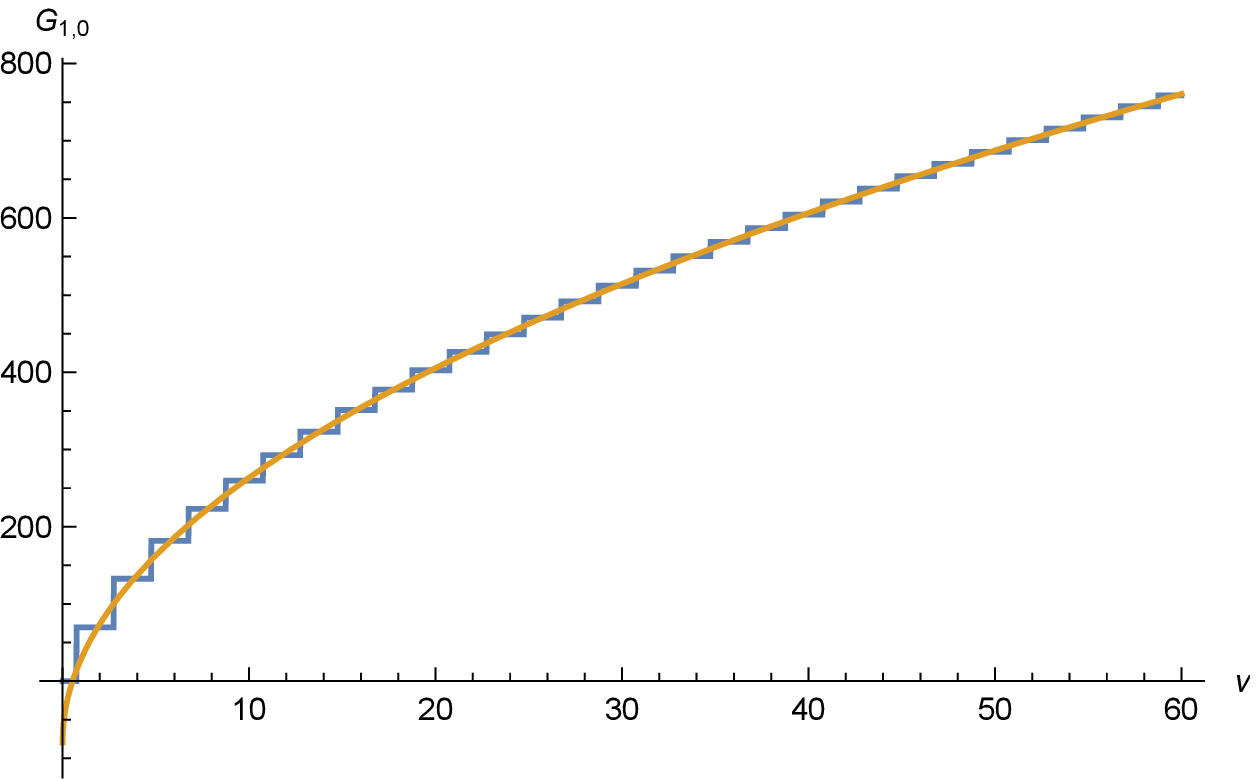}}

\ifig\diffgffzero{Error term for $G_{1,0}^{J=0}$ as a function of $\nu$. We plot $ \sqrt{\nu} \delta G_{1,0}^{J=0} =(G_{1,0}^{J=0}-{8 \Gamma(-{1 \over 4})^2 \over \sqrt \pi}\sqrt{\nu} - b_1) \times \sqrt{\nu}$ for GFF.  
Based on \leadingasymExA\ we expect the difference between $G_{1,0}$ and the fit in \gffzeromoment\ to be $O(\nu^{-1/2})$.} {\epsfxsize2.5in\epsfbox{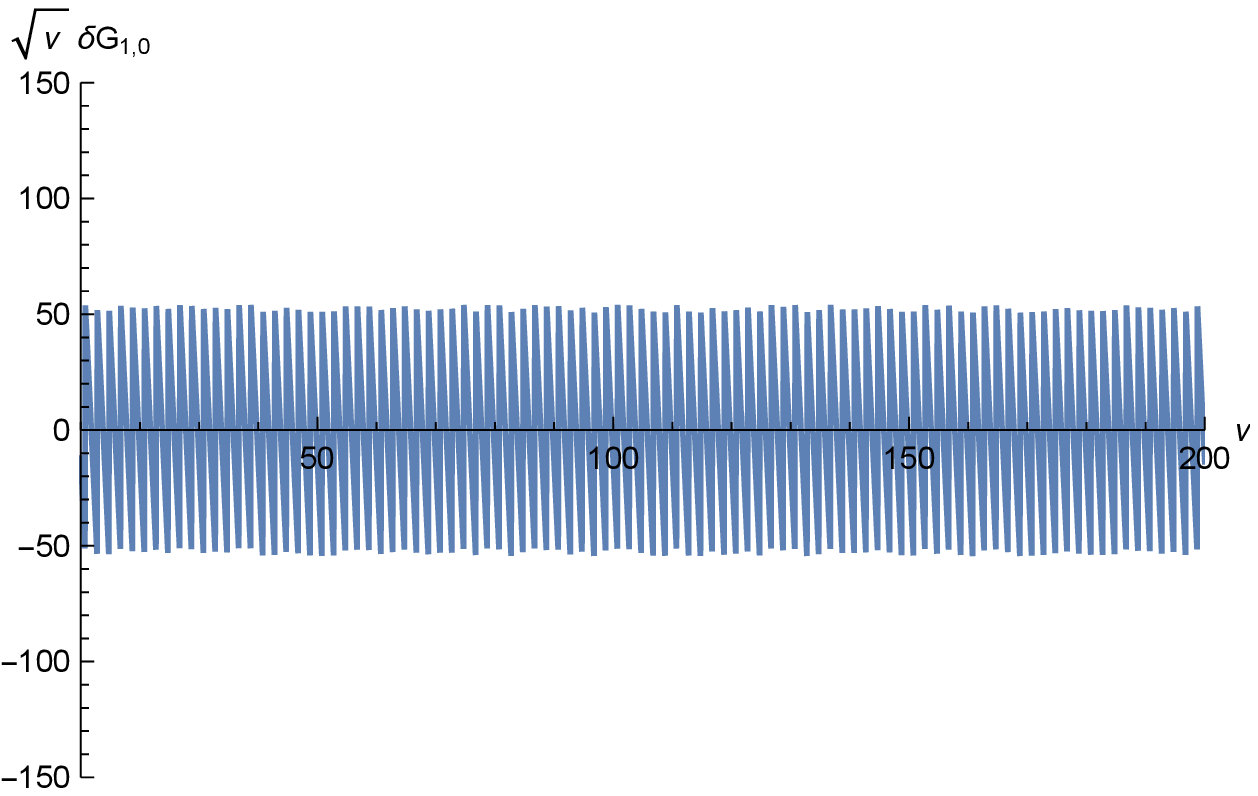}}

In the \gffzeromoment\ we plot the result for the leading asymptotic of the integrated weighted spectral density. In \diffgffzero\ we present the result for the difference between $G_{1,0}$ and the fit multiplied by $\sqrt{\nu}$. We see that the results are in perfect agreement with the formula \contribution. 

Next, let us consider the moment that is sensitive to the subleading tail $\hat \alpha_1$ in \fewtermsintheexp. One can check that the first moment in which we can access it is $G_{5,2}^{J=0}$ which takes the form
\eqn\subleadingmoment{
{1\over \nu^2} G_{5,2}^{J=0}(\nu) = {2048 \Gamma({7 \over 4})^2 \over 8505 \sqrt{\pi}} \nu^{5/2} - {608 \Gamma({7 \over 4})^2 \over 405 \sqrt{\pi}} \nu^{1/2}  + b_3 + O(\nu^{-1/2}) \ .
}
\ifig\gffsublead{Subleading tail in $G_{5,2}^{J=0}$ for GFF as a function of $\nu$. Parameters are chosen to be $d =3$, $J=0$, $\delta_{\phi} = {5 \over 8}$. 
We subtracted from both sides of \subleadingmoment\ the leading $\nu^{5/2}$ term. This way matching of the subleading tail can be seen very clearly. We fit $b_3 \approx 0.565$. We see that asymptotic formula works very well even for small $\nu$.
} {\epsfxsize3in\epsfbox{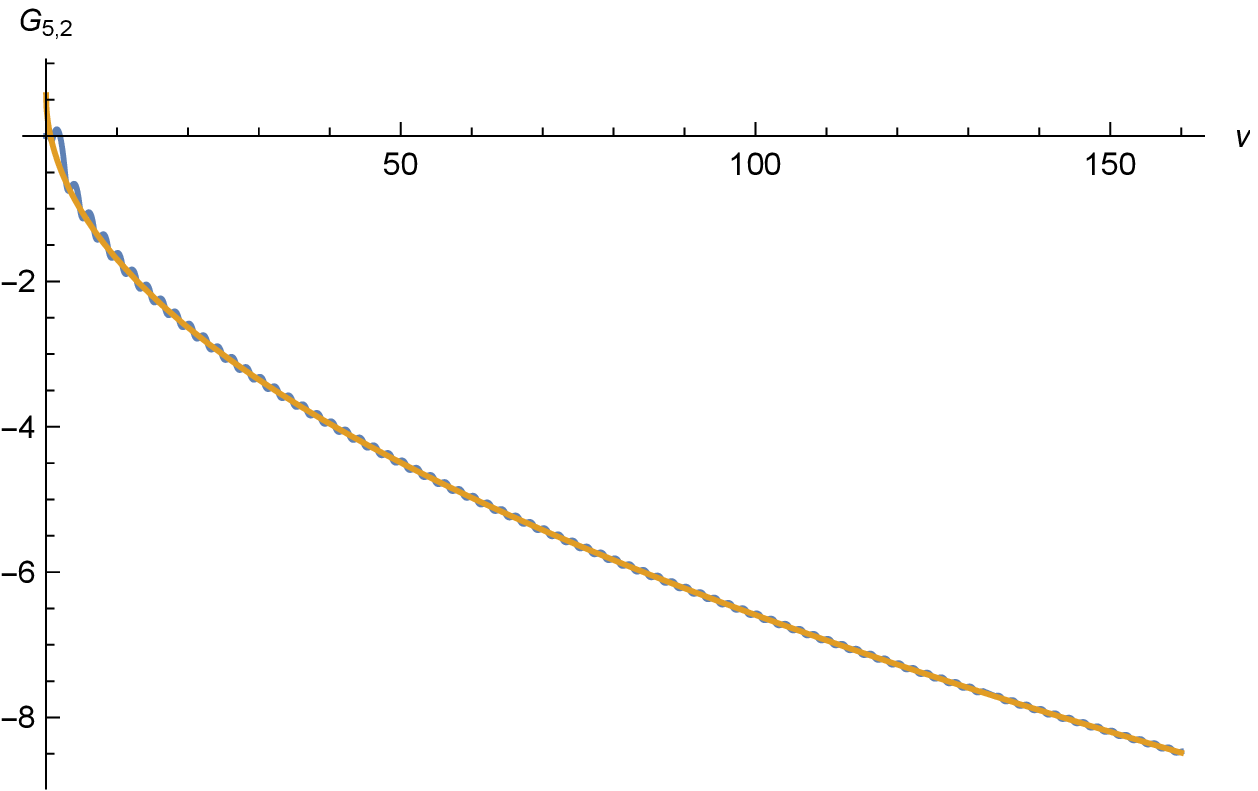}}
\noindent We subtract the leading tail ${2048 \Gamma({7 \over 4})^2 \over 8505 \sqrt{\pi}} \nu^{5/2}$ from both sides to isolate the subleading term and plot the result in the \gffsublead.

\ifig\gffsubleaderror{Error estimate for $G_{5,2}^{J=0}$ for GFF as a function of $\nu$. Parameters are chosen to be $d =3$, $J=0$, $\delta_{\phi} = {5 \over 8}$. 
We consider the difference between the LHS and the RHS in \subleadingmoment\ which we also multiply by $\sqrt{\nu}$. We see that the result agrees with \subleadingmoment .
} {\epsfxsize2.5in\epsfbox{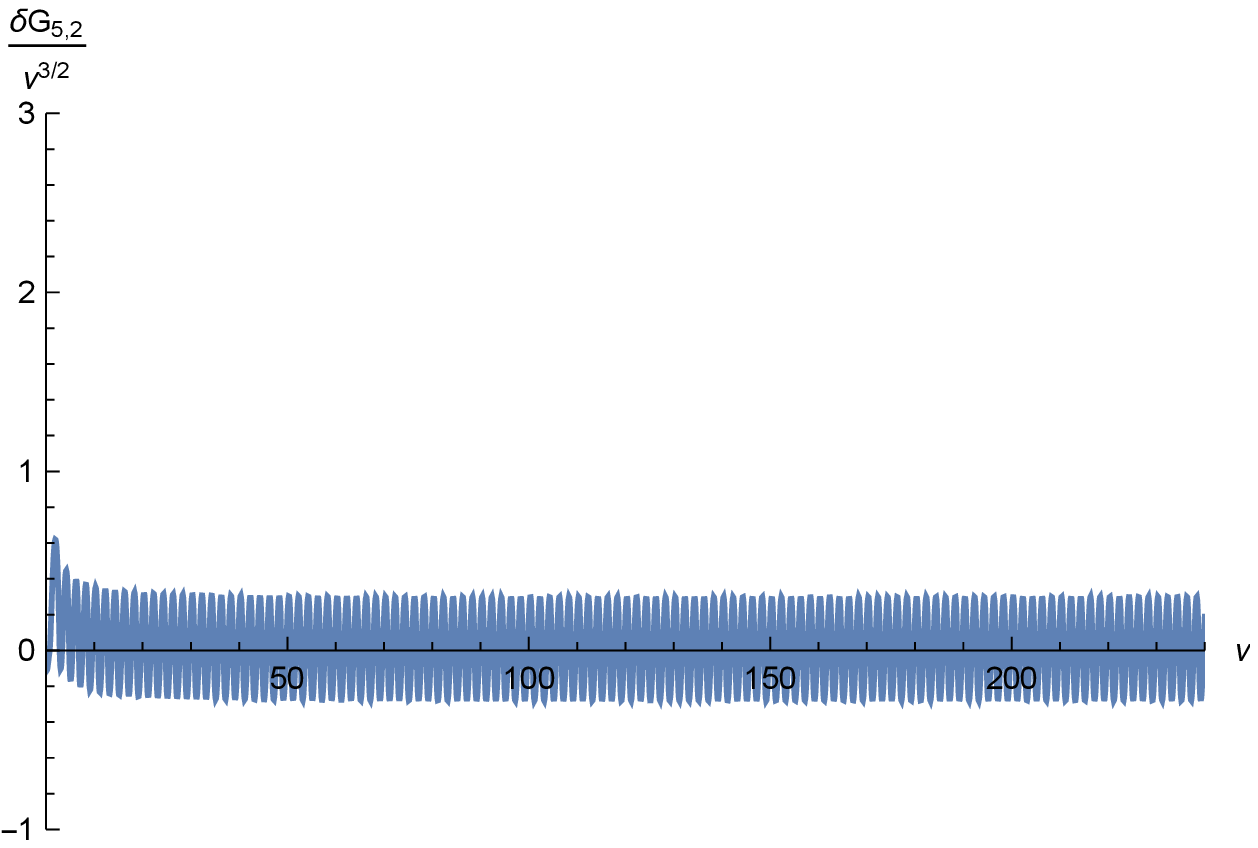}}

Finally, let us check the error estimate in \subleadingmoment . To do this we plot the difference between $G_{5,2}^{J=0}$ and the fitting function in \subleadingmoment. We also multiply it by $\nu^{1/2}$ to make it clearer. The result is plotted in the \gffsubleaderror.

Therefore, we see in this particular example that large $\Delta$ expansion works very well already for $\Delta \sim 1$. Moreover, we can clearly access subleading terms in the $t$-channel OPE by considering proper moments of the weighted spectral density.

\subsec{2d Ising Model}

Let us consider the four-point function of $\la \sigma \sigma \sigma \sigma \ra$ in the 2d Ising model of scalars with dimension $\Delta_\sigma = {1 \over 8}$. We have \BelavinVU\
\eqn\answer{\eqalign{
\la \sigma \sigma \sigma \sigma \ra &= {G(\rho, \bar \rho) \over (x_{12}^2 x_{34}^2)^{{1 \over 8}}} , \cr
G(\rho, \bar \rho) &={1 + \sqrt \rho \sqrt{ \bar \rho} \over (1 - \rho^2)^{{1 \over 4}} (1 - \bar \rho^2)^{{1 \over 4}}} .
}}

Since we are considering identical operators only even spins $J$ appear in the OPE. We have for conformal partial waves (see appendix B in \CaronHuotVEP) 
\eqn\partialwaves{\eqalign{
c_{J} (\Delta) = {c_{J, \Delta}  \over \kappa_{J+ \Delta}} &= I^{0}_{- {1 \over 4}}(\Delta + J) I^{0}_{- {1 \over 4}}(2 - \Delta + J) + I^{{1 \over 2}}_{- {1 \over 4}}(\Delta + J) I^{{1 \over 2}}_{- {1 \over 4}}(2 - \Delta + J) \cr
&- {1 \over \sqrt 2} \left( I^{{1 \over 2}}_{- {1 \over 4}}(\Delta + J) I^{0}_{- {1 \over 4}}(2 - \Delta + J) + I^{0}_{- {1 \over 4}}(\Delta + J) I^{{1 \over 2}}_{- {1 \over 4}}(2 - \Delta + J) \right) ,
}}
where
\eqn\intdef{\eqalign{
I^{p_0}_{p_1}(\beta) &= \int_0^1 d \rho {1 - \rho^2 \over 4 \rho^2} k_{\beta}(\rho) \rho^{p_0} (1 - \rho^2)^{p_1} , \cr
 k_{\beta}(\rho) &=(4 \rho)^{\beta/2} \ _2 F_1 \left({1 \over 2}, {\beta , \over 2}, {\beta + 1 \over 2}, \rho^2 \right) .
}}
The partial wave $c_{J} (\Delta) $ is symmetric under the 2d shadow transform $\Delta \to 2 - \Delta$, as expected. Since $\delta_{\sigma} = {1 \over 8} < {3 \over 4}$ we consider dispersion relations without subtractions
\eqn\dispersionrelations{
\int_0^\infty d\nu' \rho_J^{OPE}(1 +\nu') {2\nu' \over \nu'^2 + \nu^2} 
 = c_J(1 + i\nu)  + {\rm extra } .
}
Let us compute the contributions of first few terms in the RHS of \dispersionrelations . The three lightest operators in the $t$-channel OPE \OPE\ are 
\eqn\lightestoper{\eqalign{
\chi = \hat 1&: \qquad \Delta = 0, J=0, ~~~ p_{0,0} = 1 , \cr
\chi = \varepsilon&: \qquad \Delta = 1, J=0, ~~~ p_{1,0} = {1 \over 4} , \cr
\chi = T_{\mu \nu}&: \qquad \Delta = 2, J=2, ~~~ p_{2,2} = {1 \over 64} ,
}}
We will set $J=2$ in \dispersionrelations\ for concreteness, but qualitatively same conclusions hold for any spin (including $J=0$). We get the following contributions to the RHS of \dispersionrelations\ from the first three operators
\eqn\contributioninc{\eqalign{
c_{J=2}^{(\hat 1)}(1 + i\nu)  &=\sqrt 2 {128  \pi \Gamma({7 \over 8})^2 \over  \Gamma({1 \over 8})^2} \nu^{-5/2} \left( 1 - {333 \over 64} {1 \over \nu^2} + {284 565 \over 8192} {1 \over \nu^4} + ... \right) \  , \cr
c_{J=2}^{(\varepsilon)}(1 + i\nu)  &=\sqrt 2 {128  \pi \Gamma({11 \over 8})^2 \over  \Gamma(- {3 \over 8})^2} \nu^{-5/2} \left( {1 \over \nu^2}  - {655 \over 64} {1 \over \nu^4}  + ... \right) , \cr
c_{J=2}^{(T)}(1 + i\nu)  &=\sqrt 2 {\pi \Gamma(-{1 \over 8})^2 \over  \Gamma(- {15 \over 8})^2} \nu^{-5/2} \left( {1 \over \nu^4} + ... \right)
}}
Next, we evaluate tails coming from ${\rm extra }$. We follow the procedure described at the end of section 3.  The result is that extra terms contribute as follows

\eqn\contributionincnu{\eqalign{
extra(\hat 1)  &= - {128 \pi \Gamma({7 \over 8})^2 \over  \Gamma({1 \over 8})^2} \nu^{-5/2} \left( 1 - {333 \over 64} {1 \over \nu^2} + {284 565 \over 8192} {1 \over \nu^4} + ... \right) \  , \cr
extra(\varepsilon)&= {128 \pi \Gamma({11 \over 8})^2 \over  \Gamma(- {3 \over 8})^2} \nu^{-5/2} \left( {1 \over \nu^2} - {655 \over 64} {1 \over \nu^4} + ... \right)\ , \cr
extra(T) &=- {\pi \Gamma(-{1 \over 8})^2 \over  \Gamma(- {15 \over 8})^2} \nu^{-5/2} \left( {1 \over \nu^4} + ... \right) .
}}
Adding together \contributioninc\ and \contributionincnu\ we get the first three terms in the RHS of \dispersionrelations\foot{We present a numerical approximation of the coefficients not to clutter the text. Exact values are easily computable given the formulas in this section.}
\eqn\dispersionrelationsTDising{\eqalign{
\int_0^\infty d\nu' \rho_J^{OPE}(1 +\nu') {2\nu' \nu \over \nu'^2 + \nu^2} 
&= {1 \over \nu^{3/2}} \left( -4.92754 + {99.773 \over \nu^{2}} - {936.518 \over \nu^{4}}  + ... \right) \cr
&+ {a_1 \over \nu} +{ a_2 \over \nu^3} + ... \ ,
}}
which defines $\alpha_i, \delta_i$ of the first three terms in the RHS of \dispexp. Below we present some plots for the moments \contribution.

We get the following prediction for the second moment
\eqn\secondmomentdvadising{
\int_0^\nu d \nu' \nu'^2 \rho_{J=2}^{OPE}(1 + \nu') = {512 \sqrt{2} \pi^2 \over \Gamma({1 \over 8})^4} \nu^{1/2}+ b_3 + O(\nu^{-1/2}) .
}

\ifig\secondmoment{Second moment of the weighted spectral density for primary operators of spin $J=2$ in the 2d Ising model. We fit the constant $b_3 \approx 3.35$. The smooth curve is given by the RHS of \secondmomentdvadising. It is plotted against the exact integrated weighted spectral density of the 2d Ising model.}{\epsfxsize2.5in\epsfbox{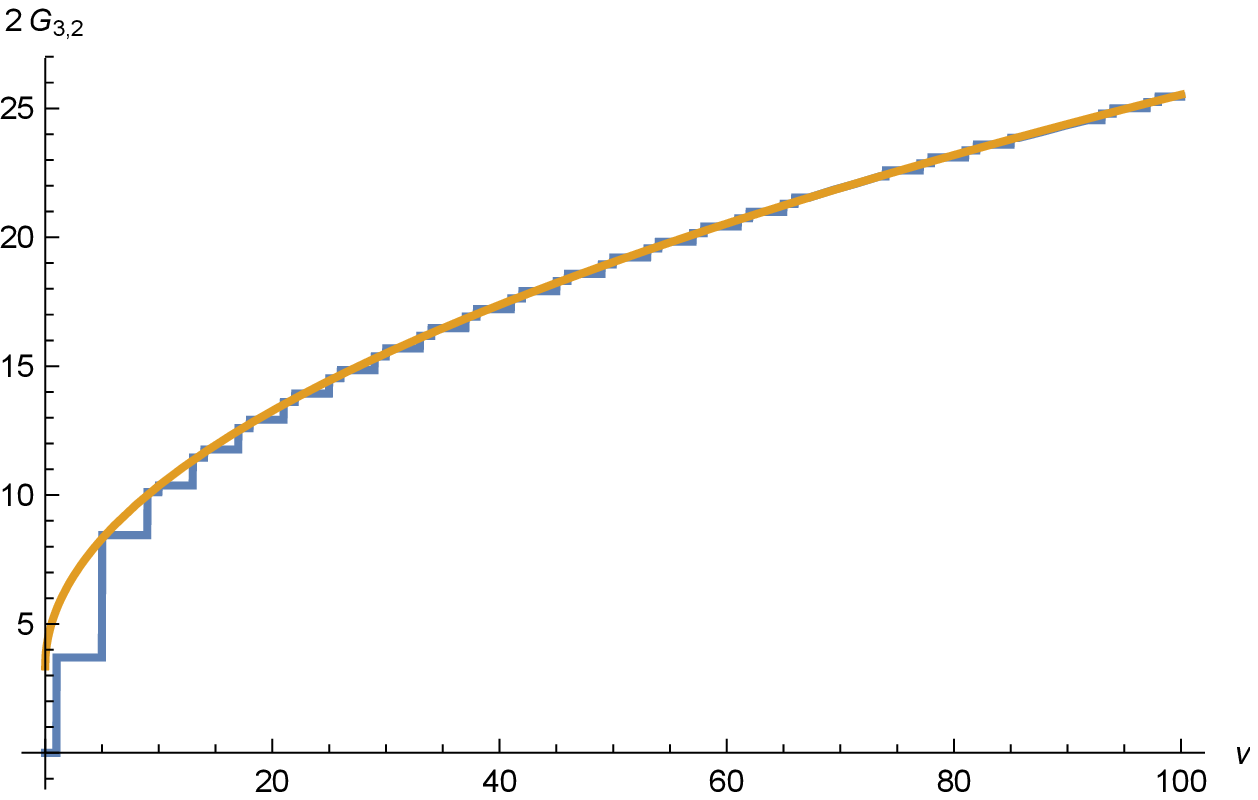}}

\ifig\secondmomenterr{Error estimate for the second moment. We consider the difference between the LHS and the smooth terms in the RHS of \secondmomentdvadising. We also multiply it by a factor $\nu^{1/2}$.  The result is a highly oscillating function of approximately constant amplitude. This is consistent with the error estimate in \secondmomentdvadising.} {\epsfxsize2.5in\epsfbox{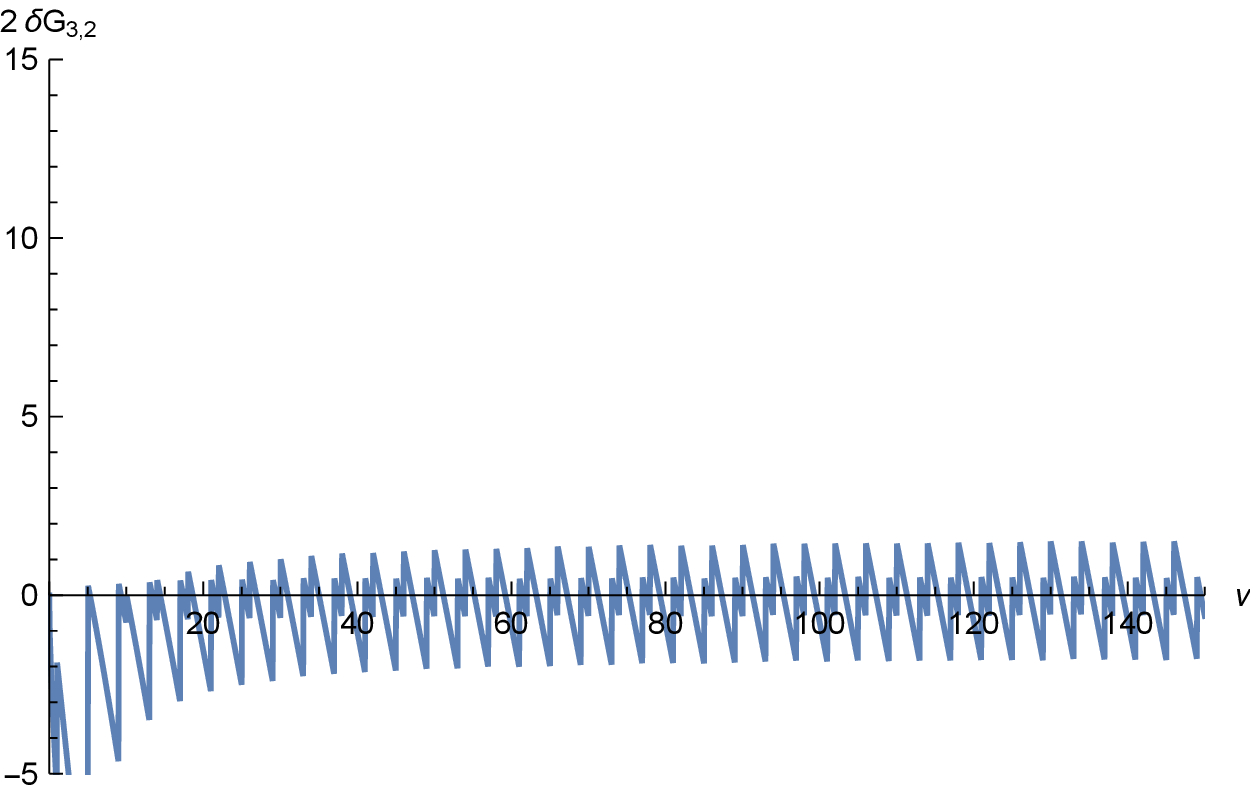}}

The plots for the leading asymptotic and the remainder term of the second moment \secondmomentdvadising\ are presented in the \secondmoment, \secondmomenterr.

At this point we would not be able to tell the difference between the 2d Ising model and a GFF field of the same dimension. Indeed, the leading asymptotic in \secondmomentdvadising\ is controlled by the unit operator. To probe the difference between different theories we consider higher moments, that are sensitive to the subleading tails. For example, consider
 $G_{10,7}^{J=2}(\nu)$
\eqn\tenseven{
{1 \over \nu^5} G_{10,7}^{J=2}(\nu) = - 2.27979 \times10^{-8} \nu^{5/2}  - 1.4288 \times 10^{-6} \nu^{1/2} + O(1) \ .
}

\ifig\ftenseven{Subleading tail in $G_{10,7}^{J=2}$ in the 2d Ising model. We plot ${1 \over \nu^5} G_{10,7}^{J=2}(\nu) + 2.27979 10^{-8} \nu^{5/2}$ versus $\nu$. } {\epsfxsize2.5in\epsfbox{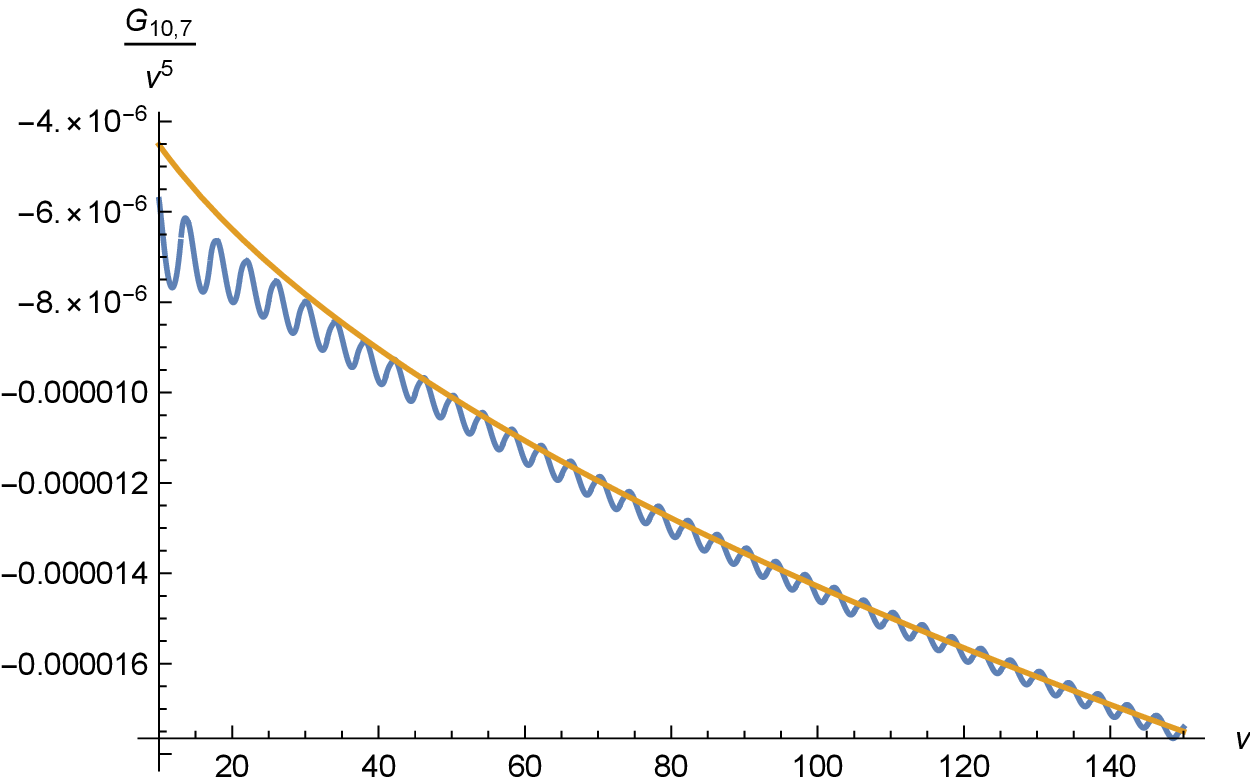}}

\ifig\tensevendiff{Error estimate for ${1 \over \nu^5} G_{10,7}^{J=2}$ in the 2d Ising model. We multiply the difference between the LHS and the smooth terms in the RHS in \tenseven\ by $\sqrt{\nu}$. Again, we get that the error is consistent with the last term in \tenseven.} {\epsfxsize2.5in\epsfbox{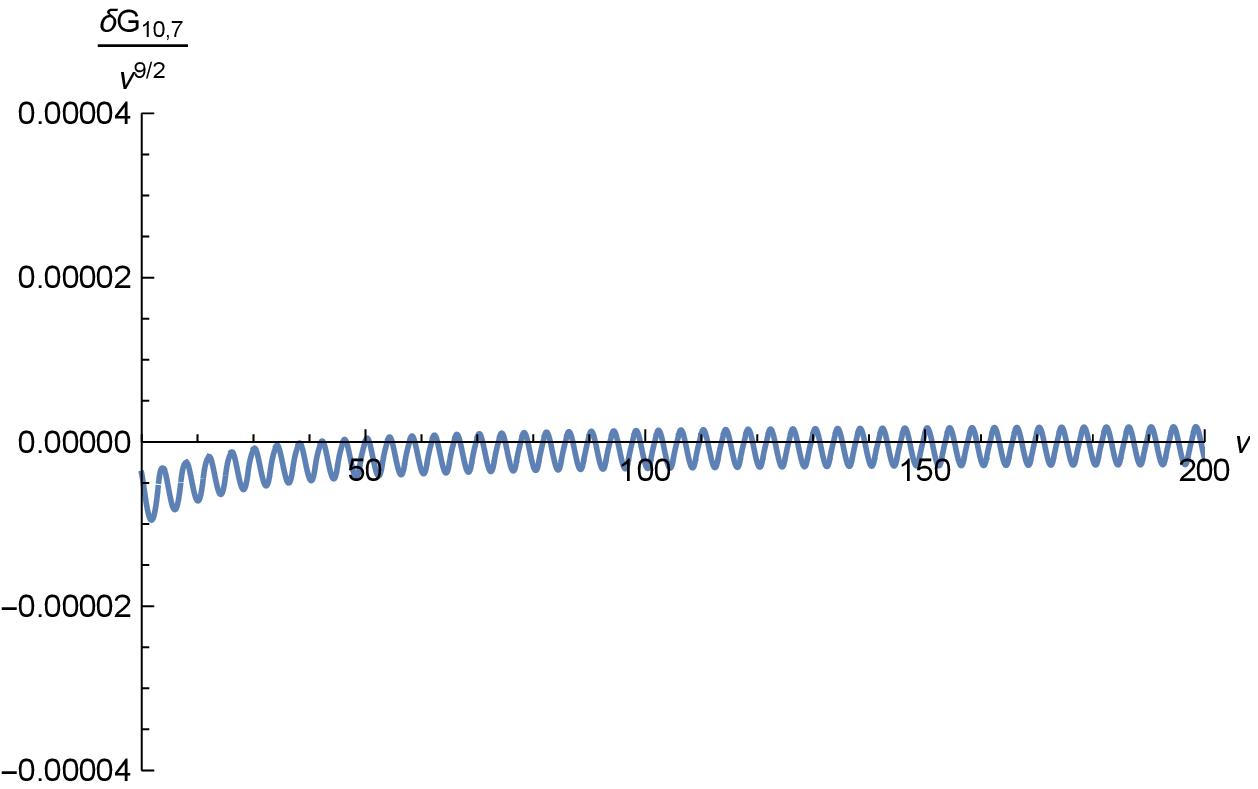}}

The plots for the subleading and remainder terms in \tenseven\ are presented in the \ftenseven, \tensevendiff.

The minimal moment sensitive to the subleading correction is $G_{7,4}^{J=2}(\nu)$. In this case the convergence in $\nu$ is slower. We believe that it might be related to the fact that in this case the corrections are enhanced due to their proximity to the poles \polesker. This problem does not arise for the moment $G_{10,7}$ above. Our prediction takes the form
\eqn\predictionsevenfour{
{1 \over \nu^2} G_{7,4}^{J=2}(\nu) = {\pi^2 \over 113400 \sqrt{2}} {\nu^{5/2} \over \Gamma({9 \over 8})^4} + 0.0332694 \nu^{1/2} + b_5 + O(\nu^{-1/2 })
}

\ifig\seventhfourth{Subleading tail in $G_{7,4}^{J=2}$ in the 2d Ising model. We plot the difference  ${1 \over \nu^2} G_{7,4}^{J=2}(\nu) - {\pi^2 \over 113400 \sqrt{2}} {\nu^{5/2} \over \Gamma({9 \over 8})^4}$ versus $\nu$. We fit $b_5 \approx 0.06365$. Again the leading behavior exactly matches the prediction. } {\epsfxsize2.5in\epsfbox{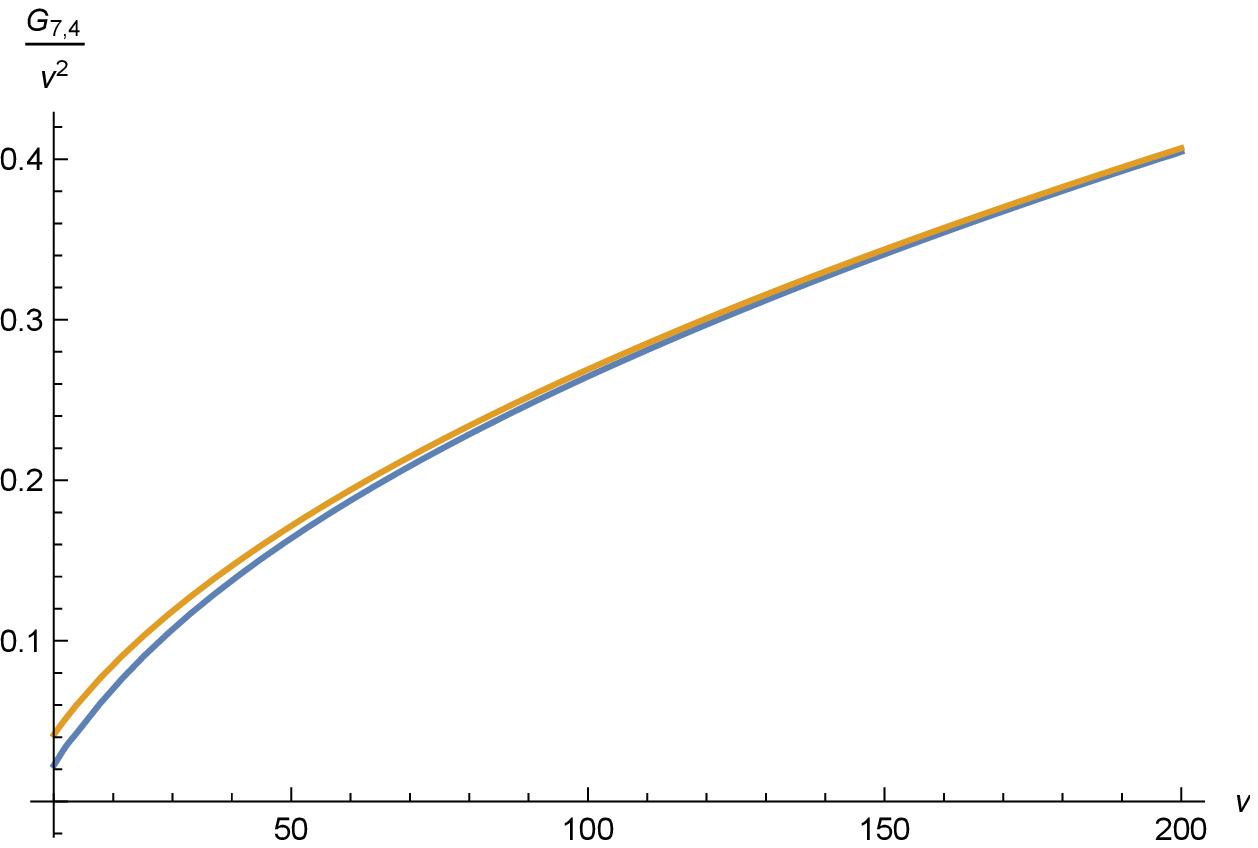}}

\ifig\seventhfourtherror{Error estimate for ${1 \over \nu^2} G_{7,4}$ in the 2d Ising model. We multiply the difference between the LHS and the RHS in \predictionsevenfour\ by $\sqrt{\nu}$. Again we get that the error is consistent with the expectation, though convergence in this case is slower.} {\epsfxsize2.5in\epsfbox{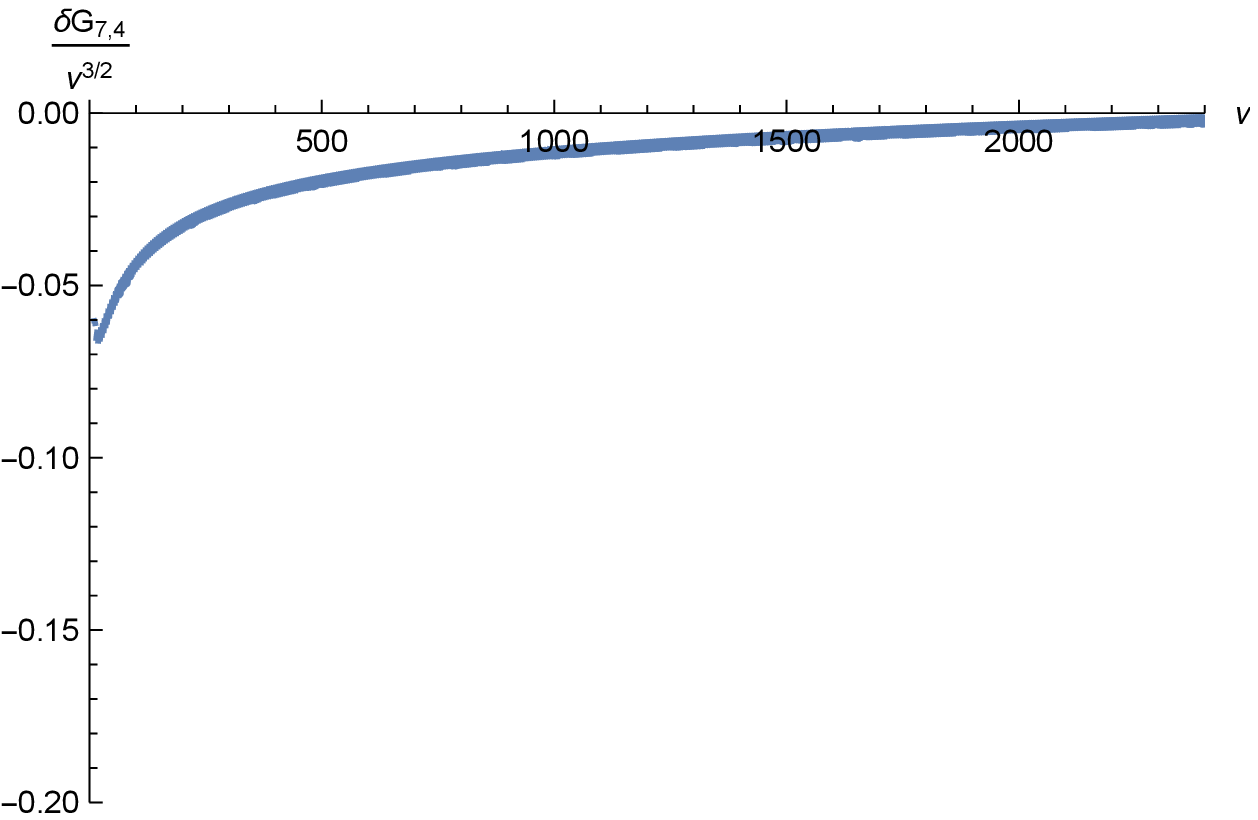}}

We subtract the leading tail from both sides of the equation \predictionsevenfour\ and plot the difference in the \seventhfourth.

Finally, we plot the difference between ${1 \over \nu^2}G_{7,4}$ and smooth terms in the RHS of \predictionsevenfour\ to estimate the error in the \seventhfourtherror.

The third term in the expansion \dispersionrelationsTDising, which includes the contribution of the stress tensor, is also easily accessible. In particular, if we consider ${1 \over \nu^5} G_{12,6}^J$ moment from \contribution\ we see that the three OPE terms from \dispersionrelationsTDising\ contribute as $\nu^{9/2}$, $\nu^{5/2}$ and $\nu^{1/2}$ correspondingly.

\ifig\twelvesix{$G_{12,6}$ in the 2d Ising model. We subtract $\nu^{9/2}$ and $\nu^{5/2}$ terms  ${1 \over \nu^5} G_{12,6}(\nu)$ to probe the sub-sub-leading $\nu^{1/2}$ term. We expect the difference to go to a constant which we fit to be $b_7 \approx 1.28 \times 10^{-5}$. Again, we get a perfect agreement with the prediction.} {\epsfxsize2.5in\epsfbox{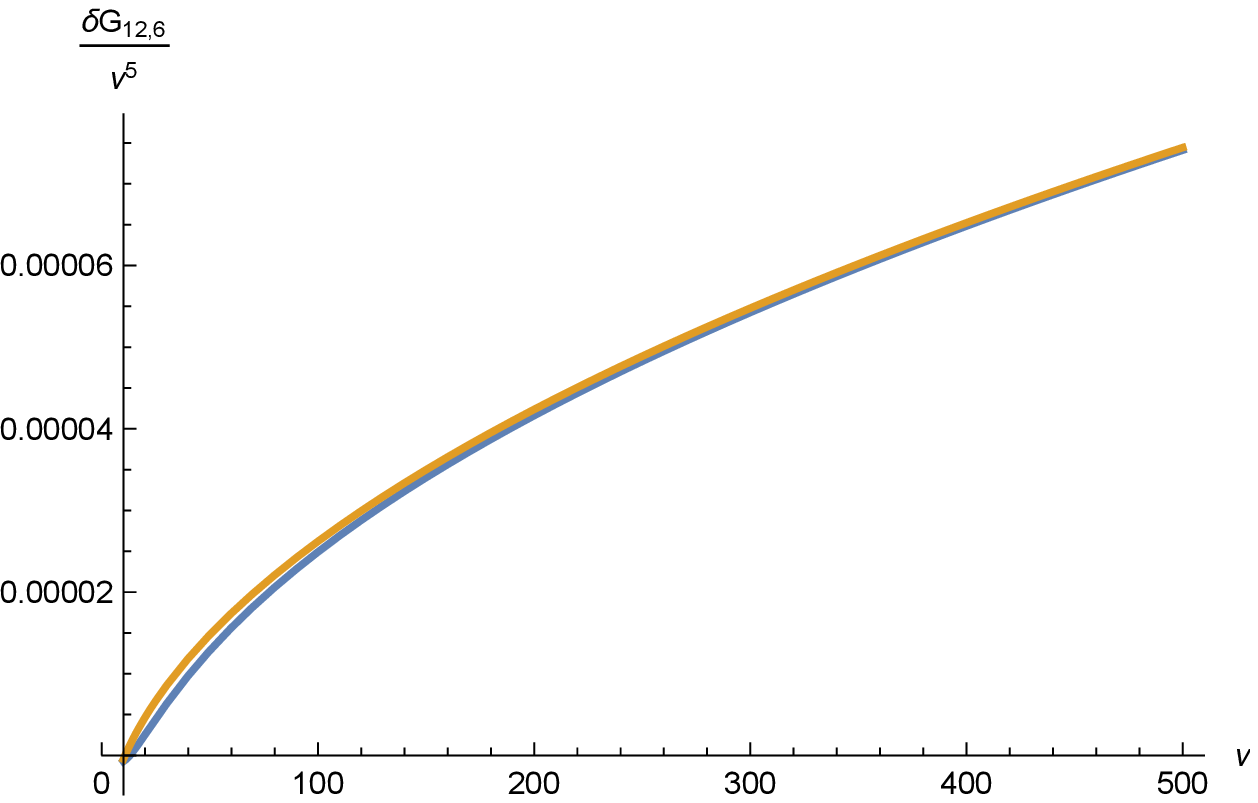}}

\subsec{3d Ising Model}

So far all explicit computations that we did with conformal blocks have been in $d=2,4$ where the expressions for them are explicitly known and relatively simple. The only result that is valid in all dimensions is the formula \cGFFd\ for the partial waves of the generalized free field or, equivalently, the contribution of the unit operator in the $t$-channel. 

Forunately, this is all we need to make non-trivial predictions for the $\la \sigma \sigma \sigma \sigma \ra$ correlator in the 3d Ising. Recall that $\Delta_{\sigma} \simeq 0.51815$ and therefore it falls into the category $\delta_{\sigma} = \Delta_\sigma - {1\over 2} \simeq 0.01815 <{1 \over 4}$ for which we can write the unsubtracted dispersion relation, but need to consider higher moments to match with the $t$-channel OPE.  The ligthest operator above the identity that contributes to the OPE is $\varepsilon$ with dimension $\Delta_{\varepsilon} \simeq 1.41$. Therefore, we get the following structure in the RHS of the dispersion relation
\eqn\dispthreedIsing{\eqalign{
&\int_0^{\infty} d \nu' \ \rho_J^{OPE} \left(3/2 + \nu' \right) {2 \nu' \nu \over  \nu'^2 + \nu^2} \cr
&= d_J \left( 1- {1 \over \cos 2 \pi \delta_\sigma} \right) \nu^{4 \delta_{\sigma} - 2} \left(1  + {\hat \alpha_1 \over \nu^2} +O(\nu^{-2 \Delta_\varepsilon}) \right) + {a_1 \over \nu} + {a_2 \over \nu^3} + ... \ ,\cr
\hat \alpha_1 &= 2 J^2 \left( \delta_\sigma - {3 \over 4} \right)+ 2 J \left(\delta_\sigma - {1 \over 2} \right)  +{8 \over 3} \delta_\sigma^3 - 8 \delta_\sigma^2 + {47 \over 6} \delta_\sigma - {17 \over 8} \ ,
}}
where $d_J$ is given in \deltadef . Notice that due to $\Delta_\varepsilon > 1$ the first subleading universal term comes from the unit operator as well. 

Therefore we can make the following predictions. The leading $\nu$ asymptotic can be extracted from the second moment \leadingasymExC
\eqn\predictionTDIB{
\int_0^\nu d \nu' \nu'^2 \rho_{J}^{OPE}({3 / 2} + \nu') = f_J   {\nu^{4\delta_\sigma} \over 4\delta_\sigma}  + b_3 + O(\nu^{4\delta_\sigma-1}) \ , 
}
where $f_J$ is defined in \leadingasymExA\ (with $\phi = \sigma$ in the present case) and $b_3$ is not computable in terms of the $t$-channel OPE. 

We can also access the subleading term in \dispthreedIsing\ by, for example, considering $G_{9,5}^J$ which takes the following form
\eqn\predictionTDIC{
\nu^{-3} G_{9,5}(\nu) = - {\Gamma(4 \delta_\sigma + 3) \over 6720 \Gamma(4 \delta_\sigma + 7) } f_J \nu^{4 \delta_\sigma + 3} \left( 1 - {(\delta_\sigma+{3\over 2})(\delta_\sigma + {5 \over 4}) \over (\delta_\sigma+{1\over 2}) (\delta_\sigma + {1 \over 4})} {\hat \alpha_1 \over \nu^2}  \right)+ b_6 + O(\nu^{4 \delta_\sigma-1}) \ .
}
where we computed the coefficients using \contribution .

Of course, in practice utility of \predictionTDIB, \predictionTDIC\ depends on how large $\nu$ has to be in order to observe the predicted behavior. Based on the examples above we believe that it should happen for small $\nu$. It would be interesting to explore this question using the numerical bootstrap.

\newsec{Meromorphic Scattering Amplitudes}

In this section we apply the same ideas to meromorphic scattering amplitudes. These arise for example in the tree-level string theory or large $N$ confining gauge theory \WittenKH.
We can use analyticity and polynomial boundedness of the scattering amplitude to express it in terms of its discontinuity. This is achieved through the standard Cauchy argument. Imagine that for a given $t$ the amplitude is polynomially bounded, namely $A(s,t) \sim f(t) (- i s)^{j(t)}$ at large $s$.\foot{Our choice of the phase will become clear below.} Then we can write a subtracted dispersion relation
\eqn\subtracted{
{1 \over n!} \p_{s}^{n} A(s,t) = \int {d s' \over 2 \pi i} {A(s',t) \over (s'-s)^{n+1}}, ~~~ n = [j(t)]+1 \ .
}
Dropping the arcs at infinity we get
\eqn\disperRelSA{\eqalign{
{1 \over n!} \p_{s}^{n} A(s,t) &= \int_{4 m^2}^\infty d s' \ \rho(s',t) \left( {1 \over (s'-s)^{n+1}}  + {(-1)^n \over (s'-u)^{n+1}} \right) , \cr
\rho(s',t)&= \sum_{J=0}^{\infty} \sum_i \delta(s'-m_i^2) \lambda_{J, m_i}^2 P_{J} \left(1+{2 t \over m_i^2 - 4 m^2}\right) \ ,
}}
where we used that the only singularities are simple poles at the positions of resonances and $m$ is the mass of external particles, which we consider to be identical. Unitarity, therefore, implies that $\rho(s',t) \geq 0$ for $t \geq 0$ and $s>4 m^2$. 

Let us assume now that at fixed $t$ and large $s$ the amplitude $A(s,t)$ admits a power-like expansion
\eqn\Reggepoles{
A(\nu,t) =\sum_i f_i(t) \nu^{j_i(t)} + ... , ~~~ s = - {t \over 2} + i \nu
}
where we introduced $\nu$ variable to connect to the discussion of CFTs in earlier sections.

This is the usual expansion in terms of the Regge trajectories that one gets by closing the $J$ contour integral in the Froissart-Gribov representation \refs{\GribovNW,\CaronHuotVEP}. It is an asymptotic expansion that  is valid away from the real axis. By going away from the real axis we get a cumulative effect from many resonances which produces a smooth power-like behavior \Reggepoles.  If the only singularities are poles we will get only powers $s^{j(t)}$. If there are Regge cuts we might get some extra slowly growing factors $(\log s)^a$. These do not affect the discussion and we keep them implicit. 

The expansion \Reggepoles\ together with \disperRelSA\ is exactly the same mathematical setup as we discussed in the previous sections. We can apply the same complex tauberian theorem as in section 5 to get an expansion of Cauchy moments of the integrated spectral density 
\eqn\CauchymomentSA{\eqalign{
F_{m}(\nu,t) &\equiv {1 \over (m-1)!} \int_0^\nu d \nu' (\nu-\nu')^{m-1} \rho(\nu' - {t / 2},t) \ .
}}
for which we get
\eqn\ebstiltSA{\eqalign{
F_m(\nu,t) &= - \sum_{i=0} f_i(t) {\sin {\pi j_i(t) \over 2} \over \pi} {\Gamma(j_i(t)+1) \over \Gamma(m + j_i(t)+1)} \nu^{m+j_i(t)} \cr
&+ \sum_{k=1}^m b_{k} {\nu^{m-k} \over (m-k)!}+O(\nu^{j_0(t)}) , ~~~  \cr
}}
where $j_0(t)$ is the leading Regge trajectory, and we used validity of the Regge expansion for ${\rm Im} [ s ] > |s|^{\eps}$, where $\eps$ is positive and fixed. As above we kept $s^\eps$ factor in the error estimate implicit.

\subsec{Veneziano Amplitude}

\noindent Consider as an example the Veneziano amplitude. We have 
\eqn\venlike{
A(s,t) = {\Gamma(-s)\Gamma(-t) \over \Gamma(-s-t)} + {\rm permutations} \ ,
}
where the external particles are taken to be massless $s+t+u = 0$. At large $s$ and fixed $t$ the amplitude admits an asymptotic expansion
\eqn\asymptexp{\eqalign{
A(s,t) &= (-s)^t \Gamma(-t) \left( 1 + {\sin \pi s \over \sin \pi (s+t)} \right)  , ~~~ {\rm Re}[s]<0 , \cr
A(s,t) &=  s^t \Gamma(-t) \left( 1 + {\sin \pi (s+t) \over \sin \pi s} \right),  ~~~ {\rm Re}[s]>0 \ ,
}}
where we used the asymptotic expansion of the Gamma function (see appendix C in \HarlowNY\ for a thorough discussion). Away from the real axis, namely for ${\rm arg}[s] \neq 0, \pi$, the oscillation terms in the brackets lead to exponentially suppressed corrections. Say, taking $s = s_0 (1+i \eps)$ the oscillating terms lead to the corrections of the type $e^{- \eps s_0}$. Neglecting those we get 
\eqn\asymexp{\eqalign{
A(s,t) &= 2 (- i s)^t \cos {\pi t \over 2} \Gamma(-t), ~~~ 0<{\rm arg}[s]<\pi , \cr
A(s,t) &= 2 ( i s)^t \cos {\pi t \over 2} \Gamma(-t), ~~~ \pi<{\rm arg}[s]<2 \pi \ .
}}
In particular, plugging \asymexp\ in the Tauberian theorem \ebstiltSA\ we get
\eqn\ebVen{
\int_0^{s} d s'  \rho(s', t) = {(s+t/2)^{t+1} \over \Gamma(t+2)} + {(s+t/2)^{t} \over 2 \Gamma(t+1)} + O(s^{t}) \ .
}
which we derived for $t>0$, where the spectral density \disperRelSA\ is positive. 

\ifig\venezianof{Integrated spectral density for the Veneziano amplitude \ebVen. We plot \ebVen\ for $t = 1.2$.} {\epsfxsize3in\epsfbox{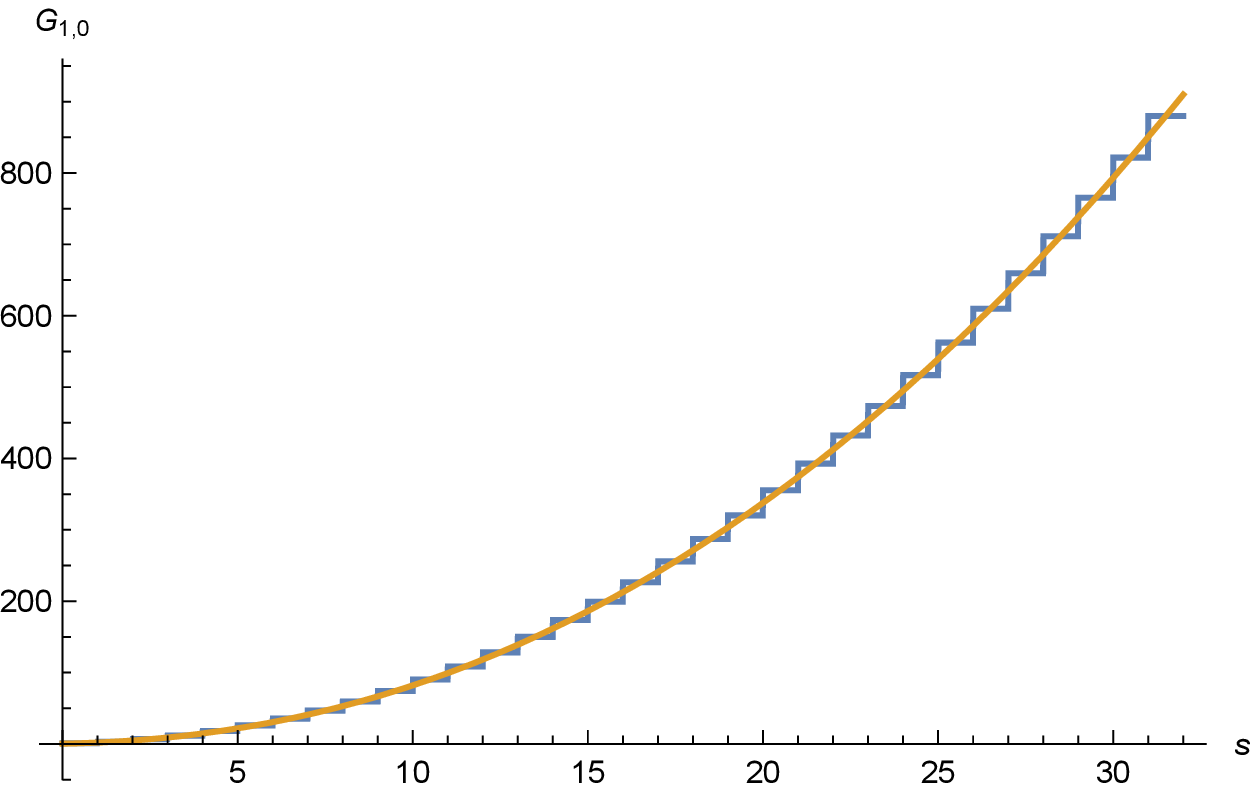}}

\ifig\venezianof{Error estimate in the Veneziano amplitude spectral density. We plot the difference ${1 \over s^{t}} \left( \int_0^{s} d s'  \rho(s', t) - {(s+t/2)^{t+1} \over \Gamma(t+2)} \right)$ for $t = 1.2$. The result is consistent with \ebVen.} {\epsfxsize2.5in\epsfbox{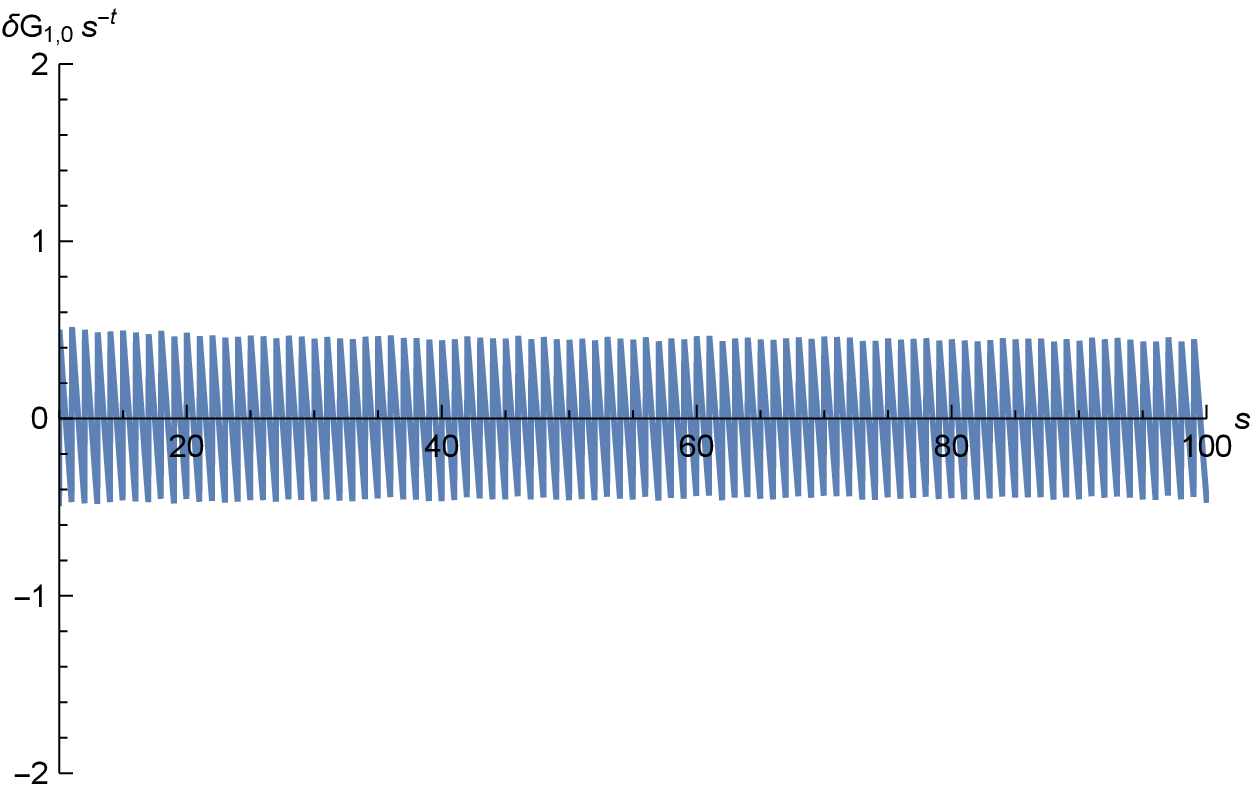}}

\noindent We plot \ebVen\ for the Veneziano amplitude in the \venezianof. Similarly, we can consider the error in \ebVen\ which we plot in the \venezianof. 

To access the next-to-leading correction we can consider $G_{4,2}$ that is given by
\eqn\predictionG{
G_{4,2}(\nu) = {1 \over 3!} \int_0^\nu d \nu' \ \nu'^2 (\nu - \nu') \rho(\nu' - {t / 2}, t) 
}
for which we get the following prediction
\eqn\predVenG{
{1 \over \nu^{2+t}}G_{4,2}(\nu) ={(1+t)(2+t) \over 6 \Gamma(5+t)} \nu^{2} +{t(1+t) \over 12 \Gamma(4+t)}  \nu + O(1) \ .
}

\ifig\venezianoCorr{Subleading Correction for $G_{4,2}$ of the Veneziano amplitude. We subtract the leading $\nu^{2}$ tail from ${1 \over \nu^{2+t}} G_{4,2}$ and compare the subleading correction $\nu$ to the one predicted by \predVenG. We set as above $t = 1.2$.} {\epsfxsize3in\epsfbox{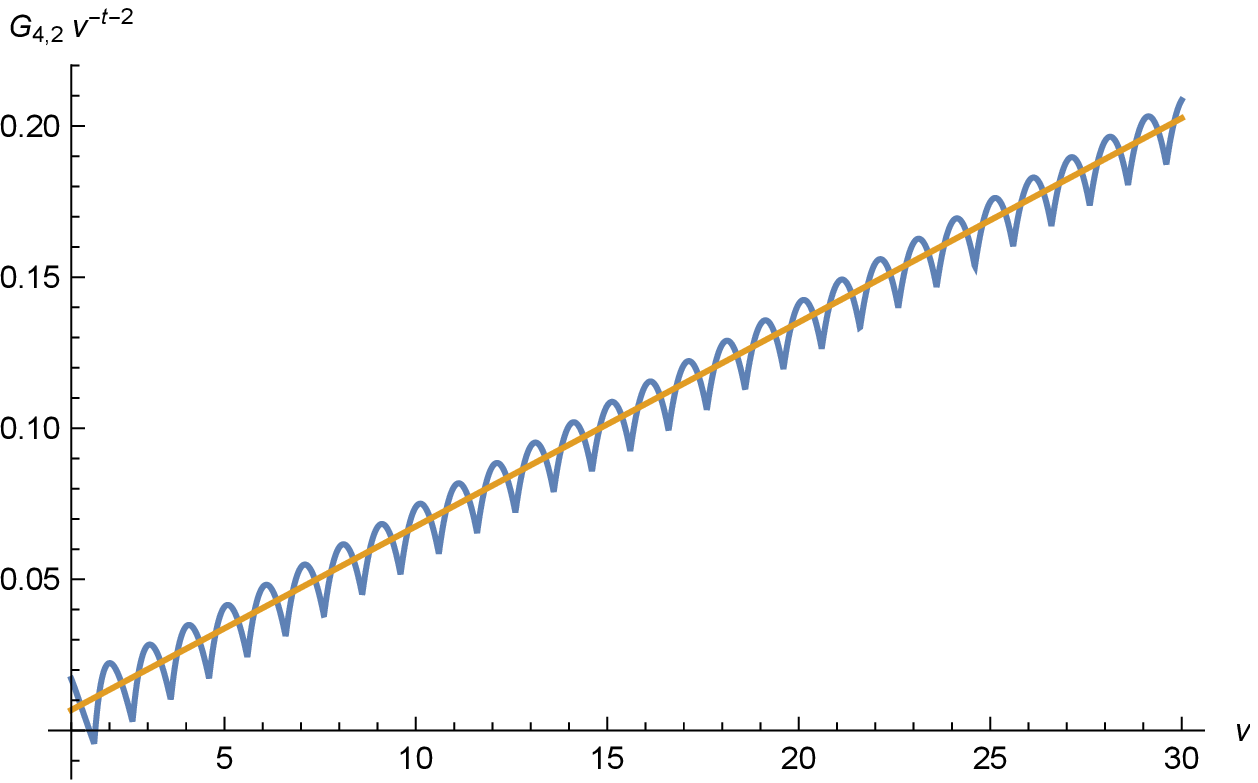}}

\ifig\venezianoCorrDiff{Error for $G_{4,2}$. We plot the difference between the LHS and the RHS in \predVenG\ for $t = 1.2$. The error is exactly as expected.} {\epsfxsize2.5in\epsfbox{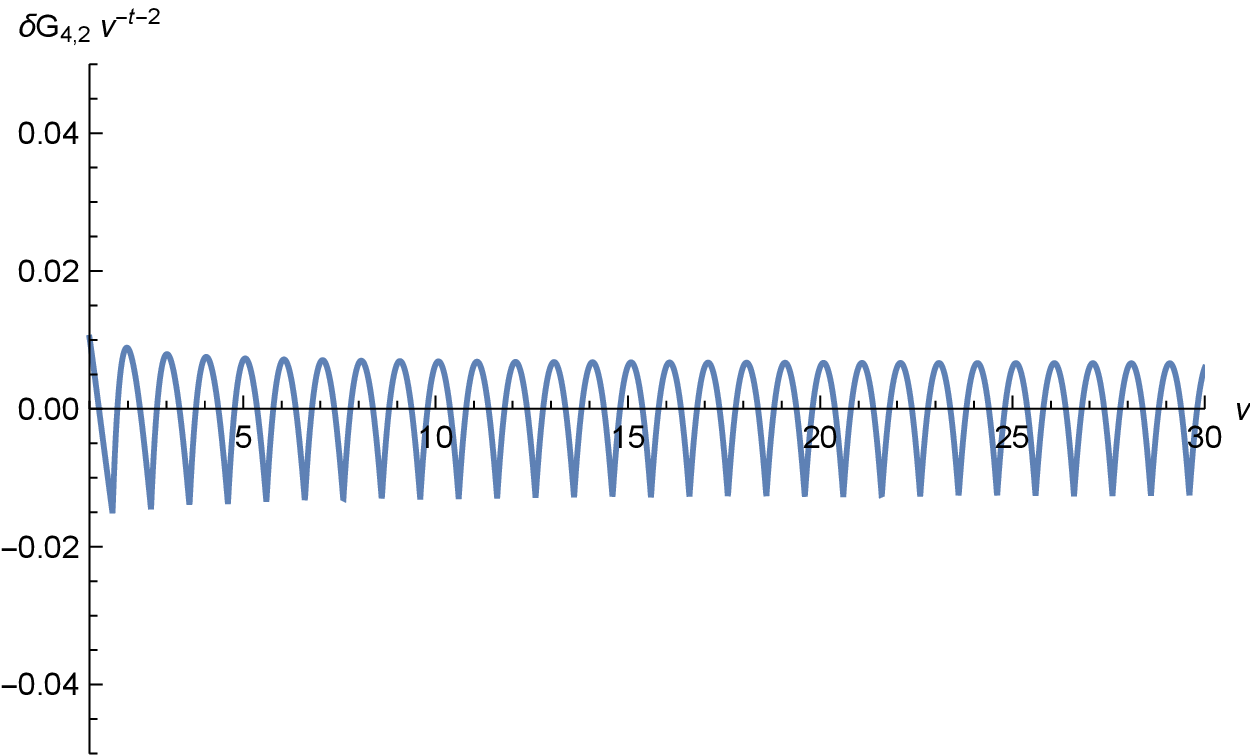}}

We plot the first subleading term of \predVenG\ in the \venezianoCorr\ and find perfect agreement. The difference between the exact $G_{4,2}$ and the first two terms in the RHS of \predVenG\ is an oscillatory function of constant amplitude. We plot it in the \venezianoCorrDiff. 

\subsec{Heuristic Derivation of Spectral Asymptotics}

Let us start with a historical comment. Finite energy sum rules (FESR) \refs{\DolenJR, \IgiZZ, \IgiZZA, \LogunovDY} are consistency conditions imposed by analyticity of scattering amplitudes. In the case of usual dispersion relations one uses analyticity and polynomial boundedness of the scattering amplitude to express the amplitude through its discontinuity. FESR are closely related to the usual dispersion relations. In this case we consider an integral of the amplitude's moment $K(s,t) A(s,t)$ over the contour ${\cal C}$ in the complex plane \usualcontour\ (the larger blue contour this time is in the complex $s$-plane). The kernel $K(s,t)$ is chosen to be analytic inside ${\cal C}$ (we also choose it to be real on the real axis). Due to analyticity of the scattering amplitude and the kernel the integral vanishes
\eqn\zeroint{
\oint_{{\cal C}} d s'  K(s',t) A(s',t) = 0.
}
Let us (without loss of generality) further restrict our consideration to the scattering of four scalar identical particles.
Permutation symmetry implies that \zeroint\ is non-zero for the odd part of the kernel only $K(s,t) = - K(u,t)$. We can rewrite \zeroint\ as follows 
\eqn\FESR{
S_n ={1 \over s_0^{n+1}} \int_0^{s_0} { d s \over \pi} \ K(s,t) {\rm Im}[A(s,t)] ={1 \over s_0^{n+1}} {1 \over 2 \pi i } \int_{{\cal C}'} d s K(s,t) A(s,t) , 
}
where $s+t+u= 4 m^2$ and ${\cal C}'$ stands for the integral over the arcs. Importantly, in this case we do not drop the contribution from the large arcs in \usualcontour.

For application of these sum rules to the pion-nucleon scattering in QCD see \DolenJR, where 
\eqn\phker{
K(s,t) = (s-u)^n .
} 
In phenomenological applications one assumes that the LHS of \FESR\ is dominated by a few low-energy resonances. One can then use the knowledge of the scattering amplitude ${\rm Im}[A(s,t)]$ to make predictions about the leading Regge asymptotic. Alternatively, one can use the knowledge of the Regge asymptotic to infer something about the properties of low-energy resonances. The basic point is that one should not add low-energy resonances contributions and the contributions of Regge poles. Adding them up would lead to a double counting as \FESR\ clearly demonstrates. This duality between resonances in one channel and the Regge trajectory exchange in the other channel is also known as the Dolen-Horn-Schmid duality. It led to the Veneziano amplitude \VenezianoYB\ and its better understanding was the original motivation of our analysis.

In the context of meromorphic amplitudes using the Regge asymptotic all the way to the real axis is not justified. Therefore we could not use the Regge limit to compute the integral over the arc.\foot{Alternatively, FESR are derived from the the so-called superconvergence sum rules \DolenJR. However, this derivation suffers from exactly the same problem for meromorphic amplitudes.} Instead, we should use the complex tauberian theorem as described above.  This puts FESR for meromorphic amplitudes on a solid mathematical ground.

Let us however offer a non-rigorous intuitive explanation of the results that we obtained using complex tauberian theorems. Consider a FESR integral \FESR\ which after switching to the $\nu$ variable takes a form
\eqn\FESRb{
\int_0^{\nu_0} d \nu' \nu'^n {\rm Im}[A(\nu',t)] = {1 \over 2 i} \int_{{\cal C}'} d \nu' \nu'^n A(\nu' ,t) .
}
Let us rewrite the RHS of \FESRb\ as follows
\eqn\RHSfesr{
{1 \over 2 i} \int_{{\cal C}'} d \nu' \nu'^n A(\nu' ,t) = {1 \over 2 i} \int_{{\cal C}'} d \nu' \nu'^n A^{Regge}(\nu' ,t) + {1 \over 2 i} \int_{{\cal C}'} d \nu' \nu'^n \left[ A(\nu' ,t) - A^{Regge}(\nu' ,t) \right] \ ,
}
where $A^{Regge}(\nu' ,t)$ is the power-like Regge asymptotic analytically continued all the way to the real axis, even though it is not a valid approximation of the amplitude in that region. The first term in the RHS of \RHSfesr\ is what produces $t$-channel predictions in the previous subsections. On the other hand, the $A(\nu' ,t) - A^{Regge}(\nu' ,t)$ term corresponds to an error estimate. The result of the theorem corresponds to an estimate $A(\nu' ,t) - A^{Regge}(\nu' ,t) = O(A^{Regge}(\nu' ,t))$ in the region of the complex plane close to the real axis. Integrating over the region where the Regge approximation is not valid (let us denote the size of this region $\Lambda \ll \nu_0$) we get an error estimate 
\eqn\FESRc{
\int_0^{\nu_0} d \nu' \nu'^n {\rm Im}[A(\nu',t)] = {1 \over 2 i} \int_{{\cal C}'} d \nu' \nu'^n A^{Regge}(\nu' ,t) + O(\Lambda \nu_0^n  A^{Regge}(\nu_0 ,t) ) .
}
This is precisely the result of our theorems, where we chose $\Lambda = s_0^\eps$ where $\eps$ is some fixed but arbitrarily small number. In practice, say for the Veneziano amplitude, we find rather an estimate $A(\nu' ,t) - A^{Regge}(\nu' ,t) = O(e^{- c {\rm Im}[\nu']}A^{Regge}(\nu' ,t))$. An extra suppression factor $e^{- c {\rm Im}[\nu']}$ leads to the disappearance of an extra factor $\Lambda$ in \FESRc\ since the integral $\int_0^{\Lambda} d a~ e^{- c a} = O(1)$ for order one number $c$ and $\Lambda \gg 1$. This is fully consistent with what we observed for the Veneziano amplitude. For CFTs we get an identical picture based on separate terms coming from the inversion formula. It is also what we get in the explicit examples of the 2d Ising and GFF. Assuming that this is a general phenomenon we would get $\Lambda(x)={\rm const}$ error estimates in the tauberian theorem of appendix B. 

The argument above also illuminates what is special about the Cauchy moments. Indeed having a factor $(\nu_0 - \nu')^m$ inside the kernel $K(s,t)$ does not change the error estimate in \FESRc. This is because effectively we have $\nu_0 - \nu' \simeq {\rm Im}[\nu']$ in the relevant integration region and again $\int_0^{\Lambda} d a \ a^m \ e^{- c a} = O(1)$. Strictly speaking since we have both the $s$- and the $u$-channel cuts we need to consider terms $(\nu_0 - \nu')^m (\nu_0 + \nu')^m$ instead or design an analytic kernel that is small on one of the cuts. The conclusion is, however, the same: we should set $\Lambda(x)={\rm const}$ in the estimates that we got from the complex tauberian theorems.

\newsec{Conclusions and Future Work}

In this paper we analyzed crossing equations in the deep Euclidean region analytically. The solution to the Euclidean crossing is organized as a ${1 \over \Delta}$ expansion of the integrated weighted spectral density \spectraldensity. The main result of the paper is the formula \contribution\ which expresses moments of the $s$-channel weighted spectral density through the $t$-channel Euclidean OPE data. 

The basic idea behind our analysis is very simple: light operators in one channel should be reproduced by an infinite sum over heavy operators in the other channel. This fact was used in \PappadopuloJK,\RychkovLCA\ to derive the asymptotic behavior of the spectral OPE data \universalasym. In this paper we developed this idea using several techniques. 

First, the asymptotic behavior \universalasym\ receives contributions from both primaries and descendants of all spins. To disentangle their contributions we considered dispersion relations of the partial wave $c_J (\Delta)$ \partialwavedec\ in the complex $\Delta$ plane. We used Lorentzian inversion formula of Caron-Huot to argue that $c_J (\Delta)$ is polynomially bounded at large $| \Delta | \gg 1$ \clargeDelta. Moreover, consistency with the OPE requires $c_J (\Delta)$ to be a meromorphic function with a specific set of poles and residues. Depending on the dimension of external operators a certain number of subtractions is required. An example of dispersion relations without subtractions is given by \disppp. 
We used the Caron-Huot's formula to develop a systematic expansion of the dispersion relations at large $\Delta$. As expected, large $\Delta$ asymptotic is mapped to small Euclidean distances. This time, however, we have only heavy primary operators of a given spin that reproduce the small distance 
asymptotic in the dual channel. This is the subject of sections 3 and 4. 

When deriving dispersion relations we encountered an infinite set of extra poles in addition to the ones that correspond to physical operators. Their presence follows from consistency of the conformal partial wave expansion with the OPE. Remarkably, we found that at large $\Delta$ the contribution of these poles can be re-summed and is controlled by the $t$-channel Euclidean OPE. In this way we obtained large $\Delta$ prediction for the physical operators only. Extra poles also introduce terms ${1 \over \Delta^{n} }$ with integer $n$ in dispersion relations. These terms are non-universal and not computable via the $t$-channel OPE. Their presence is required by consistency since they are generated by individual operators in the $s$-channel.

Going from the dispersion relations to the statement about the integrated weighted spectral density is a nontrivial task. A relevant set of results from mathematics is called {\it tauberian theorems} (see \QiaoXIF\ for a recent review of tauberian theorems in the context of the conformal bootstrap). Previously, the set of the so-called {\it real tauberian theorems} was used in the context of the conformal bootstrap \refs{\PappadopuloJK, \RychkovLCA, \QiaoXIF}. Real tauberian theorems allow one to go from the correlator (or dispersion relations) to the statement about the leading asymptotic behavior of the weighted spectral density. However, they do not allow one to study the corrections, the reason being that real tauberian theorems predict only very weak logarithmic constraints on the corrections to the leading asymptotic of the weighted spectral density, see e.g. \resulttauberian. In this paper we noted that for physical theories the expansion of correlators (or a dispersion relation) naturally takes place in the complex plane. Using the asymptotic behavior in the complex domain one can significantly strengthen the predictions about the large $\Delta$ behavior of the integrated spectral density. The relevant set of results is known as {\it complex tauberian theorems} \refs{\Postnikov, \SubhankulovR, \Subhankulov}.\foot{Complex tauberian theorems that we used are small variations of an extensive analysis by Subhankulov \SubhankulovR.} We showed how one can use these results to solve crossing equations systematically in the ${1 \over \Delta}$ expansion in  any unitary CFT. This is the subject of sections 2 and 5.

Equipped with this understanding we tested our predictions in section 6 on generalized free fields and 2d Ising model. In all cases we find a complete agreement with the general theory and moreover we observe that the large $\Delta$ expansion works surprisingly well already at small $\Delta$. This makes it potentially interesting in the context of numerical bootstrap. We made predictions for the 3d Ising model \predictionTDIB, \predictionTDIC. As in the discussion of the finite energy sum rules in the context of pion-nucleon scattering \DolenJR, we can hope to observe crossing symmetry at work already for a few primary operators of given spin! 

More generally, we expect complex tauberian theorems to find a wide range of new applications outside of conformal bootstrap. We considered one such application - meromorphic scattering amplitudes - in section 7.

There are many future directions one might pursue. Let us list a few:

$\bullet $ Strictly speaking, our derivation applies only to $J>1$ (since we used the Lorentzian inversion formula).  It would be interesting to understand how to extend the argument to $J=0,1$. In the special examples considered in section 6 we found that our formulas work in these cases as well. 

$\bullet$  It will be useful to explore more examples.\foot{It would be curious to explore crossing in the Euclidean regime in non-unitary theories as well, see e.g. \GromovHUT. } For instance, to see whether the formula \contribution\ always works well even at small $\nu$, as we observed in GFF and 2d Ising.

$\bullet$ It will be interesting to consider the case of non-identical operators and, in particular, the limit when a pair of operators become heavy. For example, we should recover the familiar dispersion relations for the thermal Green's function, see e.g. \CaronHuotNS. 

$\bullet$ Another interesting direction is to consider large spin, large twist limit of the dispersion relations, i.e. $\Delta \gg 1$ with ${J \over \Delta}$ fixed. We expect a very similar story to hold in this case.

$\bullet$ One can also consider external operators with spin.

$\bullet$ Further, one can ask what are the implications of our results in the context of holography \refs{\MaldacenaRE,\WittenQJ,\GubserBC}. Primary operators of fixed spin $J$ and asymptotically large dimension are expected to be dual to black holes. Therefore, our results imply that average matrix elements between a large black hole and two light particles should be universal, as expected based on general arguments \refs{\GiddingsGJ,\FitzpatrickHU}. This question was recently addressed using the 2d CFT techniques in \DasCNV.\foot{ We believe it should be possible to make the analysis of \DasCNV\ rigorous and systematic using the complex tauberian theorem for Laplace transform discussed in the present paper. Similarly, it would be interesting to apply our discussion of the Laplace transform to partition functions, see e.g.  \refs{\CardyIE,\HellermanBU,\CollierCLS}.}  This universality seems to be closely related to the eingenstate thermalization hypothesis \LashkariVGJ. It would be interesting to understand if our analysis could be used to say something nontrivial about the landscape of consistent UV completions of gravity.\foot{On a side note it is amusing to imagine a hypothetical character, say Alice, who  learned QFT via the conformal bootstrap. By focusing on a single correlator and solving crossing in different regimes she might think that solving bootstrap equations is a rather mundane task. By studying crossing equations for multiple correlators she would observe that things get more constrained. That there are sometimes isolated islands in the OPE data space and that solving crossing equations might not be that easy after all \refs{\KosBKA, \KosYSD}. It is not clear if she would discover something fundamentally special about $d \leq 6$ or need of  supersymmetry to make the gap in the spectrum of higher spin operators to be large \HeemskerkPN\ in a large $N$ CFT. Her intuition would be in a stark contrast with the one of Bob who spent a lot of time thinking about Lagrangians and string compactifications. Alice and Bob would come up with very different notions of landscape. For example, the string landscape of Bob would require supersymmetry to have a CFT with a gravity dual, see e.g. \OoguriPDQ. The bootstrap landscape of Alice would have in addition a possibility of infinitely many sporadic solutions of crossing which are impossible to find by methods of Bob. In this vast bootstrap landscape of theories with Einstein gravity duals there is a subset which admits a much better theoretical control and is much easier to discover due to its special properties (like supersymmetry). On the other hand, it could happen that as Alice includes more and more operators in her system of crossing equations (and higher-dimensional analogs of modular invariance) she would discover that indeed the only solutions with semi-classical gravity duals that survive are the ones of Bob. Whether it is true is an open question of fundamental importance. } At the moment from studying crossing equations we do not have any evidence for that, nor do we have a slightest idea what the bootstrap landscape is. Here be dragons.

$\bullet$ The relevance of complex tauberian theorems for physics goes far beyond the conformal bootstrap. We can use it whenever we have a dispersion relation. These are ubiquitous in Quantum Field Theory, study of scattering amplitudes or thermal physics. For example, complex tauberian theorems is a rigorous way to connect high frequency expansion of conductivity to the Euclidean OPE \refs{\KatzRLA,\HartnollAPF}. Relatedly, it would be interesting to understand if one can use complex tauberian theorems to gain new insights into the QCD sum rules \ShifmanJV.\foot{ It would be interesting to understand if there is a relation of our analysis with \VladimirovXX, if any.}

\newsec{Acknowledgments}

We are grateful to F. Alday, A. Dymarsky, D. Jafferis, R. Mahajan, D. Mazac, S. Pal, M. Paulos, S. Rychkov and G. Veneziano for useful discussions. The work of BM is supported in part by NSFCAREER grant PHY-1352084.

\appendix{A}{Complex Tauberian Theorem for Laplace Transform}

In this appendix we prove a complex tauberian theorem for Laplace transform that we used in section 2. The proof is basically a review of results of \SubhankulovR\ where many extra details can be found.  

We will write $O(x)$ to estimate the magnitude of different quantities. Let us remind the reader that 
 \eqn\bigO{
f(x) = O( g(x) ), ~~~ x \to \infty \quad (x \to a)
}
iff there exist numbers $M, x_0$ ($M,\delta$) s.t. 
\eqn\defO{
|f(x)| < M |g(x)|, \qquad \forall ~ x > x_0 \quad (\forall ~ |x-a| < \delta)
}

We start with the following useful lemma \refs{\Postnikov, \SubhankulovR}.

\noindent {\bf Lemma \ 1:} Let $0 < \sigma < \Lambda$. Then for arbitrary real $\nu$ we have an estimate
\eqn\lemmaOne{
{1 \over 2 \pi} \int_{- \Lambda}^{\Lambda} dt {(\Lambda^2 - t^2)^2 \over \sigma + i t} e^{(\sigma + i t) \nu} =(\Lambda^2 + \sigma^2)^2 \theta(\nu) + O(e^{\sigma \nu} {\rm min}[\Lambda^4 , \Lambda^2 \nu^{-2}]) .
}
where $\theta(\nu)$ is the Heaviside function.

Consider say $\nu \geq 0$. We set $z = \eta + i t $ and consider a closed contour ${\cal C}$ in the $z$-plane that consists of vertical segment $[- \Lambda, \Lambda]$ at $\eta = 0$ and a part of the circle $K$ centered at $z = - \sigma$, with the radius $R = | \sigma + i \Lambda |$. In this way we get
\eqn\circle{\eqalign{
{1 \over 2 \pi i} \oint_{{\cal C}} {(\Lambda^2 + z^2 )^2 \over \sigma + z} e^{(\sigma + z) \nu} &= {1 \over 2 \pi} \int_{- \Lambda}^{\Lambda} dt {(\Lambda^2 - t^2)^2 \over \sigma + i t} e^{(\sigma + i t) \nu} + {1 \over 2 \pi i} \int_{K} {(\Lambda^2 + z^2 )^2 \over \sigma + z} e^{(\sigma + z) \nu} \cr
&= (\Lambda^2 + \sigma^2)^2 \ ,
}}
where in the last line we evaluated the integral by taking the residue at $z =- \sigma$. We can estimate the integral over $K$ in two different ways. First, we have
\eqn\estimateone{
{1 \over 2 \pi i} \int_{K} {(\Lambda^2 + z^2 )^2 \over \sigma + z} e^{(\sigma + z) \nu} = O \left({\Lambda^4 \over R} e^{\sigma \nu} \int_K | d z | \right) = O(e^{\sigma \nu} \Lambda^4) \ ,
}
where we used that $|z| \leq R+ \sigma \leq 2 \Lambda$, $|\sigma + z | = R$, $| e^{z \nu} | \leq 1$ along $K$ for $\nu >0$. Another estimate comes from writing $e^{z \nu} = {1 \over \nu^2} {d^2 \over d z^2} e^{z \nu} $ and integrating by parts.  For $\nu < 0$ we construct the contour ${\cal C}$ by attaching to the vertical segment a small part of the same circle.

The integral in \lemmaOne\ could be of course computed exactly. The point of \lemmaOne\ is that it provides a convenient estimate of the correction to $\theta(\nu)$ for arbitrary $\nu$ which will be very useful in proving tauberian theorems. 

To see utility of the lemma above recall that we are studying the Laplace transform (which is just the OPE expansion in case of CFTs)
\eqn\OPE{
{\cal L}_b(s) \equiv \int_0^{\infty} e^{- s u} d b(u)
}
and we would like to derive some statements about the integrated spectral density
\eqn\intspectraldensity{
F_b (x) \equiv \int_0^x db(u) \ .
}
Using the lemma above we establish the following useful statement.

\noindent {\bf Lemma \ 2:} Assume that ${\cal L}_b(s)$ converges absolutely for ${\rm Re}[s]>0$ and let $0 < \sigma < \Lambda$. Then
\eqn\lemmaTwo{\eqalign{
 F_b (x) &= {1 \over 2 \pi (\Lambda^2 + \sigma^2)^2} \int_{- \Lambda}^{\Lambda} dt {(\Lambda^2 - t^2)^2 \over \sigma + i t} {\cal L}_b(\sigma + i t) e^{(\sigma + i t) x} \cr
&+ O\left(\int_0^\infty e^{\sigma(x-u)}{\rm min}[1 , \Lambda^{-2} (x-u)^{-2} ] | db(u) | \right) .
}}
This lemma expresses the integrated spectral density $F_b (x)$ in terms of the Laplace integral $ {\cal L}_b(s)$ plus a correction. This lemma follows from applying \lemmaOne\ to the integral in the RHS of the first line in \lemmaTwo\ and exchanging two integrations.

Having this two lemmas we are ready to prove an almost relevant theorem (a simplified version of theorem 2.3.1 in \SubhankulovR).

\noindent {\bf Theorem I:} Let the functions $\phi(u)$ and $\psi(u)$ be defined for $u \geq 0$. We assume that they are non-decreasing and positive-definite. Moreover, we assume that $\phi(u)$ locally does not grow faster than a power, namely there exist a positive constant $b$ such that
\eqn\assumptionone{
u \phi'(u) < b \phi(u) , ~~~ u \geq u_0 \ .
}
We also assume that Laplace transforms $F_{\phi}(s)$ and $F_{\psi}(s)$ satisfy
\eqn\condition{
F_{\psi}(s) =F_{\phi}(s) + O(|s|^\alpha), ~~~ s = \sigma + i t
}
in the region 
\eqn\region{
| t | \leq c \sigma^{\omega}, ~~~ 0 \leq \omega \leq 1 .
}
The strength of the result will depend on the value of $\omega$. The larger is the complex domain (smaller $\omega$'s) in which the estimate \condition\ holds, the better is the bound. The case relevant for CFTs is $\omega = 0$.

Let us also assume that
\eqn\simplifying{
\psi(0) - \phi(0) = 0 \ .
}
This is just a technical assumption that does not play any important role.

Then for every $m \geq 0$ we have
\eqn\statement{\eqalign{
\int_0^x (x-u)^m d \psi(u) = \int_0^x (x-u)^m d \phi(u) + O\left({\phi(x)\over x} x^{\omega (m+1)} \right) \cr
+ O({\rm max}[x^{m- \alpha}, (x^{- \omega})^{\alpha - m}] ),  ~~~ m \neq \alpha \ .
}}
If $m=\alpha$ the estimate in the second line becomes $\ln x$. Let us go through the proof of \statement. For further details see \SubhankulovR.

\subsec{Estimate for $\psi(u)$}

It is convenient to integrate \assumptionone\ to get
\eqn\realbound{
{\phi(v) \over \phi(u)} <  \left({v \over u} \right)^b .
}
We will use this estimate extensively below.

Let us first prove that 
\eqn\stepone{\eqalign{
\psi(u) = O( \phi(u) ) .
}}
To show this it is crucial that $\psi(u)$ is non-decreasing and positive. We set $\sigma = {1\over u}$ and do the following estimate
\eqn\estim{\eqalign{
\psi(u) &= O( \sigma \psi(u) \int_u^\infty e^{- \sigma v} d v) = O( \sigma \int_u^\infty \psi(v)  e^{- \sigma v} d v) =O( \sigma \int_0^\infty \psi(v)  e^{- \sigma v} d v)  \cr
&= O(\sigma \int_0^\infty \phi(v)  e^{- \sigma v} d v) + O(\sigma^{1+\alpha}) ,
 }}
 where we used \condition\ to switch from the Laplace transform of $\psi$ to the one of $\phi$. Then we can use the power-like bound on $\phi$ \realbound\ to estimate
 \eqn\boundphi{\eqalign{
 &O(\sigma \int_0^\infty \phi(v)  e^{- \sigma v} d v)  = O(\phi(u)) + O(\sigma \int_u^{\infty} \phi(v) e^{- \sigma v} d v) \cr
 &= O(\phi(u)) + O({\sigma \phi(u) \over u^b} \int_u^{\infty} e^{- \sigma v} v^b d v ) = O(\phi(u)),
 }}
 where we used that $\sigma = {1 \over u}$.
 
 As a last step note that \condition\ is only meaningful if the second term in the RHS is small compared to the first. Therefore, we can drop $O(\sigma^{1+\alpha})$ in the last part of \estim.

\subsec{Apply Lemma 2}

Next we use lemma 2 from above to write
\eqn\steptwo{\eqalign{
\psi(x) - \phi(x) &= O\left( \int_0^\infty | d \psi(u) - d \phi(u) | e^{\sigma (x-u)} {\rm min}[1 , \Lambda^{-2}  (u-x)^{-2}] \right) \cr
&+{1 \over 2 \pi (\Lambda^2 + \sigma^2)^2} \int_{- \Lambda}^{\Lambda} dt {(\Lambda^2 - t^2)^2 \over \sigma + i t} e^{(\sigma + i t) x} [{\cal L}_{\psi}( \sigma + i t ) - {\cal L}_{\phi}( \sigma+i t ) ].
}}
In bounding these terms we set $\sigma = {1 \over x}$ and $\Lambda= c \sigma^{\omega}$. We also think of $x$ as being large. 

Below we will use a series of estimates to show that (see equation 2.3.9 in \SubhankulovR) 
\eqn\estimate{\eqalign{
&O\left( \int_0^\infty | d \psi(u) - d \phi(u) | e^{\sigma (x-u)} {\rm min}[1 , \Lambda^{-2}  (u-x)^{-2}]  \right) = O\left({\phi(x) \over \Lambda x}\right) \cr
&+ O\left({1 \over \Lambda} {\rm max}_{{x \over 2} \leq v \leq 2x} | \int_{- \Lambda}^{\Lambda} d t (1 - {|t| \over \Lambda}) e^{ i t v} ({\cal L}_{\psi}({1 \over v} + i t) - {\cal L}_{\phi}({1 \over v} + i t) ) | \right) .
}}
The argument goes as follows. The idea is to split the $u$ integral as $\int_{0}^{x-y}$  $+ \int_{x-y}^{x+y} + \int_{x+y}^{\infty}$, where 
\eqn\yyy{
y = \left({x \over \Lambda} \right)^{1/2} \ll x.
} 
The point of this splitting is that for $0<u<x-y$ and $u> x+y$ we have $ {\rm min}[1 , \Lambda^{-2}  (u-x)^{-2}] = \Lambda^{-2}  (u-x)^{-2}$, whereas for $x-y < u < x + y$ it is not necessarily true.

 Let us first bound the $d \phi(u) $ terms.  We can use integration by parts to estimate
\eqn\estimminsqOne{
{1 \over \Lambda^2} \int_{0}^{x-y} {d \phi(u) \over (x-u)^2} = O\left( {\phi(x) \over \Lambda^2 y^2} \right) = O\left( {\phi(x) \over \Lambda x} \right) .
}
And we use power-like bound to show that
\eqn\estimminsqTwo{\eqalign{
&{1 \over \Lambda^2} \int_{x+y}^{\infty} {e^{\sigma(x-u)}d \phi(u) \over (x-u)^2} = O\left( {1 \over \Lambda^2 }  \int_{x+y}^{\infty} {e^{- \sigma u}d \phi(u) \over (x-u)^2}  \right) \cr
&=O\left( {\phi(x) \over \Lambda^2 y^2}) + O( {\sigma \over \Lambda^2 y^2} \int_{x+y}^\infty d u \ e^{- \sigma u} \phi(u) \right) \cr
&=O\left( {\phi(x) \over \Lambda^2 y^2}\right) + O\left( {\sigma \over \Lambda^2 y^2} {\phi(x+y) \over (x+y)^b} \int_{x+y}^\infty d u \ e^{- \sigma u} u^b \right) \cr
&= O\left( {\phi(x) \over \Lambda x} \right) ,
}}
where we again used \realbound .

Finally we want to show that
\eqn\finalestimate{
\int_{x-y}^{x+y} e^{- \sigma u} {\rm min}[1 , \Lambda^{-2}  (u-x)^{-2}]  d \phi(u) = O( {\phi(x) \over \Lambda x} ) .
}
The basic observation is that for ${x \over 2} < \tilde x < 2 x$ we can write
\eqn\observation{
\phi\left(\tilde x + {1 \over \Lambda}\right) - \phi(\tilde x) < \phi (\tilde x) \left( \left( {\tilde x + {1 \over \Lambda} \over \tilde x } \right)^b - 1 \right) = O\left({\phi(\tilde x) \over \Lambda \tilde x}\right).
}
This bound is not very surprising and is another way of saying that $\phi(x)$ grows locally at most like a power for purposes of estimates.

Now we split $\int_{x-y}^{x+y} d u$ into many intervals of the size $\Lambda$ and to each of them we will apply \observation . We get
\eqn\splitting{\eqalign{
&\int_{x-y}^{x+y} e^{- \sigma u} {\rm min}[1 , \Lambda^{-2}  (u-x)^{-2}]  d \phi(u) = O\left( {\phi(x) \over \Lambda x} \right) = i_1 + i_2 + i_3 , \cr
i_1 &= {1 \over \Lambda^2} \sum_{k=2}^{y \Lambda} \int_{x - {k \over \Lambda}}^{x-{k-1 \over \Lambda}} {d \phi(u) \over (x-u)^2} , \cr
&= O\left(\sum_{k=2}^{y \Lambda} {1 \over k^2} \int_{x - {k \over \Lambda}}^{x-{k-1 \over \Lambda}} d \phi(u)\right) = O\left({\phi( x) \over \Lambda x} \sum_{k=2}^{y \Lambda}\right) = O\left({\phi( x) \over \Lambda  x}\right) , \cr
i_2 &=\int_{x - {1 \over \Lambda}}^{x+{1 \over \Lambda}} d \phi(u) = O\left({\phi( x) \over \Lambda  x}\right) , \cr
i_3 &= {1 \over \Lambda^2} \sum_{k=2}^{y \Lambda} \int_{x + {k-1 \over \Lambda}}^{x+{k \over \Lambda}} {d \phi(u) \over (x-u)^2} = O\left({\phi( x) \over \Lambda  x}\right) ,
}}
where we used that $y \Lambda = (x \Lambda)^{1/2} \gg 1$ since $\omega < 1$. 

For $d \psi(u)$ terms we also split the integral as $\int_{0}^{x-y} + \int_{x-y}^{x+y} + \int_{x+y}^{\infty}$ and then bound separately each of the terms. 

For integrals $\int_{0}^{x-y}$ and $\int_{x+y}^\infty$ we integrate by parts, use \stepone\ and estimates for $\phi(u)$ from above. Therefore we are left with the estimate
\eqn\lastestimate{
\int_{x-y}^{x+y} e^{- \sigma u} {\rm min}[1 , \Lambda^{-2}  (u-x)^{-2}]  d \psi(u) = i_1 + i_2 + i_3 \ ,
}
where we split the integral as we did above for $\phi$. In this integral we cannot integrate by parts and simply use \stepone\ because of $1$ inside $ {\rm min}[1 , \Lambda^{-2}  (u-x)^{-2}] $, which leads to boundary terms in the integration by parts $O(\phi(x))$. 

To circumvent this difficulty we need another auxiliary lemma (lemma 2.1.6 in \SubhankulovR). It states that for $x/2 <\tilde x < 2 x$ (this interval could be changed) and functions that satisfy the conditions that we used above we have
\eqn\lastlemma{\eqalign{
&\psi\left(\tilde x + {1 \over \Lambda}\right) - \psi\left(\tilde x -  {1 \over \Lambda}\right) = O\left({\phi(x) \over \Lambda x}\right) \cr
&+ O\left( \left| {1 \over \Lambda}  \int_{- \Lambda}^{\Lambda} d t ( 1 - {|t| \over \Lambda}) e^{i \tilde x t} \left( {\cal L}_{\psi}(\tilde \sigma + i t) - {\cal L}_\phi(\tilde \sigma + i t) \right) \right| \right)
}}
where $\tilde \sigma = {1 \over \tilde x}$. In this way to estimate \lastestimate\ we just repeat the analysis for $\phi$ while keeping the difference of Laplace transform terms. 

The way this lemma works is as follows. Start with the following relation
\eqn\relationint{
{1 \over \Lambda} \int_{- \Lambda}^{\Lambda} \left( 1 - {| t | \over \Lambda} \right) e^{i \nu t}= \left( {\sin {\Lambda \nu \over 2} \over {\Lambda \nu \over 2}} \right)^2 . 
}
We can use it to write
\eqn\lemmastepA{\eqalign{
\psi\left(\tilde x + {1 \over \Lambda}\right) - \psi\left(\tilde x -  {1 \over \Lambda}\right) = \int_{\tilde x - {1 \over \Lambda}}^{\tilde x + {1 \over \Lambda}} d \psi(u) = O\left( \int_0^{\infty} e^{- \tilde \sigma u}  \left( {\sin {\Lambda (\tilde x - u) \over 2} \over {\Lambda (\tilde x - u)  \over 2}} \right)^2 d \psi(u) \right) .
}} 
To estimate the RHS we use \relationint\
\eqn\lemmastepB{\eqalign{
&{1 \over \Lambda} \int_{- \Lambda}^{\Lambda} d t \left( 1 - {|t| \over \Lambda}\right) e^{i \tilde x t} \left( {\cal L}_{\psi}(\tilde \sigma + i t) - {\cal L}_\phi(\tilde \sigma + i t) \right) \cr 
&=  \int_0^{\infty} e^{- \tilde \sigma u}  \left( {\sin {\Lambda (\tilde x - u) \over 2} \over {\Lambda (\tilde x - u)  \over 2}} \right)^2 (d \psi(u) - d \phi(u) ) .
}}
Estimating all the terms that involve $\phi$ by methods identical to above we arrive at \relationint .

At this point let us reiterate an important intermediate result
\eqn\intermediateimp{\eqalign{
\psi(x) - \phi(x) &= O\left({\phi(x) \over \Lambda x}\right) \cr
&+ O\left({1 \over \Lambda} {\rm max}_{{x \over 2} \leq v \leq 2x} | \int_{- \Lambda}^{\Lambda} d t (1 - {|t| \over \Lambda}) e^{ i t v} ({\cal L}_{\psi}({1 \over v} + i t) - {\cal L}_{\phi}({1 \over v} + i t) ) | \right) \cr
&+{1 \over 2 \pi (\Lambda^2 + \sigma^2)^2} \int_{- \Lambda}^{\Lambda} dt {(\Lambda^2 - t^2)^2 \over \sigma + i t} e^{(\sigma + i t) x} [{\cal L}_{\psi}( \sigma + i t ) - {\cal L}_{\phi}( \sigma+i t ) ] .
}}

To prove the desired statement for $m=1$ we simply need to estimate the integrals that involve difference of Laplace transforms in \intermediateimp .

\subsec{Estimate For the Difference ${\cal L}_{\psi} - {\cal L}_{\phi}$}

Next, we estimate the rest of the terms that involve the difference of Laplace transforms. We get
\eqn\estimateoneST{\eqalign{
&{1 \over 2 \pi (\Lambda^2 + \sigma^2)^2} \int_{- \Lambda}^{\Lambda} dt {(\Lambda^2 - t^2)^2 \over \sigma + i t} e^{(\sigma + i t) x} [{\cal L}_{\psi}( \sigma + i t ) - {\cal L}_{\phi}( \sigma+i t ) ] \cr
&= O\left( \int_{- \Lambda}^{\Lambda} dt \ {| {\cal L}_{\psi}( \sigma + i t ) - {\cal L}_{\phi}( \sigma+i t ) | \over |\sigma + i t|} \right) = O\left( \int_0^{\Lambda} |\sigma + i t|^{\alpha-1} d t \right) \cr
&=  O( {\rm max}[\sigma^{\alpha}, \Lambda^{\alpha}] ) = O({\rm max}[x^{- \alpha}, \Lambda^{\alpha}] ) .
}}
and similarly
\eqn\estimatetwoST{\eqalign{
&O\left({1 \over \Lambda} {\rm max}_{{x \over 2} \leq v \leq 2x} | \int_{- \Lambda}^{\Lambda} d t (1 - {|t| \over \Lambda}) e^{ i t v} ({\cal L}_{\psi}({1 \over v} + i t) - {\cal L}_{\phi}({1 \over v} + i t) ) | \right)\cr
&=O\left( {1 \over \Lambda}  {\rm max}_{{x \over 2} \leq v \leq 2x}  \int_{- \Lambda}^{\Lambda} d t  | {\cal L}_{\psi}({1 \over v} + i t) - {\cal L}_{\phi}({1 \over v} + i t) | )=  O({\rm max}[x^{- \alpha}, \Lambda^{\alpha}] \right).
}}

Therefore, we showed that 
\eqn\resultsofarST{
\psi(x) - \phi(x) =  O \left({\phi(x) \over \Lambda x} \right)  + O({\rm max}[x^{- \alpha}, \Lambda^{\alpha}]) .
}
Recall that $\sigma = {1 \over x}$ and $\Lambda = c \sigma^{\omega}$, from which $m=0$ claim of the theorem follows.

\subsec{Higher Cauchy Moments}

For higher $m$'s the theorem is proved by induction. Imagine it holds for $m$'th moment and let us try to prove it for $(m+1)$'th moment. Consider $m$'th Cauchy moment
\eqn\cauchymoment{
\Phi_m (x) = {1 \over m!} \int_0^x (x -u)^m d [\psi(u) - \phi(u)] .
}
Differentiating $m$ times by parts we get (here we use the condition $\phi(0) - \psi(0) = 0$)
\eqn\resultnew{
{\cal L}_\psi (s) - {\cal L}_\phi (s) = s^{m+1} \int_0^\infty du \ e^{-s u} \Phi_m (s)
}
from which an estimate 
\eqn\estimatenew{
H(s) = \int_0^\infty du \ e^{-s u} \Phi_m (s) = O(|s|^{\alpha - m -1}) \ .
}
immediately follows. We then apply lemma 2 to get
\eqn\lemmaagain{\eqalign{
 (\Lambda^2 + \sigma^2)^2 \int_0^{x} d u \ \Phi_m(u) = O\left(\int_0^{\infty} | \Phi_m(u)| e^{(x-u)\sigma} {\rm min}[\Lambda^4, \Lambda^2 (x-u)^{-2}] \right) \cr
+ {1 \over 2 \pi} \int_{- \Lambda}^{\Lambda} dt {(\Lambda^2 - t^2)^2 \over \sigma + i t} H(\sigma + i t) e^{(\sigma + i t) x}
}}
Then we apply the $m$'th step estimate to the first line of \lemmaagain\ and \estimatenew\ to estimate the second line in \lemmaagain . From this theorem for $(m+1)$-th moment follows.

\subsec{Important Ingredients}
Sign-definiteness of  $d \psi$ and $d \phi$ is clearly very important for the proof. Also the power-like bound on local behavior of $\phi(u)$ is extensively used. Other assumptions, e.g. the ones stated in theorem 2.3.1 \SubhankulovR, seem to be technical details that could be easily relaxed.
 
\subsec{Case of CFTs}

In the case of CFTs as we discussed in the main text we have the $t$-channel OPE expansion for the Laplace transform
\eqn\startingpoint{
{\cal L}_{\psi}(\beta) = {1 \over  \beta^{2 \Delta_{\phi} } } \sum_{\Delta_i} c_{\Delta_i} \beta^{ \Delta_i}  + ... \ ,
}
for any $|\beta| \ll 1 $. Therefore we can set $\omega =0$ in the previous section and take $\Lambda$ to be constant.  Let us review the $m=0$ part of the theorem. We take $d \phi(u)$ to be simply a set of powers $u^\alpha$ with proper coefficients so that in the difference ${\cal L}_{\psi}(\beta) - {\cal L}_{\phi}(\beta)$ all the singular terms cancel.\foot{A slightly better prescription is to choose $d \phi(u) = \sum_k c_k u^k \theta(0<u<1) + \sum_\alpha d_\alpha u^{\alpha} \theta(u \geq1) $. In this way $\alpha$ could be arbitrarily small.} Therefore we get the OPE expansion
\eqn\differenceCFT{
{\cal L}_{\psi}(\beta) - {\cal L}_{\phi}(\beta) = c_\alpha \beta^\alpha + ... \ , ~~~ \alpha > 0 ,
}
where by ellipses we denoted higher order terms in the OPE. Usually, the $t$-channel OPE is formulated in terms of $(1-z)^h (1-\bar z)^{\bar h}$. While for us $z = \bar z = e^{- \beta}$ and we expand each term $(1-z)^h (1-\bar z)^{\bar h}$ for small $\beta$ and then swap the small $\beta$ expansion with the sum over operators. This is possible due to the absolute convergence of the series. 
Since we have a convergent OPE expansion around $s=0$ we can make a better estimate of the integrals \estimateoneST\ and \estimatetwoST\ using the OPE for small but constant $\Lambda$. We get the following estimate of the relevant integrals\foot{A very similar statement is theorem 2.3.2 in \SubhankulovR. Instead of regularity used in theorem 2.3.2 we used here the fact that we have a convergent OPE expansion for ${\cal L}_{\psi}(s) - {\cal L}_{\phi}(s)$.}
\eqn\integralestimate{\eqalign{
&{1 \over 2 \pi (\Lambda^2 + \sigma^2)^2} \int_{- \Lambda}^{\Lambda} dt {(\Lambda^2 - t^2)^2 \over \sigma + i t} e^{(\sigma + i t) x} (\sigma + i t)^{\alpha} \cr
&= {\sin \pi \alpha \ \Gamma(\alpha) \over \pi x^{\alpha}} \left( 1 + O({1 \over x^2}) \right)+ {8 e \Lambda^{\alpha - 3} \over \pi x^3} \cos ({\pi \alpha \over 2} + x \Lambda) + ..., \cr
}}
The leading asymptotic is different for integer $\alpha$ in which case the first term in the RHS \integralestimate\ vanishes. The leading asymptotic is then captured by the second line in \integralestimate. Let us now estimate the second integral
\eqn\secondintegralestimate{\eqalign{
&{1 \over \Lambda} {\rm max}_{{x \over 2} \leq v \leq 2x}  | \int_{- \Lambda}^{\Lambda} d t (1 - {|t| \over \Lambda}) e^{ i t v}  ({1 \over v} + i t)^{\alpha} | = O\left( {1\over x^{1+\alpha}} \right) + O\left({1 \over x^2}\right) = O\left({1 \over x}\right) .
}}
Using this better estimates we have
\eqn\resultCFT{
\psi(x) - \phi(x) =  O\left({\phi(x) \over x} \right).
}
We then set up an induction. As we consider higher Cauchy moments more and more terms in the OPE become effectively singular. We simply add those extra terms to the naive spectral density and repeat the argument. The endpoint of this reasoning is \complextauberian. Powers of $E$ in the RHS of \complextauberian\ is simply the contribution of $\phi$ in the Cauchy moments that we discussed in this section.

\appendix{B}{Complex Tauberian Theorem for Stieltjes Transform}

In this section we prove a complex tauberian theorem for the Stieltjes transform used in section 5. We extensively use methods of \Subhankulov, where similar theorems had been proved.

\ifig\gammacontour{$\gamma_+$ (solid line) and $\gamma_-$ (dashed line) integration contour in the $z$-plane.} {\epsfxsize2.5in\epsfbox{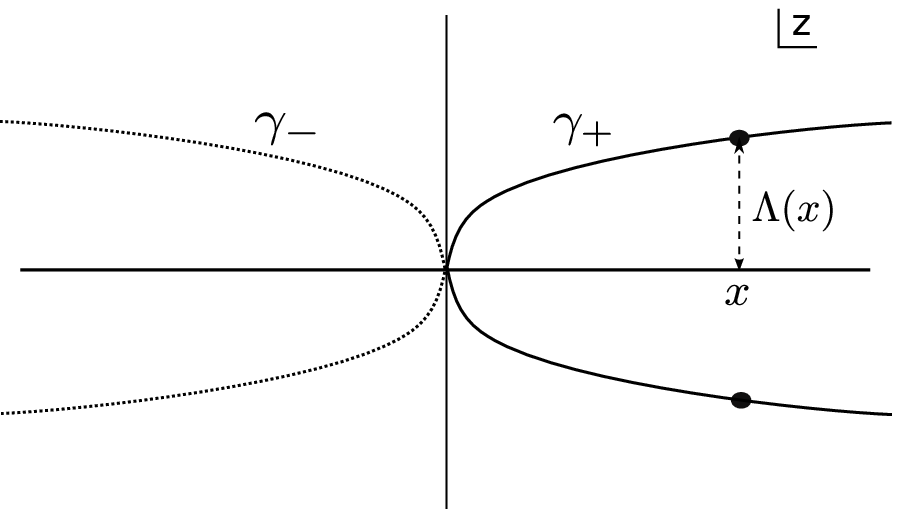}}

Let us define curves $\gamma_{\pm}$ by 
\eqn\gammadef{\eqalign{
&\gamma_{\pm} = \{ z = x+i y; ~ |y| =  \Lambda(x), \pm x \geq 0 \} 
}}
where $\Lambda(x)$ is a positive-definite function of $x$ (in particular it could be constant)
s.t. curves $\gamma_{\pm}$ are smooth (including at $x=0$),
 see \gammacontour.  By $G$ we will denote a complex region outside of $\gamma_{\pm}$
\eqn\Gdef{
G = \{ z = x+i y; ~ |y| \geq \Lambda(x) \}.
}

Then the following theorem is true.

{\bf Theorem II:} Suppose we are given two functions $\rho(\nu), \phi(\nu)$ s.t.
\eqn\rhorho{\eqalign{
\rho(\nu), \phi(\nu) \geq 0 \ .
}}
For $|\nu|>a$ the function $\phi(\nu)$ is smooth, $|\phi(\nu)|$ is monotonically decreasing with $|\nu|$ and locally polynomially bounded
\eqn\phipower{
\alpha |\phi(\nu)| < - |\nu\phi'(\nu)| < \beta |\phi(\nu)|, \qquad |\nu| > a, \quad  \alpha < \beta < 0.
}
Furthermore, the following condition holds
\eqn\ebcond{
\int_{-\infty}^\infty d\nu { \rho(\nu) \over \nu - z} -\int_{-\infty}^\infty d\nu { \phi(\nu) \over \nu - z} = R(z)  , \qquad 
z \in G \ ,
}
where the function $R(z)$ is analytic in the complex region $G$ and
\eqn\Rzinfty{
R(z) = O(|z|^{-\omega}), \qquad |z| \to \infty, \quad z \in G, \quad \omega > m
}
Then Cauchy moments of $\rho$ and $\phi$ are related by ($m=1,2,\dots$)
\eqn\ebth{\eqalign{
 F_{m}(x) &= \Phi_{m}(x) + \sum_{k=1}^m b_k {x^{m-k} \over (m-k)!}  +O(y^{m} \phi(x))  + O(y^m x^{m  -\omega}) \cr 
 &+ O\left(y^{(m+1)} |\phi(-x)| \over x \right)
}}
where $y \equiv \Lambda(x)$ and we defined 
\eqn\Cauchydef{\eqalign{
&F_m(\nu) =   {1\over (m-1)!} \int_0^\nu d\nu' (\nu - \nu')^{m-1} \rho(\nu') \ , \cr 
&\Phi_m(\nu) = {1\over (m-1)!} \int_0^\nu d\nu' (\nu - \nu')^{m-1} \phi(\nu') \ ,
}}
and the constants $b_k$ are determined by $R(z)$.

Before giving a proof of the theorem, let us make two comments in the context of dispersion relations. The spectral density $\rho(\nu)$ is in general not a smooth function and e.g. may contain delta-function components. While the function $\phi(\nu)$ can be thought of as naive spectral density given by a sum of powers. In particular, the condition \phipower\ says that it behaves like a power of $\nu$. 

Both $\phi(\nu)$ and $R(z)$ are defined by the $t$-channel expansion. The splitting of the $t$-channel OPE into $\phi(\nu)$ and $R(z)$ is completely arbitrary.  Every term in the large $z$ expansion of $R(z)$ can be rewritten as a power of $\nu$ term in $\phi(\nu)$. Thus, we can make $\omega$ as large as we like in \Rzinfty.

\subsec{Lemma}

The following lemma will be useful in the proof of the theorem \refs{\Postnikov, \Subhankulov}. Consider a finite part of the contour $\gamma_+$ defined as $\Gamma_x = \{ z'=x'+iy' \in \gamma_+; ~ x'<x \}  $. Then we have 
\eqn\lemma{
{1\over 2\pi i } \int_{\Gamma_x} {dz' \over \nu - z'} = \theta(0<\nu < x)  + {y\over \pi} {\rm Re} {1\over \nu-z} + 
O\left( y^2 \over (\nu-x)^2 + y^2  \right)
} 
where $z = x + iy$ and $ y = \Lambda(x)$. Indeed, we have
\eqn\lemmaproof{\eqalign{
&{1\over 2\pi i } \int_{\Gamma_x} {dz' \over \nu - z'} = \theta(0<\nu < x)  + {1\over 2\pi i} \int_{K} {dz' \over \nu - z'}  = \cr 
&=\theta(0<\nu < x)  + {\nu-x \over \pi}  \int_0^y {dy' \over (\nu-x)^2 +y'^2}  \ ,
}}
where in the first line we added and subtracted an integral over a vertical segment $K = \{ z = x+iy';~ |y'| < y = \Lambda(x) \} $ to close the contour $\Gamma_x$. The integral on the RHS is, of course, easy to do explicitly. However, it will prove useful to estimate it as follows instead
\eqn\fesrint{\eqalign{
&(\nu-x) \int_0^y {dy' \over (\nu-x)^2 +y^2}  = {y(\nu-x) \over (\nu-x)^2 + y^2} + 2(\nu-x) \int_0^y dy'{ y'^2 \over [(\nu-x)^2 +y^2]^2} = \cr 
& =   {y(\nu-x) \over (\nu-x)^2 + y^2} + 2\int_0^{y/(\nu-x)} {dy' \over 1+y'^2} {y'^2 \over 1+y'^2} = \cr
&= {y(\nu-x) \over (\nu-x)^2 + y^2}  +O\left(  {y^2 \over (\nu-x)^2 + y^2}   \int_0^{y/(\nu-x)} {dy' \over 1+y'^2}\right)
}}
where in the first equality we integrated by parts and in the last equality we substituted a monotonically increasing function ${y'^2 \over 1+y'^2} $ by its value at the upper limit. The remaining integral in the third line of \fesrint\ is a bounded function. Therefore, we get \lemma.

The virtues of the formula \lemma\ are twofold. First, it relates the Stieltjes kernel $1\over \nu-z$ to the indicator function $\theta(0<\nu < x)$, needed to obtain Cauchy moments \Cauchydef. Second, the remainder terms on the RHS of \lemma\ are again given by the Stieltjes kernel.\foot{The last term in \lemma\ is the imaginary part of the Stieltjes kernel.} This will allow us to estimate the remainder terms using the original condition \ebcond.

\subsec{First Cauchy Moment}
\noindent We start with the proof of \ebth\ for $m=1$. Integrating \ebcond\ over $\Gamma_x$ and using \lemma\ we obtain
\eqn\Fone{
F_1(x) - \Phi_1(x)  + {y\over \pi} {\rm Re}~ R(z) + 
O\left( y^2 \int_{-\infty}^\infty d\nu { |\rho(\nu) - \phi(\nu)| \over (\nu-x)^2 + y^2 } \right) = 
{1\over 2\pi i } \int_{\Gamma_x} dz' R(z')
}
The $R(z)$ term on the LHS is $O(x^{\e - \omega})$. The RHS is $b_1 + O(|z|^{1-\omega})$. Indeed, for instance the integral over the arc in the upper-half plane is
\eqn\RzRHS{\eqalign{
&\int_0^z dz' R(z')  = \int_0^\Lambda dz' R(z')+ \int_{\Lambda}^z  dz' R(z')  = \cr 
&= \int_0^\Lambda dz' R(z') + \int_{\Lambda}^z  dz' \left( {r_1 \over z'^\omega} + {r_2 \over z'^{\omega+1} } + \dots \right) = const + O(|z|^{1-\omega})
}}
Therefore, \Fone\ becomes 
\eqn\Foneone{\eqalign{
&F_1(x) - \Phi_1(x) - b_1 = O(x^{1-\omega}) + \cr 
&+O\left( y^2 \int_{-\infty}^\infty d\nu { |\rho(\nu) | \over (\nu-x)^2 + y^2 } \right) + 
O\left( y^2 \int_{-\infty}^\infty d\nu { |\phi(\nu) | \over (\nu-x)^2 + y^2 } \right)
}}
First, we estimate the integral with $\phi$
\eqn\phiintest{\eqalign{
&O\left( y^2 \int_{-\infty}^\infty d\nu {|\phi(\nu)| \over (\nu-x)^2 + y^2 } \right) = \cr 
& O\left( y^2\left[ \int_{-\infty}^{-x} + \int_{-x}^{-a} + \int_{-a}^a + \int_a^{x/2} + \int_{x/2}^{x-y} + \int_{x-y}^{x+y} + \int_{x+y}^\infty \right]  d\nu {|\phi(\nu)| \over (\nu-x)^2 + y^2 } \right) = \cr 
& = i_1 + \dots + i_7
}}
Due to monotonicity of $|\phi(\nu)|$ we have
\eqn\ione{\eqalign{
&i_1 = O\left( y^2 \int_{-\infty}^{-x}  d\nu {|\phi(\nu)| \over (\nu-x)^2 + y^2 } \right)  =O\left( y^2 \int_{x}^{\infty}  d\nu {|\phi(-\nu)| \over (\nu+x)^2  } \right)  = O\left( y^{2} |\phi(-x)| \over x \right), \cr 
&i_5 = O\left( y^2 \int_{x/2}^{x-y}  d\nu {|\phi(\nu)| \over (\nu-x)^2 + y^2 } \right)  = O\left( y^2 |\phi(x/2)| \int_{x/2}^{x-y}{d\nu \over (\nu-x)^2 + y^2 } \right) = O(y\phi(x)) \cr 
&i_6 = O\left( y^2 \int_{x-y}^{x+y}  d\nu {|\phi(\nu)| \over (\nu-x)^2 + y^2 } \right) = O\left( \int_{x-y}^{x+y}  d\nu |\phi(\nu)| \right) = O(y \phi(x)), \cr 
&i_7 = O\left( y^2 \int_{x+y}^{\infty}  d\nu {|\phi(\nu)| \over (\nu-x)^2 + y^2 } \right)  = O\left( y^2 \phi(x) \int_{x+y}^\infty {d\nu\over (\nu-x)^2 } \right) = O(y \phi(x))
}}
We also have 
\eqn\ithree{
i_3 = O\left( y^2 \int_{-a}^{a}  d\nu {|\phi(\nu)| \over (\nu-x)^2 + y^2 } \right) = O(y^2/x^2)  
}
Further, to estimate $i_2, i_4$ we use that \phipower\ implies
\eqn\phipowerint{
\left( \mu \over \nu \right)^\alpha < {\phi(\mu) \over \phi(\nu) } < \left( \mu \over \nu \right)^\beta, \qquad |\mu| > |\nu| > a
}
and therefore
\eqn\itwo{\eqalign{
&i_2 =  O\left( y^2 \int_{-x}^{-a}  d\nu {|\phi(\nu)| \over (\nu-x)^2 + y^2 } \right)   =  O\left( y^2 \int_{a}^{x}  d\nu {|\phi(-\nu)| \over (\nu+x)^2 } \right)   = \cr 
& =  O\left( y^2 { |\phi(-x)| \over x^{2+\alpha}}  \int_{a}^{x}  d\nu ~ \nu^\alpha \right)   = O\left( y^{2 } |\phi(-x) |\over x \right) , \cr 
&i_4 = O\left( y^2 \int_{a}^{x/2}  d\nu {|\phi(\nu)| \over (\nu-x)^2 + y^2 } \right)    = O\left( y^2 {\phi(x) \over x^{\alpha+2} }  \int_{a}^{x/2}  d\nu ~ \nu^\alpha \right) = O\left( y^{2 } {\phi(x) \over x}  \right).
}}
Collecting \ione\ - \itwo\ we obtain
\eqn\phiest{
O\left( y^2 \int_{-\infty}^\infty d\nu {|\phi(\nu)| \over (\nu-x)^2 + y^2 } \right) = O\left( y\phi(x)  \right) + O\left( y^{2 } |\phi(-x) | \over x \right) + O(y^2/x^{2}).
}
Finally, we need to estimate the $\rho$ integral in \Foneone. Since $\rho(\nu) \geq 0$ for all $\nu$ we have by taking the imaginary part of \ebcond\ and using \phiest
\eqn\psiest{\eqalign{
&O\left( y^2 \int_{-\infty}^\infty d\nu {|\rho(\nu)| \over (\nu-x)^2 + y^2 } \right)=O\left( y^2 \int_{-\infty}^\infty d\nu {\phi(\nu)\over (\nu-x)^2 + y^2 } \right) + O(y|R(z)|) = \cr 
&=O\left( y\phi(x)  \right) + O\left( y^2 |\phi(-x) | \over x \right) + O(y^2/x^{2}) + O(y x^{- \omega})
}}

Therefore, \phiest\ and \psiest\ imply for \Foneone\ 
\eqn\eblead{
F_1(x) - \Phi(x) - b_1 = O(y \phi(x)) + O\left(y^2 |\phi(-x)|\over x \right) +O(x^{1-\omega} ) \ .
}
This finishes the proof of $m=1$ case of \ebth.

\subsec{Higher Cauchy Moments}
\noindent Integrating \ebcond\ by parts we have
\eqn\Ftwo{
\int_{-\infty}^\infty d\nu {F_1(\nu) - \Phi_1(\nu) - b_1 \over (\nu-z)^2} = R(z) 
}
where we also added $0$ as $b_1$ term in the integral. Integrating this from $z$ to $\infty$ along $\gamma_+$ we get
\eqn\Ftwotaubcond{
\int_{-\infty}^\infty d\nu {F_1(\nu) - \Phi_1(\nu) - b_1 \over \nu-z} = \int_z^\infty dz' R(z' ) = O(|z|^{1-\omega}) 
}
Notice that $F_1(\nu) - \Phi_1(\nu) - b_1 \to 0 $ as $\nu \to \infty$ due to \eblead, so that the integral on the LHS of \Ftwotaubcond\ converges. Integrating \Ftwotaubcond\ over $\Gamma_x$ we obtain
\eqn\Ftwogamma{\eqalign{
&F_2(x) - b_1 x - {y\over \pi} {\rm Re} ~ \int_z^\infty dz' R(z') + 
O\left( y^2 \int_{-\infty}^\infty d\nu {|F_1(\nu) - \Phi_1(\nu) - c_1| \over (\nu-x)^2 + y^2 } \right)  = \cr 
& = - {1\over 2\pi i } \int_{\Gamma_x} dz' \int_{z'}^\infty dz'' R(z'')
}}
Using \eblead\ we estimate similarly to \phiest
\eqn\Otwoest{\eqalign{
&O\left( y^2 \int_{-\infty}^\infty d\nu {|F_1(\nu) - \Phi_1(\nu) - b_1| \over (\nu-x)^2 + y^2 } \right) = \cr 
&= O\left( y^2 \int d\nu {|L(\nu) \phi(\nu)| + |L(\nu)^2 \nu^{-1} \phi(-\nu)| + |\nu|^{1-\omega} \over (\nu-x)^2 + y^2 } \right)  = \cr 
& = O(y^2\phi(x)) +O\left(y^3 |\phi(-x)| \over x \right) +O(y x^{1-\omega}) .
} }
Therefore \Ftwogamma\ gives
\eqn\ebtwo{
F_2(x) - \Phi_2(x)  -b_1 x - b_2= O(y^2\phi(x)) + O\left(y^3|\phi(-x)| \over x \right) +O(x^{2-\omega} ) \ ,
}
where the constant $b_2$ comes from the finite $u$ part of the integral on the RHS of \Ftwogamma, similarly to \RzRHS. This proves \ebth\ for $m=2$.
Iterating this argument we obtain the tauberian theorem \ebth\ for all $m=1,2,\dots$.

\subsec{Odd Densities}

The theorem above is not quite what we need in bootstrap applications. Instead we would like to consider parity odd densities that satisfy
\eqn\odddens{
\rho(-\nu) = - \rho(\nu), ~~~ \phi(-\nu) = - \phi(\nu) \ .
}
Most of the proof goes intact apart from application of the tauberian condition \ebcond\ in \psiest. Indeed in this case we have
\eqn\newthing{
y^2 \int_{-\infty}^\infty d\nu {|\rho(\nu)| \over (\nu-x)^2 + y^2 } = y^2 \int_{-\infty}^\infty d\nu {\rho(\nu) \over (\nu-x)^2 + y^2 } + 2 y^2 \int_{-\infty}^0 d \nu {-\rho(\nu) \over (\nu-x)^2 + y^2 } .
}
For the first term we could use the estimates above but the second term should be estimated separately. We get 
\eqn\estimate{\eqalign{
&y^2 \int_{-\infty}^0 d \nu {-\rho(\nu) \over (\nu-x)^2 + y^2 }=O(y^2/x^2) +  O \left( y^2  \int_0^\infty d \nu {\nu \rho(\nu) \over (\nu+x)^2 + y^2} \right) \cr
&= O \left( y^2  \int_0^\infty d \nu {\nu \rho(\nu) \over \nu^2 + x^2} \right) + O(y^2/x^{ 2})  ,
}}
where we used the fact that $\nu, x>0$ and \odddens. Note that 
\eqn\estB{
\int_{-\infty}^{\infty} d \nu {\rho(\nu) \over \nu - i x} = \int_0^{\infty} d \nu \rho(\nu) \left( {1 \over \nu - i x} + {1 \over \nu + i x}  \right) = 2 \int_0^\infty d \nu {\rho(\nu) \nu \over \nu^2 + x^2 } \ ,
}
where we again used \odddens .

Therefore we have
\eqn\estimateC{
y^2 \int_{- \infty}^0 d \nu {- \rho(\nu) \over (\nu-x)^2 + y^2} = O \left( y^2  \int_{-\infty}^{\infty} d \nu {\rho(\nu) \over \nu - i x}  \right) .
}

To estimate this we can use the tauberian condition for $z = i x$. Therefore we get
\eqn\estimateD{
y^2 \int_{- \infty}^0 d \nu {- \rho(\nu) \over (\nu-x)^2 + y^2} = O \left( y^2  | R(i x) | \right)   + O \left( y^2 \int_{-\infty}^{\infty} d \nu {\phi(\nu) \over \nu - i x}  \right) .
}

Now we can estimate the last integral
\eqn\lastestim{\eqalign{
&O \left( y^2  \int_{-\infty}^{\infty} d \nu {\phi(\nu) \over \nu - i x}  \right) = O \left( y^2 \int_{0}^{\infty} d \nu {\phi(\nu) \nu \over \nu^2 + x^2}  \right) \cr
&= i_1 + i_2 + i_3 \ ,
}}
where we split the integral into $\int_0^{a} + \int_{a}^{x} + \int_{x}^\infty$.  Let us estimate each integral using the usual techniques
\eqn\estione{\eqalign{
i_1 &= O \left( y^2 \int_{0}^{a} d \nu {\phi(\nu) \nu \over \nu^2 + x^2}  \right) = O({y^2 / x^2})  , \cr
i_2 &=O \left( y^2 \int_{a}^{x} d \nu {\phi(\nu) \nu \over \nu^2 + x^2}  \right) = O\left({y^2 \phi(x) \over x^{2+\alpha}} \int_a^x d \nu \nu^{1+\alpha} \right) = O(y^2 \phi(x) ) , \cr
i_3 &=O \left( y^2 \int_x^{\infty} d \nu {\phi(\nu) \nu \over \nu^2 + x^2}   \right) = O\left({y^2 \phi(x) \over x^\beta} \int_x^{\infty} d \nu {\nu^{1+\beta} \over \nu^2 + x^2}\right) =  O( y^2  \phi(x) )
}}
Thus, we get the following estimate 
\eqn\estimatefinal{
y^2 \int_{-\infty}^0 d \nu {-\rho(\nu) \over (\nu-x)^2 + y^2 } = O( y^2 x^{  - \omega} ) + O(y^2/x^{ 2} ) +  O(y^2 \phi(x) ) \ .
}
The conclusion is that the estimate in this case takes the form
\eqn\eblead{
F_1(x) - \Phi(x) - b_1 = O(y^2\phi(x)) + O(x^{1-\omega} ) \ ,
}
where as usual $y \equiv \Lambda(x)$. For higher $m$ the argument is identical the one discussed in appendix B.3.  This theorem is what we leads to the statement \ebstilt.

\subsec{Construction of $\Phi_m$}

Let us understand better how to construct $\Phi_m$. Consider the following ansatz for the subtraction density
\eqn\ansatz{
\rho_{J}^{naive}(\nu) =\theta( 0 < \nu < 1) \sum_{i} \tilde \alpha_i \nu^i + \theta(\nu > 1) \sum_{i} \alpha_i \nu^{- \delta_i - 1} {\cos {\pi \delta_i \over 2} \over \pi} .
}
It has the following large $\nu$ expansion
\eqn\dispexpSUB{ \eqalign{
\int_0^{\infty} d \nu' \  \rho_{J}^{naive}(\nu) {2 \nu' \nu \over  \nu'^2 + \nu^2} &= \sum_i \alpha_i \left( \nu^{- \delta_i} + {2 \cos {\pi \delta_i \over 2} \over \pi} \sum_{k=0}^{\infty} {(-1)^k \over (\delta_i -1-2k ) \nu^{1+2k}} \right) \cr
&+\sum_{i} \tilde \alpha_i \sum_{k=0}^{\infty} {(-1)^k \over (i + 2 + 2k) \nu^{1+2k}} \ .
}}
This takes care of all non-integer powers that appear in the OPE. In addition, it contributes to the non-universal terms in the dispersion relations, namely to ${1 \over \nu^{1 + 2k}}$ in \dispexp. To cancel those we can add terms $\theta( 0 < \nu < 1) \tilde \alpha_i \nu^i$ to the naive density and fix the coefficients so that they cancel the RHS of \dispexp\ to any required order. We can then apply the theorem and compute $\Phi_m(\nu)$. Note that non-integer powers of $\nu$ that we are interested at only come from the term $\nu^{- \delta_i}$. One can use the explicit form of $\phi(\nu)$ \ansatz\ to check estimates from the proof as well as to analyze the contribution of operators that produce $\log \nu$ terms.

A careful reader might have noticed that $\phi(\nu) = \rho_{J}^{naive}(\nu) $ above is not necessarily positive for $0<u<1$, whereas in the assumptions of the theorems we assumed that it is. It is trivial to check that the behavior of $\phi(\nu)$ on a finite interval is completely immaterial for the proof apart from clattering it a bit, see again \refs{\SubhankulovR, \Subhankulov} . 

\subsec{Case With Subtractions}

Similarly, we need a version of the theorem for the case with subtractions \dispthree. It is useful to consider the following identity \Subhankulov
\eqn\identity{
\int_0^z d \tilde z \ {(z- \tilde z)^{n-1} \over (\nu - \tilde z)^{n+1}}  = {1 \over n} \left({z \over \nu}\right)^n {1 \over \nu -  z} \ .
}
Let us rewrite the dispersion relation \dispthree\ as follows
\eqn\rewriteDR{
\int_{- \infty}^\infty d \nu {\rho(\nu) \over (\nu - z)^{N+1} }  = {1 \over \Gamma(N+1)} \p_{z}^N c_J(z) ,
}
where we defined the full density at negative $\nu$ through $\rho(-\nu) = - \rho(\nu)$. Applying \identity\ to \rewriteDR\ we get

\eqn\newDR{
\int_{- \infty}^\infty d \nu {\rho(\nu) \over \nu^N (\nu - z)} = {1 \over z^N} c_J(z) - \sum_{k=0}^{N-1} {1 \over z^{N-k} (k)!} \p_z^{k} c_J(0)
}
where only even $k$ contribute since $c_J(z)$ is an even function of $z$. Therefore, for even $N$ we get the following structure of dispersion relations at large $z $

\eqn\structureNEW{
\int_{- \infty}^\infty d \nu {\rho^{OPE}(\nu) \over \nu^N (\nu - z)} = \sum_{i} \alpha_i z^{-N-\delta_i} + \sum_{k=1}^{\infty} d_k z^{-2k} \ ,
}
where again $\alpha_i$ are computable in terms of the OPE and $\rho^{OPE}(-\nu) = -\rho^{OPE}(\nu)$. The naive spectral density that will reproduce the RHS is of course exactly the same as before. 

In principle, one can try to derive a separate tauberian theorem for the kernel \structureNEW, see e.g. \Subhankulov. For us however it suffices to consider $\tilde \rho (\nu)= {\rho^{OPE}(\nu) \over \nu^N}$ and apply the theorem II. Note also that we have the following identity between the moments of $\rho$ and $\tilde \rho$
\eqn\identityU{
F_{m}(\nu) =(-1)^N {(N+m-1)! \over (m-1)!} \tilde G_{N+m,N}(\nu),
}
where recall that $G_{m,k}$ was defined in \combin. This completes the consideration of the case with subtractions.

The symmetry of $\tilde \rho$ is different for odd and even $N$'s $\tilde \rho(-\nu) = (-1)^{N+1} \tilde \rho(\nu) $. For even $N$ we can directly use the formulas from our analysis of the unsubtracted dispersion relations. Consider a term $\alpha_i z^{-N-\delta_i}$ in \structureNEW. Using \contribution\ and \identityU\ (and being careful about factors of $i$ in \structureNEW, \dispthree) we recover the result \ebstilt. For odd $N$ the conclusion is the same. It had to be the case by self-consistency of the whole construction, namely for a given external scaling dimension we could have considered dispersion relations with different numbers of subtractions, but this should not affect the result for the leading asymptotic. This is indeed the case.

\appendix{C}{Tauberian Optimality Example}

Here we would like to understand properties of the Laplace transform in the complex $\beta$ plane of the spectral density \subkhint
\eqn\density{
f(E) = \left( 1 + \sin [(\log E)^2 ] \right) \theta(E -1) .
}
The relevant integral to study is the following
\eqn\Lbeta{
{\cal L}(\beta) = \int_0^\infty dE e^{-\beta E} f(E) = {e^{-\beta} \over \beta} + \int_1^\infty dE ~ e^{-\beta E} \sin [(\log E)^2 ] .
}

To analyze the second integral it is convenient to use the standard Mellin representation for $e^{- \beta E}$
\eqn\cahen{
e^{-x} = \int_{- \eps - i \infty}^{- \eps + i \infty} {d \delta \over 2 \pi i} \Gamma( - \delta ) x^{\delta} \ , ~~~ \eps > 0 .
}
Convergence of this integral depends on the argument of $x$. Denoting $\delta = \delta_R + i \delta_I$ we get the asymptotic behavior
\eqn\limitcahen{
\Gamma(- \delta) x^{\delta} \sim |\delta_I|^{\delta_R - 1/2} e^{\delta_I {\rm arg}[x]- {\pi \over 2} |\delta_I|} .
}
In particular, the integral converges only for $|{\rm arg}[x] |<{\pi \over 2}$. 

We then have for the second term in \Lbeta\ ($x = \log E$)
\eqn\integralSov{\eqalign{
& \int_1^\infty dE ~ e^{-\beta E} \sin [(\log E)^2 ]  = \int_0^\infty dx ~ e^x \sin (x^2) e^{-\beta e^x} = \cr 
& =  \int_{-\eps - i \infty}^{-\eps + i \infty} {d\delta \over 2\pi i} \Gamma(- \delta) \beta^\delta \int_0^\infty dx ~e^{(\delta + 1)x} \sin(x^2) =  \cr 
& ={1\over 2} \sqrt{\pi \over 2} \int_{-\eps - i \infty}^{-\e + i \infty} {d\delta \over 2\pi i} \Gamma(- \delta) \beta^\delta \times \cr 
& \times \left[ \cos\left( (\delta + 1)^2 \over 4 \right) \left( 1 + 2 C\left( \delta + 1 \over \sqrt{2\pi} \right) \right) + 
 \sin\left( (\delta + 1)^2 \over 4 \right) \left( 1 + 2 S\left( \delta + 1 \over \sqrt{2\pi} \right) \right)    \right] ,
}}
where $C(x)$, $S(x)$ denote the corresponding Fresnel integrals
\eqn\Fresnel{
C(z) \equiv \int_0^{z} d t \cos {\pi t^2\over 2}, ~~~ S(z) \equiv \int_0^{z} d t \sin {\pi t^2\over 2} \ .
}
In doing the $x$ integral we assumed $\eps > 1$ for convergence. The result however is regular at $\eps = -1$ and we can shift the $\delta$ contour to estimate the asymptotic of the integral \Lbeta. 

Let us analyze the convergence properties of the integral \integralSov\ in the complex $\beta$ plane. We set $\beta = |\beta| e^{i \phi}$. The danger is that now we have an extra $e^{i \phi \delta}$ factor which blows up in the lower half-plane.  Evaluating the asymptotic of the integrand we get
\eqn\asympt{
e^{- \phi \delta_I} e^{- {\pi \over 2} |\delta_I |} e^{{(1+\delta_R) |\delta_I | \over 2}} |\beta|^{\delta_R+ i \delta_I} .
}
The integral converges for
\eqn\convergenceSov{
\phi + {1+ \delta_R \over 2}  < {\pi \over 2} \ .
}
Therefore we cannot evaluate the integral by simply shifting the contour to the right, namely increasing $\delta_R$.

It does however allow us to evaluate the leading asymptotic for $\phi = 0$. From the first three poles at $\delta = 0,1,2$ we get 
\eqn\integralest{\eqalign{
&\int_1^\infty dE ~ e^{-\beta E} \sin [(\log E)^2 ]  = c_0 + c_1 \beta + c_2 \beta^2 + \dots , \cr
c_0 &= {1\over 2} \sqrt{\pi \over 2}\left[ \cos\left( 1 \over 4 \right) \left( 1 + 2 C\left(  1 \over \sqrt{2\pi} \right) \right) + 
 \sin\left( 1 \over 4 \right) \left( 1 + 2 S\left(  1 \over \sqrt{2\pi} \right) \right)    \right] , \cr
}}
and similar expressions for $c_1$ and $c_2$.
Pushing contour to further poles is not possible because it would violate \convergenceSov\ $\delta_R^{max} = \pi -1$. As we increase $\phi$ the range of maximal allowed $\delta_R$  decreases. Therefore we cannot use the estimate above anymore. By studying numerically the integral we observed that the next term in the expansion is of the type $\beta^{\pi - 1} \cos (\log \beta)^2 L(\beta)$ where $L(\beta)$ is slowly varying function. We also observed that this term captures the asymptotic behavior correctly for complex $\beta$ as well. As we increase the argument of $\beta$ the term $\beta^{\pi - 1} \cos (\log \beta)^2 L(\beta)$ becomes dominant and as $\beta$ becomes imaginary it behaves as ${1 \over \beta} L( i \beta)$ which is again consistent with our numerical observations.

\listrefs
\bye